\documentclass[a4paper, 12pt]{article}

\usepackage[english]{babel}

\usepackage[margin=1in]{geometry}

\usepackage{amsmath}
\usepackage{amssymb}
\usepackage{amsthm}
\usepackage{thmtools}
\usepackage{thm-restate}
\usepackage{graphicx}
\usepackage[colorlinks=true, allcolors=blue]{hyperref}
\usepackage[capitalise]{cleveref}
\usepackage{xspace}
\usepackage[dvipsnames]{xcolor}
\usepackage{tikz}
\usepackage{enumerate}
\usepackage{subcaption}
\usepackage{booktabs}
\usetikzlibrary{arrows.meta}
\usetikzlibrary{calc}

\definecolor{myred}{RGB}{213,0,0}
\definecolor{mygray}{RGB}{187,187,187}
\definecolor{myorange}{RGB}{230,159,0}
\definecolor{myblue}{RGB}{0,114,178}
\definecolor{mygreen}{RGB}{0,158,115}

\newtheorem{theorem}{Theorem}
\newtheorem{lemma}[theorem]{Lemma}
\newtheorem{corollary}[theorem]{Corollary}
\newtheorem{observation}[theorem]{Observation}
\newtheorem{definition}[theorem]{Definition}
\newtheorem{claim}[theorem]{Claim}
\newtheorem{remark}[theorem]{Remark}

\newenvironment{claimproof}[1][\proofname]{
  
  \begin{proof}[#1]}{\end{proof}}

\newcommand{\opt}{\text{opt}\xspace}
\newcommand{\kopt}{\text{$k$-\opt}\xspace}

\newcommand{\R}{\mathbb{R}}

\newcommand{\C}{\mathcal{C}}
\newcommand{\calS}{\mathcal{S}}

\newcommand{\cv}{\beta}
\newcommand{\ce}{\alpha}
\newcommand{\cs}{\mu}
\newcommand{\cn}{\nu}
\newcommand{\vect}[1]{\mathbf{#1}}
\newcommand{\labv}{\boldsymbol{\ell}}
\DeclareMathOperator{\bin}{\{0,1\}}
\DeclareMathOperator{\bins}{\{1,2\}}

\newcommand{\vg}{vertex\xspace}
\newcommand{\eg}{path\xspace}
\newcommand{\sg}{star\xspace}
\renewcommand{\ng}{node\xspace}
\newcommand{\Vg}{Vertex\xspace}
\newcommand{\Eg}{Path\xspace}
\newcommand{\Sg}{Star\xspace}
\newcommand{\Ng}{Node\xspace}
\newcommand{\br}[1]{\bar{#1}}

\definecolor{defblue}{rgb}{0, 0.4, 0.796}
\newcommand{\defi}[1]{\textcolor{defblue}{\emph{#1}}}

\title{A Near-Complete Resolution of the Exponential-Time Complexity of \kopt for the Traveling Salesman Problem}
\author{Sophia Heimann\thanks{Hertz Chair for Algorithms and Optimization, University of Bonn, Germany (sheimann@uni-bonn.de)}, 
        Hung P. Hoang\thanks{Algorithms and Complexity Group, Faculty of Informatics, TU Wien, Austria, (phoang@ac.tuwien.ac.at) funded by the Austrian Science Foundation (FWF, project Y1329 START-Programm)}, 
        Stefan Hougardy\thanks{Research Institute for Discrete Mathematics and Hausdorff Center for Mathematics, University of Bonn, Germany (hougardy@or.uni-bonn.de) funded by the Deutsche Forschungsgemeinschaft (DFG, German Research Foundation) under Germany's Excellence Strategy -- EXC-2047/1 -- 390685813}}

\begin{document}
\maketitle

\begin{abstract}
The \kopt algorithm is one of the simplest and most widely used heuristics for solving the traveling salesman problem. Starting from an arbitrary tour, the \kopt algorithm improves the current tour in each iteration by exchanging up to $k$ edges. The algorithm continues until no further improvement of this kind is possible.
For a long time, it remained an open question how many iterations the \kopt algorithm might require for small values of $k$, assuming the use of an optimal pivot rule. In this paper, we resolve this question for the cases $k = 3$ and $k = 4$ by proving
that in both these cases an exponential number of iterations may be needed even if an optimal pivot rule is used. Combined with a recent result from~\cite{HHH2024}, this provides a complete answer for all $k \ge 3$ regarding the number of iterations the \kopt algorithm may require under an optimal pivot rule. In addition we establish an analogous exponential lower bound for the $2.5$‑\opt algorithm, a variant that generalizes $2$‑\opt and is a restricted version of $3$‑\opt. All our results hold for both the general and the metric traveling salesman problem.
 \end{abstract}

\section{Introduction}
Local search is a general and powerful technique for solving combinatorial optimization problems.
The idea is simple: start with an arbitrary solution and, in each iteration, improve the current solution 
through small local changes. Once no further such improvement is possible, return the current solution.
Prominent examples of successful applications of local search include the simplex algorithm for linear programming~\cite{Dan1990}, the \kopt algorithm for the traveling salesman problem~\cite{Lin1965}, and the multiple-swap heuristic for $k$-means clustering~\cite{KMNPSW2004}.

Despite their simplicity, local search algorithms can be difficult to analyze. 
The simplex algorithm, for instance, is remarkably efficient in practice, yet
it is a longstanding open problem in the theory of linear programming, 
whether any pivot rule for the simplex algorithm can guarantee polynomial running time. 
For many natural pivot rules, the worst-case running time of the simplex algorithm is known to be exponential~\cite{KM1972,Jer1973,AC1978,GS1979,DFH2023,Bla2025}.

In this paper, we study the running time of the \kopt algorithm for the traveling salesman problem (TSP). Given an edge‑weighted complete graph, the TSP asks for a Hamiltonian cycle--also called a tour--of minimum total length. The problem is NP‑hard and remains so for the well‑studied metric TSP, where the edge weights satisfy the triangle inequality~\cite{Kar1972}. To solve the TSP in practice, the 
\kopt algorithm is one of the most commonly used heuristics. 
Starting from an arbitrary tour, it iteratively improves the tour by exchanging up to $k$ edges,
terminating when a local optimum is reached, i.e., when no such improvement is possible.
In 1956 Flood~\cite{Flo1956} was the first who proposed the $2$-\opt algorithm as an effective method 
for solving traveling salesman problems. Johnson and McGeoch~\cite{JM2002} conducted an 
extensive experimental study, showing that $2$-\opt and $3$-\opt can solve random Euclidean instances with $100{,}000$ points within seconds, producing solutions that on average are $4.8\%$, respectively $3.0\%$ away from an optimum solution. For instances with $n$ points, they observe a running time of $O(n \log^2 n)$ to reach a local optimum. Thus, the situation here is similar to the simplex algorithm: We observe very good running times in practice and would like to show that this can be justified theoretically. 
Similar as in the case of the simplex algorithm it is known that there exist pivot rules for the \kopt
algorithm that require exponential running time for all $k\ge2$~\cite{CKT1999, Englert2014}. However, 
to obtain these results artificially bad pivot rules have been designed. 
It remains the question if polynomial running time can be obtained for \kopt if a better 
pivot rule is chosen, e.g., steepest descent. In this paper we will show 
that the answer to this question is \emph{no} for $k\ge 3$ and all possible pivot rules. 
To formulate this statement more precisely, we use the notion of the all‑exp property.

A local search algorithm has the \emph{all-exp property} if there exist infinitely many pairs of instances and initial solutions for which the algorithm requires an exponential number of iterations to reach a local optimum--regardless of the pivot rule used.
The simplex algorithm cannot have the all-exp property; this follows from a result of Kalai and Kleitman~\cite{KK1992}. For the \kopt algorithm the all-exp property has been proven in 1989 by 
Krentel~\cite{Kre1989} if $k$ is sufficiently large. 
Hoang and Hougardy~\cite{HH2023} showed that Krentel's proof works for all $k \ge 14{,}208$.
In 2024, Heimann, Hoang, and Hougardy~\cite{HHH2024} significantly improved this bound, showing that the \kopt algorithm has the all-exp property for all $k \ge 5$. This leaves exactly those cases open that are the most interesting ones from a practical point of view, namely the cases $k=2$, $k=3$, and $k=4$.

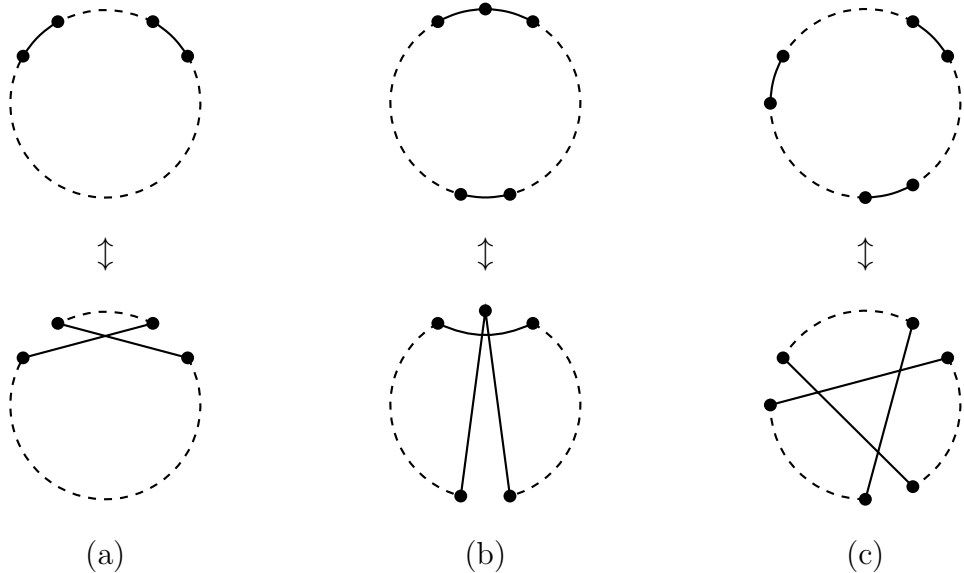
\begin{figure}
    \centering
    \begin{tikzpicture}[vertex/.style = {circle,fill=black,inner sep=0.6mm},
                        dashededge/.style={dashed, thick},     
                        solidedge/.style={thick}]     
        \def\radius{1.25} 
    
        \begin{scope}
            \foreach \i/\j in {1/1,  2/2, 3/4, 4/5} {
                \coordinate (P\i) at ({\radius*cos(30*\j)}, {\radius*sin(30*\j)});
                \node[vertex] at (P\i) {};
            }
            \draw[solidedge] (P1) to[bend right=12.5] (P2);
            \draw[solidedge] (P3) to[bend right=12.5] (P4);
            \draw[dashededge] (P2) to[bend right=25] (P3);
            \draw[dashededge] (30:\radius) arc[start angle=30, end angle=-210, radius=\radius];
            \draw (0,-2) node {$\updownarrow$};
        \end{scope}

        \begin{scope}[shift={(0,-4)}]
            \foreach \i/\j in {1/1,  2/2, 3/4, 4/5} {
                \coordinate (P\i) at ({\radius*cos(30*\j)}, {\radius*sin(30*\j)});
                \node[vertex] at (P\i) {};
            }
            \draw[solidedge] (P1) -- (P3) (P2) -- (P4);
            \draw[dashededge] (P2) to[bend right=25] (P3);
            \draw[dashededge] (30:\radius) arc[start angle=30, end angle=-210, radius=\radius];
            \draw (0,-2) node {(a)};
        \end{scope}

        \begin{scope}[shift={(5,0)}]
            \foreach \i/\j in {1/2,  2/3, 3/4, 4/-2.5, 5/-3.5} {
                \coordinate (P\i) at ({\radius*cos(30*\j)}, {\radius*sin(30*\j)});
                \node[vertex] at (P\i) {};
            }
            \draw[solidedge] (P1) to[bend right=12.5] (P2) to[bend right=12.5] (P3);
            \draw[solidedge] (P5) to[bend right=12.5] (P4);
            \draw[dashededge] (60:\radius) arc[start angle=60, end angle=-75, radius=\radius];
            \draw[dashededge] (120:\radius) arc[start angle=120, end angle=255, radius=\radius];
            \draw (0,-2) node {$\updownarrow$};
        \end{scope}

        \begin{scope}[shift={(5,-4)}]
            \foreach \i/\j in {1/2,  2/3, 3/4, 4/-2.5, 5/-3.5} {
                \coordinate (P\i) at ({\radius*cos(30*\j)}, {\radius*sin(30*\j)});
                \node[vertex] at (P\i) {};
            }
            \draw[solidedge] (P4) -- (P2) -- (P5);
            \draw[solidedge] (P1) to[bend left=25] (P3);
            \draw[dashededge] (60:\radius) arc[start angle=60, end angle=-75, radius=\radius];
            \draw[dashededge] (120:\radius) arc[start angle=120, end angle=255, radius=\radius];
            \draw (0,-2) node {(b)};
        \end{scope}

        \begin{scope}[shift={(10,0)}]
            \foreach \i/\j in {1/1,  2/2, 3/5, 4/6, 5/9, 6/10} {
                \coordinate (P\i) at ({\radius*cos(30*\j)}, {\radius*sin(30*\j)});
                \node[vertex] at (P\i) {};
            }
            \draw[solidedge] (P1) to[bend right=12.5] (P2) (P3) to[bend right=12.5] (P4) (P5) to[bend right=12.5] (P6);
            \draw[dashededge] (30:\radius) arc[start angle=30, end angle=-60, radius=\radius];
            \draw[dashededge] (-90:\radius) arc[start angle=-90, end angle=-180, radius=\radius];
            \draw[dashededge] (60:\radius) arc[start angle=60, end angle=150, radius=\radius];
            \draw (0,-2) node {$\updownarrow$};       
        \end{scope}

        \begin{scope}[shift={(10,-4)}]
            \foreach \i/\j in {1/1,  2/2, 3/5, 4/6, 5/9, 6/10} {
                \coordinate (P\i) at ({\radius*cos(30*\j)}, {\radius*sin(30*\j)});
                \node[vertex] at (P\i) {};
            }
            \draw[solidedge] (P1) -- (P4) (P2) -- (P5) (P3) -- (P6);
            \draw[dashededge] (30:\radius) arc[start angle=30, end angle=-60, radius=\radius];
            \draw[dashededge] (-90:\radius) arc[start angle=-90, end angle=-180, radius=\radius];
            \draw[dashededge] (60:\radius) arc[start angle=60, end angle=150, radius=\radius];
            \draw (0,-2) node {(c)};
        \end{scope}
    \end{tikzpicture}
    \caption{All three local changes are valid for 3-\opt, but only (a) is valid for 2-\opt. 2.5-\opt allows (a) and (b) but not (c). In other words, each iteration of 2-, 2.5-, and 3-\opt changes edges incident to at most 4, 5, and 6 vertices, respectively. }
    \label{fig:opt}
\end{figure}

\subsection{Our contribution}
As Yannakakis observed in~\cite{Yan1997}, commenting on Krentel’s proof: “Further reductions to say $k = 2$ or $k = 3$ would require substantially new ideas.” 
In this paper, we present such ideas that allow us to resolve two of the three remaining open cases:

\begin{restatable}[Main Theorem 1]{theorem}{thmmain}
    The $3$-\opt algorithm and the $4$-\opt algorithm for the traveling salesman problem have the all-exp property.
    \label{thm:main}    
\end{restatable}

As observed by Krentel~\cite{Kre1989} any instance of the traveling salesman problem can be made metric by adding a sufficiently large constant to all edge weights. This will not change the behavior of the \kopt algorithm. Together with the result~\cite{HHH2024} we therefore
resolve the all-exp property for \kopt with respect to $k$ almost completely:
 
\begin{theorem}
    For constant $k\ge 3$ the \kopt algorithm for the metric traveling salesman problem has the all-exp property.
\end{theorem}

Thus, only the case $k = 2$ remains open. 
The approach that we use to prove~\cref{thm:main} for $k = 3$ and $k = 4$ does not appear to extend to $k = 2$. 
However, with substantial changes to our reduction, we can prove the all‑exp property for 2.5‑\opt.
The $2.5$-\opt algorithm has been introduced by Bentley in 1992~\cite{Ben1992} (called $2H$-\opt there)
as a special case of $3$-\opt and a generalization of $2$-\opt. 
In $2.5$-\opt three edges are allowed to be exchanged but two of them need to be incident.
See~\cref{fig:opt} for an illustration of $2$-\opt, $2.5$-\opt, and $3$-\opt. In practice, $2.5$-\opt produces better tours than $2$‑\opt without incurring the running time overhead of $3$‑\opt~\cite{JG1997}.
We can show that also this restricted version of $3$-\opt has the all-exp property:

\begin{restatable}[Main Theorem 2]{theorem}{thmmaintwo}
    The $2.5$-\opt algorithm for the traveling salesman problem has the all-exp property.
    \label{thm:main2}    
\end{restatable}

With the same argument as above, this result also holds for the metric TSP. 

Our proofs of~\cref{thm:main} and~\cref{thm:main2} use as a starting point the all-exp result of Michel and Scott~\cite{Michel_Max_Cut_4} for the Max-Cut problem in graphs of maximum degree 4 with the flip-neighborhood.
This result was also the starting point in~\cite{HHH2024} to prove the all-exp property 
for \kopt for $k\ge 5$. However, it seems to be impossible, to achieve a value better than $k \geq 5$ with the reduction used in~\cite{HHH2024}. We will present a different and arguably simpler construction that 
works for $k \in \{3,4\}$. In addition we show that our result leads to a simpler proof for $k\ge 5$.
Compared to~\cite{HHH2024} a crucial difference in our construction is that we do not explicitly model  
to which side of the cut a vertex in a Max-Cut instance belongs to. Instead we model this implicitly in our construction: This results in fewer edges that have to be exchanged in the corresponding tours. 
We need some additional effort to also settle the case $k=2.5$. Especially, we will exploit 
the ``irregular behavior'' of some gadgets used in the construction. In all preceding
constructions this irregular behavior would invalidate the proof. But here we found a way to profit from it. 
Moreover, we extend the result of Michel and Scott~\cite{Michel_Max_Cut_4} for Max-Cut to instances with a certain star decomposition structure and make use of an explicit characterization of how the cut evolves throughout any improving sequence for these instances.  

\subsection{Related work} To formalize the question of finding local optima Johnson, Papadimitriou, and Yannakakis~\cite{JPY1988} introduced in 1988 the class PLS (polynomial-time local search). They defined a corresponding PLS-reduction and established the notion of PLS-completeness as a measure of the hardness of
finding local optima. Later, Papadimitriou, Schäffer, and Yannakakis~\cite{PSY1990} defined a stronger reduction, called tight PLS-reduction, and showed that the tight PLS-completeness of a problem implies the all-exp property for its standard local search algorithm. Thus, proving the tight PLS-completeness of a local search problem is a possible way to establish the all-exp property for a local search algorithm. However, results obtained this way are usually weaker than direct proofs of the all-exp property. For example, in case of the Max-Cut problem with the flip-neighborhood tight PLS-completeness is known for graphs of unbounded maximum degree~\cite{SY1991} while
the all-exp property is known for graphs of maximum degree four~\cite{Monien_Max_Cut_4, Michel_Max_Cut_4}.
Similarly, in~\cite{HHH2024} the PLS-completeness of \kopt was proven for $k\ge 17$ while the all-exp property has been established for $k\ge 5$. 

A possible way to explain the good running time of \kopt in practice is provided by results on its smoothed complexity. 
In this model Englert, Röglin, and Vöcking~\cite{ERV2016} proved a polynomial bound on the expected number of 
iterations that $2$-\opt needs to reach a local optimum. Their result has been improved for $2$-\opt in the Euclidean case by Manthey and van Rhijn~\cite{MR2025}. Giannakopoulos, Grosz, and Melissourgos~\cite{GGM2024, GGM2025} extended these results showing a polynomial smoothed complexity of \kopt for $k \ge 3$.

Besides its running time, the approximation ratio of a local search algorithm is also highly relevant.
In case of the \kopt algorithm we almost have a complete picture: $2$-\opt for metric TSP on $n$ vertices has approximation ratio $\sqrt{n/2}$ 
and this result is tight~\cite{HZZ2020}. For constant $k\ge 3$ the \kopt algorithm for metric TSP has approximation ratio 
$O(n^{1/k})$ and this result is tight if the Erd\H os girth conjecture holds~\cite{Zho2020}. 
For Euclidean TSP the approximation ratio of the \kopt algorithm is $\Theta(\log n / \log \log n)$
for all constant $k\ge 2$~\cite{BHZ2023}.

\subsection{Outline of the paper} 
Section~\ref{sec:prelims} introduces the necessary background, including key terms and notations.
Section~\ref{sec:overall_construction} then provides a high-level overview of our constructions.
The next two sections are dedicated to the all-exp property of the 3- and 4-\opt algorithms: Section~\ref{sec:construction_3} details the construction, while Section~\ref{sec:correctness_3} establishes its correctness.
Section~\ref{sec:k_5} discusses how our construction above can also be used to show the all-exp property for $k \geq 5$, serving as an alternative proof to the result in~\cite{HHH2024}.
The proof for $k=2.5$ follows a structure similar to that for $k \in \{3,4\}$: we start with the construction in Section~\ref{sec:construction_2_5} and follow with the correctness in Section~\ref{sec:correctness_2_5}.
Finally, Section~\ref{sec:conclusion} presents concluding remarks.

\section{Preliminaries}
\label{sec:prelims}
We write $[n]$ for the set $\{1, \dots, n\}$.
For a vector $\vect{s}$ and an index $i$, we denote by $\vect{s}[i]$ the value at the $i$th coordinate of $\vect{s}$.

We denote the degree of a vertex $x$ in a graph $G$ with $d_G(x)$.
The \defi{girth} of $G$ is the size of the smallest cycle in $G$.
A $uv$-path is a path that has the vertices $u$ and $v$ as endpoints.
We call a path of length $t$ (i.e., consisting of $t$ edges) a \defi{$t$-path}.
A \defi{2-path 2-cover} of a graph $G$ is a set of edge disjoint paths such that its union is $G$, each path has length at most two, and each vertex of $G$ is contained in exactly two paths in the set. 
Note that we allow paths of length~$0$, i.e., isolated vertices.
If $G$ has a 2-path 2-cover, we say $G$ is \defi{2-path 2-coverable}. 
Clearly, a 2-path 2-coverable graph must have maximum degree four. But not all graphs of maximum degree four
are 2-path 2-coverable: the complete graph on four vertices is a simple counterexample. 

\subsection{Local search problems}
A \defi{local search problem} $P$ is an optimization problem that consists of a set of instances $D_{P}$, a finite set of (feasible) solutions $F_{P}(I)$ for each instance $I\in D_{P}$, an objective function $f_{P}$ that assigns an integer value to each instance $I\in D_{P}$ and solution $s\in F_P(I)$, and a neighborhood $N_{P}(s,I)\subseteq F_{P}(I)$ for each solution $s\in F_{P}(I)$. 
The size of every solution $s \in F_{P}(I)$ must be bounded by a polynomial in the size of $I$. 
The goal is to find a \defi{locally optimal solution} for a given instance $I$; that is, a solution $s \in F_{P}(I)$, such that no solution $s' \in N_{P}(s,I)$ yields a better objective value than $f_P(s,I)$.
Formally, this means, for all $s'\in N_{P}(s,I)$, $f_{P}(s,I)\leq f_{P}(s',I)$ if $P$ is a minimization problem, and $f_{P}(s,I)\geq f_{P}(s',I)$ if $P$ is a maximization problem.

The \defi{standard local search algorithm} for an instance $I$ proceeds as follows.
It starts with some initial solution $s \in F_{P}(I)$.
Then it iteratively visits a neighbor with strictly better objective value, until it reaches a local optimum. 
If a solution has more than one better neighbor, the algorithm has to choose one by some prespecified rule, often referred to as a \defi{pivot rule}.

\begin{definition}[All-exp property]
    A local search problem~$P$ has the \defi{all-exp} property, if there are infinitely many pairs of an instance~$I$ of $D_P$ and an initial solution $s \in F_P(I)$, for which the standard local search algorithm always needs an exponential number of iterations 
    for all possible pivot rules. (By exponential we mean a growth of at least $\Omega(c^n)$ for some constant $c>1$ and instances of size $n$.)
\end{definition}

An alternative way to define the all-exp property is to consider the \defi{transition graph} of an instance $I$
of a local search problem $P$. This graph has as vertices all the solutions $F_{P}(I)$ and there exists a directed edge $(s, s')$ if the solution $s'$ is in the neighborhood of $s$ and has strictly better 
objective value than $s$. By definition, the transition graph is acyclic and a directed path of length $l$ in the transition graph corresponds to $l$ iterations of the standard local search algorithm. A local optimum corresponds to a sink in the transition graph. 
A local search problem has the all-exp property if and only if there exist
infinitely many pairs of instances and initial solutions such that all paths in the transition graph
from the initial solution to a sink have exponential length. 

\subsection{\texorpdfstring{\boldmath TSP/\kopt}{TSP/k-opt}}
A \defi{Hamiltonian cycle} (resp., a \defi{Hamiltonian path}) of a graph is a cycle (resp., a path) that contains all vertices of the graph.
In this paper, we also refer to a Hamiltonian cycle as a \defi{tour}.

A TSP instance is a tuple $(G, c)$, where $G$ is a complete undirected graph $(V, E)$, and $c:E\to \R$ is a function that assigns a  weight to each edge of $G$.
The goal is to find a tour of $G$ that minimizes the sum of the edge weights in the tour, we call this sum the \defi{weight of the tour}.
Note that the TSP in the literature is usually defined with nonnegative edge weights, which is not the case in our definition above; however, the definitions with and without this weight restriction are equivalent in the context of \kopt, since we can simply add a quantity to all edge weights to make them nonnegative without changing the transition graph.

A \defi{swap} is a tuple $(E_1, E_2)$ of subsets $E_1, E_2 \subseteq E$ with $|E_1| = |E_2|$.
If $|E_1| \leq k$ for some $k$, then we call it a \defi{$k$-swap}.
A \defi{2.5-swap} is a swap $(E_1, E_2)$, where either $|E_1| = 2$ or $E_1$ contains exactly three edges, two of which are incident.
For $E_1 \subseteq E(G')$, performing a swap $(E_1, E_2)$ from a subgraph $G'$ of $G$ refers to the act of removing $E_1$ from $G'$ and adding $E_2$ to $G'$. 
We also call it swapping $E_1$ for $E_2$ in $G'$.
Given a tour~$\tau$, a swap $(E_1, E_2)$ is \defi{improving} for $\tau$, if after swapping $E_1$ for $E_2$ in $\tau$, we obtain a tour with lower weight.
For a subgraph $G''$ of $G$, we say that a swap $(E_1, E_2)$ \defi{involves} $\ell$ edges in $G''$, if $\ell=|E_1\cap E(G'')|+|E_2\cap E(G'')|$. 
Note that here we do not require $E_1\subseteq E(G'')$; in our later context, this occurs when $E_1$ is a set of edges across many gadgets, whereas $G''$ is one gadget.

A \defi{($k$-)swap sequence} is a sequence $L = (S_1, \dots, S_{\ell})$, such that each $S_i$ is a ($k$-)swap.
$L$ is \defi{improving} for a tour~$\tau$, if performing the swaps in the order in the sequence decreases the total edge weight at every step.

The local search problem TSP/\kopt corresponds to TSP with the \kopt neighborhood (that is, the neighbors of a tour~$\tau$ are those that can be obtained from~$\tau$ by an improving $k$-swap).
The \kopt algorithm is then the standard local search algorithm for this problem, and an execution of the algorithm corresponds to an improving $k$-swap sequence.

\subsection{Max-Cut/Flip}
A Max-Cut instance is a tuple $(H, w)$, 
where $H$ is an undirected graph $(V,E)$ and $w: E \to \R$ is a function assigning weights to the edges of $H$.
A \defi{cut} $(V_1, V_2)$ of $H$ is a partition of the vertices of $H$ into two disjoint sets $V_1$ and $V_2$, which we call the \defi{$1$-} and \defi{$2$-sets} of the cut, respectively.
The \defi{cut-set} of a cut $(V_1, V_2)$ is the set of edges~$xy \in E$ such that $x \in V_1$ and $y \in V_2$.
The goal of Max-Cut is to find a cut to maximize the \defi{value} of the cut, that is the total weight of the edges in the cut-set.

Given a Max-Cut instance and an initial cut, the \defi{flip} of a vertex is a move of that vertex from a set of the cut to the other set.
The flip of a vertex is \defi{improving}, if it results in an increase in the value of the cut.
For a cut~$\sigma$, its \defi{flip neighborhood} is the set of all cuts obtained from~$\sigma$ by an improving flip.
The Max-Cut/Flip problem is the local search problem that corresponds to the Max-Cut problem with the flip neighborhood.
We call its standard local search algorithm the \defi{Flip algorithm}.
A \defi{flip sequence} is a sequence $(v_1, \dots, v_{\ell})$ of vertices of $H$.
A flip sequence is \defi{improving} if flipping the vertices in the order 
given by the sequence increases the cut value at each step.
In other words, a maximal improving flip sequence corresponds to an execution of the Flip algorithm.

Schäffer and Yannakakis~\cite{SY1991} and independently Haken~\cite{Hak1989} proved that 
Max-Cut/Flip has the all-exp property. 
Monien and Tscheuschner~\cite{Monien_Max_Cut_4} showed the all-exp property for Max-Cut/Flip even for graphs with bounded degree.

\begin{theorem}[\cite{Monien_Max_Cut_4}]
\label{thm:all_exp_maxcut}
    Max-Cut/Flip has the all-exp property for graphs of maximum degree four.
\end{theorem}

Michel and Scott~\cite{Michel_Max_Cut_4} recently presented an alternative proof for \cref{thm:all_exp_maxcut}.
Interestingly, their construction is highly structured and exhibits a unique property: With a suitable initial cut, there is exactly one improving flip sequence, and this sequence has exponential length.

Note that \cref{thm:all_exp_maxcut} is tight with respect to the maximum degree, since the Flip algorithm on graphs with maximum degree at most three always terminates after a polynomial number of iterations~\cite{Poljak1995} (the result is proved only for cubic graphs and integer weight functions, but 
it easily extends to real valued weight functions and to graphs of maximum degree three).

\section{Overall construction idea}
\label{sec:overall_construction}
In this section, we give a high-level description of our construction. 
For each value of $k$, we transform a suitable Max-Cut/Flip instance of maximum degree four to a TSP/\kopt instance and show that our transformation preserves in a certain sense local search steps in both instances.

\subsection{Construction for \texorpdfstring{\boldmath $k \in \{3,4\}$}{k in \{3,4\}}}
Let $(H, w)$ be a Max-Cut instance such that $H$ is 2-path 2-coverable.
In order to avoid confusion with the vertices and edges in the TSP instance later on, we use \defi{$H$-vertices} and \defi{$H$-edges} for the vertices and edges of $H$.

In our construction of the corresponding TSP instance, we use two types of gadgets, \vg gadgets and \eg gadgets.
Each gadget has a few \defi{portals}, which are vertices used to attach itself to other gadgets.
We distinguish two types of portals: PV-portals, which connect a \eg gadget with a \vg gadget, and PP-portals, which connect two \eg gadgets.
The PV-portals are grouped into pairs, called \defi{sides}.
Each \vg gadget has two sides, while \eg gadgets come with three variants, \eg-1, \eg-2, and \eg-3 gadgets, which have one, two, and three sides, respectively.

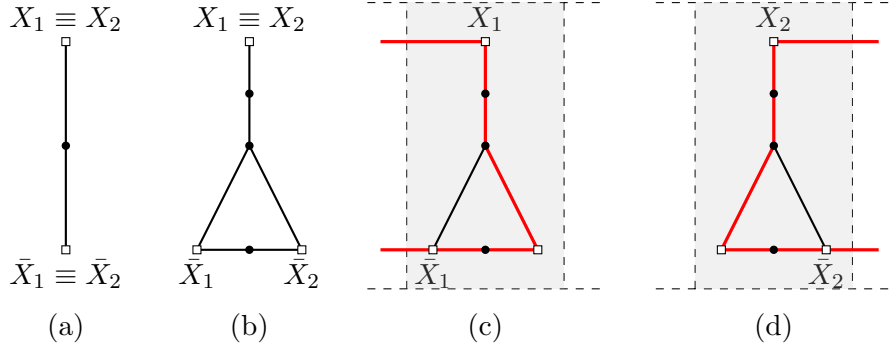
\begin{figure}[t!]
    \centering
    
\begin{tikzpicture}[scale=0.69,
    vertex/.style={fill, circle, radius=0.8mm},
    pvportal/.style={rectangle, draw, fill=white, inner sep=1.5pt},
]
  \small
  \def\defvertices{%
    \coordinate (Xb1) at (0,0);
    \coordinate (a) at (1,0);
    \coordinate (Xb2) at (2,0);
    \coordinate (b) at (1,2);
    \coordinate (c) at (1,3);
    \coordinate (X) at (1,4);
  }
  \def\defbox{
    \draw[dashed] (-0.5,-0.75) -- (-0.5,4.75) (2.5,4.75) -- (2.5,-0.75) (-1.25,-0.75) -- (3.25, -0.75) (-1.25,4.75) -- (3.25, 4.75);
    \fill[mygray, opacity=0.2] (-0.5,-0.75) -- (-0.5,4.75) -- (2.5,4.75) -- (2.5,-0.75) -- (-0.5,-0.75) ;
  }
  \newcommand{\drawverts}{%
    \foreach \v in {a,b,c} { \fill[vertex] (\v) circle; }
    \foreach \v in {Xb1,Xb2,X} { \node[pvportal] at (\v) {}; }
  }

  \begin{scope}
    \coordinate[label=above: $X_1\equiv X_2$] (top) at (1,4);
    \coordinate (mid) at (1,2);
    \coordinate[label=below: $\br{X}_1\equiv \br{X}_2$] (bot) at (1,0);
    \draw[thick] (top) -- (mid) -- (bot);
    \foreach \v in {top,bot} { \node[pvportal] at (\v) {}; }
    \fill[vertex] (mid) circle;
    \draw (1,-1.5) node {(a)};
  \end{scope}

  \begin{scope}[shift={(3.5,0)}]
    \defvertices
    \coordinate[label=below: $\br{X}_1$] (dummy1) at (Xb1);
    \coordinate[label=below: $\br{X}_2$] (dummy2) at (Xb2);
    \coordinate[label=above: $X_1 \equiv X_2$] (dummy3) at (X);
    \draw[thick] (Xb1) -- (a) -- (Xb2) -- (b) -- (Xb1) (b) -- (c) -- (X);
    \drawverts
    \draw (1,-1.5) node {(b)};
  \end{scope}

  \begin{scope}[shift={(8,0)}]
    \defvertices
    \coordinate[label=below: $\br{X}_1$] (dummy1) at (Xb1);
    \coordinate[label=above: $X_1$] (dummy3) at (X);
    \defbox
    \draw[thick] (Xb1) -- (b);
    \draw[very thick, red] (-1,0) -- (Xb1) -- (a) -- (Xb2) -- (b) -- (c) -- (X) -- (-1,4);
    \drawverts
    \draw (1,-1.5) node {(c)};
  \end{scope}
  
  \begin{scope}[shift={(13.5,0)}]
    \defvertices
    \coordinate[label=below: $\br{X}_2$] (dummy1) at (Xb2);
    \coordinate[label=above: $X_2$] (dummy3) at (X);
    \defbox
    \draw[thick] (Xb2) -- (b);
    \draw[very thick, red] (3,0) -- (Xb2) -- (a) -- (Xb1) -- (b) -- (c) -- (X) -- (3,4);
    \drawverts
    \draw (1,-1.5) node {(d)};
  \end{scope}
\end{tikzpicture}

    \caption{The two \vg gadgets used in our construction: (a) stick gadget for $k = 3$ and (b) buoy gadget for $k = 4$. $\{X_1, \br{X}_1\}$ and $\{X_2, \br{X}_2\}$ are the 1- and 2-sides of the gadget, respectively. (Note that some portals of different sides may be identical to each other.)
    The other two panels illustrate the two ways how a buoy gadget can be engaged: either (c) there are tour edges connecting the vertices of the 1-side (i.e., $X_1, \br{X}_1$) with the \eg gadget attached to that side or (d) there are tour edges connecting the vertices of the 2-side with the \eg gadget attached to that side.}
    \label{fig:vertex_gadgets_high_level}
\end{figure}

The purpose of a \vg gadget is to model which set of the cut an $H$-vertex belongs to.
In particular, we have a \vg gadget for each $H$-vertex.
The two sides of the \vg gadget are called the 1- and 2-sides, corresponding to the 1- and 2-sets of a cut of $H$.
In any tour that we will encounter during an execution of the \kopt algorithm, the \vg gadget is \defi{engaged} at exactly one side (i.e., there are tour edges between it and the \eg gadget attached on that side, while there are no tour edges between it and that on the other side).
See \cref{fig:vertex_gadgets_high_level} for an illustration.

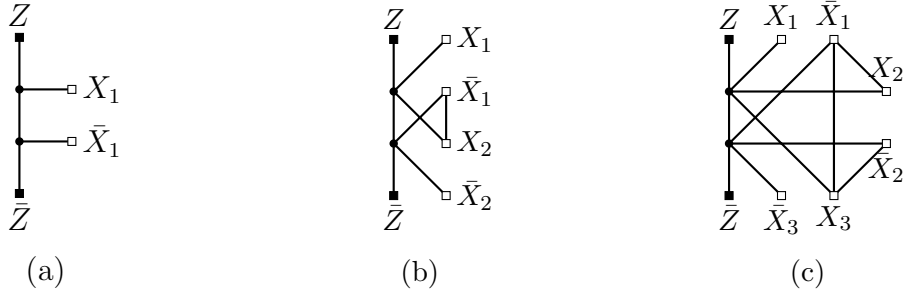
\begin{figure}[ht!]
    \centering
    \tikzset{
      vertex/.style={fill, circle, radius=0.8mm},
      pvportal/.style={rectangle, draw, fill=white, inner sep=1.5pt},
      ppportal/.style={rectangle, draw, fill=black, inner sep=1.5pt},
    }
    \begin{subfigure}[t]{0.3\textwidth}
    \centering
        \begin{tikzpicture}[scale=0.69]
            \def\defvertices{
                \coordinate[label=right: $X_1$] (X) at (1,1);
                \coordinate[label=right: $\br{X}_1$] (X') at (1,0);
                \coordinate[label=above: $Z$] (Z) at (0,2);
                \coordinate[label=below: $\br{Z}$] (Z') at (0,-1);
                \coordinate[] (S) at (0,1);
                \coordinate[] (S') at (0,0);
            }
            \def\drawvertices{
                \fill[] (S) circle (0.8mm);
                \fill[] (S') circle (0.8mm);
                \node[pvportal] at (X) {};
                \node[pvportal] at (X') {};
                \node[ppportal] at (Z) {};
                \node[ppportal] at (Z') {};
            }

            \defvertices
            \draw[thick] (Z) -- (S) -- (X) (X') -- (S') -- (Z') (S) -- (S');
            \draw (0.5, -2.5) node {(a)};
            \drawvertices
        \end{tikzpicture}
    \end{subfigure}
    \begin{subfigure}[t]{0.3\textwidth}
    \centering
       \begin{tikzpicture}[scale=0.69]
        \small
        \def\defvertices{
        \coordinate[label=right: $X_1$] (X) at (1,3);
        \coordinate[label=right: $\br{X}_2$] (Y') at (1,0);
        \coordinate[label=above: $Z$] (Z) at (0,3);
        \coordinate[label=right: $\br{X}_1$] (X') at (1,2);
        \coordinate[label=right: $X_2$] (Y) at (1,1);
        \coordinate[label=below: $\br{Z}$] (Z') at (0,0);
        \coordinate[] (R) at (0,2);
        \coordinate[] (R') at (0,1);
        }
        \def\drawvertices{
        \fill[] (R) circle (0.8mm);
        \fill[] (R') circle (0.8mm);
        \node[pvportal] at (X) {};
        \node[pvportal] at (X') {};
        \node[pvportal] at (Y) {};
        \node[pvportal] at (Y') {};
        \node[ppportal] at (Z) {};
        \node[ppportal] at (Z') {};
        }
        
        \defvertices
        \draw[thick] (Z) -- (R) -- (R') -- (Z') (R) -- (X) (R') -- (X') -- (Y) -- (R) (R') -- (Y');
        \draw (0.5, -1.5) node {(b)};
        \drawvertices

    \end{tikzpicture}
    \end{subfigure}
    \begin{subfigure}[t]{0.3\textwidth}
    \centering
       \begin{tikzpicture}[scale=0.69]
        \small
        \def\defvertices{
        \coordinate[label=above: $X_1$] (X) at (1,3);
        \coordinate[label=above: $\br{X}_1$] (X') at (2,3);
        \coordinate[label=above: $X_2$] (Y) at (3,2);
        \coordinate[label=below: $\br{X}_2$] (Y') at (3,1);
        \coordinate[label=below: $X_3$] (T) at (2,0);
        \coordinate[label=below: $\br{X}_3$] (T') at (1,0);
        \coordinate[label=above: $Z$] (Z) at (0,3);
        \coordinate[label=below: $\br{Z}$] (Z') at (0,0);
        \coordinate[] (R) at (0,2);
        \coordinate[] (R') at (0,1);
        }
        \def\drawvertices{
        \fill[] (R) circle (0.8mm);
        \fill[] (R') circle (0.8mm);
        \node[pvportal] at (X) {};
        \node[pvportal] at (X') {};
        \node[pvportal] at (Y) {};
        \node[pvportal] at (Y') {};
        \node[pvportal] at (T) {};
        \node[pvportal] at (T') {};
        \node[ppportal] at (Z) {};
        \node[ppportal] at (Z') {};
        }
        
        \defvertices
        \draw[thick] (Z) -- (R) -- (R') -- (Z') (R) -- (X) (R') -- (X') -- (Y) -- (R) (R') -- (Y') -- (T) -- (R) (R') -- (T') (X') -- (T);
        \draw (1.5, -1.5) node {(c)};
        \drawvertices
        \end{tikzpicture}
    \end{subfigure}
    
    \caption{(a) \Eg-1 gadget, (b) \eg-2 gadget, and (c) \eg-3 gadget. Edge weights are omitted. The sides are $\{X_1, \br{X}_1\}$, $\{X_2, \br{X}_2\}$, and $\{X_3, \br{X}_3\}$. White squares and black squares are the PV- and PP-portals, respectively.}
    \label{fig:path_gadgets_high_level}
\end{figure}

The purpose of a \eg gadget is to calculate the contribution of the $H$-edges to the objective.
For $t \in \{1,2,3\}$, a \eg-$t$ gadget can model any path on $t$ $H$-vertices.
(We say that these $t$ $H$-vertices are \defi{related} to the \eg gadget).
To some of the edges in a path-2 and path-3 gadget we assign weights that are related to the weights of the 
edges in $H$ between the corresponding $H$-vertices. 
See \cref{fig:path_gadgets_high_level} for an overview of the three types of \eg gadgets.
Since $H$ is 2-path 2-coverable, these gadgets are sufficient to ensure that every \vg gadget can be attached to exactly two \eg gadgets.
Each of the $t$ sides of the \eg gadget is attached to one side of the related \vg gadget; by \defi{attaching}, we mean the process of identifying the two vertices of the side of the \eg gadget with the two vertices of the corresponding side of the \vg gadget.
In addition to the attachments between the sides of \vg gadgets and \eg gadgets, we also have 
a cyclic connection between all \eg gadgets via the PP-portals. 
In particular, each \eg gadget has two PP-portals.
We line up the \eg gadgets in a cyclic sequence, and for every two adjacent \eg gadgets, we identify a PP-portal of each of these gadgets with each other.
See Figure~\ref{fig:schematic-construction} for a schematic representation of this construction for an example graph.

\begin{figure}
    \centering
    \tikzset{
  vertex/.style   ={circle,fill=black,inner sep=1.5pt, outer sep = -2pt},
  greenedge/.style={mygreen,thick},
  orangeedge/.style ={myorange,very thick},
  blueedge/.style ={blue,very thick}
}

\hspace*{-16pt}
\begin{tikzpicture}[every node/.style={font=\scriptsize}]
\begin{scope}
    \node[vertex,label=above right: $a$] (a) at (-1,0) {};
    \node[vertex,label=above right: $b$] (b) at ( 0,0) {};
    \node[vertex,label=above right: $c$] (c) at ( 1,0) {};
    \node[vertex,label=above right: $d$] (d) at ( 2,0) {};
    \node[vertex,label=above right: $e$] (e) at ( 0,1) {};
    \node[vertex,label=above right: $f$] (f) at ( 0,-1) {};
    \draw[thick] (d)--(c)--(b)--(a)--(f)--(b)--(e);
    \coordinate[label=above: (a)] (dummy) at (0.5,-3);
\end{scope}

\begin{scope}[xshift=4.5cm]
    \node[vertex] (a) at (-1,0) {};
    \node[vertex] (b) at ( 0,0) {};
    \node[vertex] (c) at ( 1,0) {};
    \node[vertex] (d) at ( 2,0) {};
    \node[vertex] (e) at ( 0,1) {};
    \node[vertex] (f) at ( 0,-1) {};
    \draw[thick] (d)--(c)--(b)--(a)--(f)--(b)--(e);

    \node[vertex,fill= red,label=above: \textcolor{red}{$P_1$}] (e2) at ( 0,1.2) {};
    \node[vertex,fill= red,label=right: \textcolor{red}{$P_6$}] (d2) at ( 2.2,0) {};
    \node[vertex,fill= red] (a2) at ( -1.1,-0.2) {};
    \node[vertex,fill= red] (f2) at ( -0.2,-1.1) {};
    \draw[thick, red]  (a2) -- node[midway,below left]  {$P_3$} (f2);

    \node[vertex,fill= red] (e3) at ( -0.2,1) {};
    \node[vertex,fill= red] (a3) at ( -1,0.2) {};
    \draw[thick, red] (e3) .. controls (-0.2,0.2) .. node[midway,above left]  {$P_2$} (a3);

    \node[vertex,fill= red] (f3) at ( 0.2,-1) {};
    \node[vertex,fill= red] (c2) at ( 0.9,-0.2) {};
    \node[vertex,fill= red] (c3) at ( 1.1,-0.2) {};
    \node[vertex,fill= red] (d3) at ( 2,-0.2) {};
    \draw[thick, red]  (c3) -- node[midway,below]  {$P_5$} (d3);
    \draw[thick, red] (f3) .. controls (0.2,-0.2) .. node[midway,below right]  {$P_4$} (c2);
    \coordinate[label=above: (b)] (dummy) at (0.5,-3);

\end{scope}

\begin{scope}[xshift=8cm]

\def\r{3mm}   
\def\pr{2pt}    

\coordinate[label=center: $a$] (S1) at (1,0);
\coordinate[label=center: $b$] (S2) at (2.8,0);
\coordinate[label=center: $c$] (S3) at (4.6,0);
\coordinate[label=center: $d$] (S4) at (6.4,0);
\coordinate[label=center: $e$] (S5) at (2.8,1.5);
\coordinate[label=center: $f$] (S6) at (2.8,-1.5);

\foreach\number in {1,...,6} \draw[thick] ($(S\number)+(-\r,-\r)$) rectangle ($(S\number)+(\r,\r)$);

\foreach \angle in {90,270} \coordinate (S1-\angle) at ($(S1)+(\angle:\r)$);
\foreach \angle in {90,270} \coordinate (S2-\angle) at ($(S2)+(\angle:\r)$);
\foreach \angle in {0,180}  \coordinate (S3-\angle) at ($(S3)+(\angle:\r)$);
\foreach \angle in {0,180}  \coordinate (S4-\angle) at ($(S4)+(\angle:\r)$);
\foreach \angle in {0,180}  \coordinate (S5-\angle) at ($(S5)+(\angle:\r)$);
\foreach \angle in {0,180}  \coordinate (S6-\angle) at ($(S6)+(\angle:\r)$);

\coordinate[label=center: \textcolor{red}{$P_6$}] (C6) at (7.7,0);
\coordinate[label=center: \textcolor{red}{$P_5$}] (C5) at (5.5,-1.5);
\coordinate[label=center: \textcolor{red}{$P_4$}] (C4) at (4.0,-1.5);
\coordinate[label=center: \textcolor{red}{$P_3$}] (C3) at (1,-1.5);
\coordinate[label=center: \textcolor{red}{$P_2$}] (C2) at (1,1.5);
\coordinate[label=center: \textcolor{red}{$P_1$}] (C1) at (5,1.5);

\foreach\number in {1,...,6} \draw[red, thick] (C\number) circle (\r);

\foreach \angle in {90,180,270}       \coordinate (C6-\angle) at ($(C6)+(\angle:\r)$);
\foreach \angle in {0,45,135,180}     \coordinate (C5-\angle) at ($(C5)+(\angle:\r)$);
\foreach \angle in {0,90,135,180,270} \coordinate (C4-\angle) at ($(C4)+(\angle:\r)$);
\foreach \angle in {0,90,180,270}     \coordinate (C3-\angle) at ($(C3)+(\angle:\r)$);
\foreach \angle in {0,90,180,270,315} \coordinate (C2-\angle) at ($(C2)+(\angle:\r)$);
\foreach \angle in {0,90,180}         \coordinate (C1-\angle) at ($(C1)+(\angle:\r)$);

\draw[greenedge] (S4-180)--(C5-45) (C5-135)--(S3-0) (C4-180)--(S6-0) (S1-90)--(C2-270) (C2-315)--(S2-90) (C2-0)--(S5-180);

\draw[orangeedge] (C6-180)--(S4-0) (C4-90)--(S3-180) (C4-135)--(S2-270) (S6-180)--(C3-0) (C3-90)--(S1-270) (S5-0)--(C1-180);

\draw[blueedge, out=90, in = 0] (C6-90) to (C1-0);
\draw[blueedge, out=-90, in = 0] (C6-270) to (C5-0);
\draw[blueedge] (C5-180) to (C4-0);
\draw[blueedge, out=-150, in = -30] (C4-270) to (C3-270);
\draw[blueedge, out=120, in = 240] (C3-180) to (C2-180);
\draw[blueedge, out=30, in = 150] (C2-90) to (C1-90);

\foreach\redpoint in {C1-180, C2-0, C2-270, C2-315, C3-0, C3-90, C4-90, C4-135, C4-180, C5-45, C5-135, C6-180} \fill[red] (\redpoint) circle (\pr);
\foreach\bluepoint in {C1-90, C1-0, C2-90, C2-180, C3-180, C3-270, C4-0, C4-270, C5-0, C5-180, C6-90, C6-270} \fill[blue] (\bluepoint) circle (\pr);
\foreach\blackpoint in {S1-90, S1-270, S2-90, S2-270, S3-0, S3-180, S4-0, S4-180, S5-0, S5-180, S6-0, S6-180}  \fill (\blackpoint) circle (\pr);

    \coordinate[label=above: (c)] (dummy) at (4,-3);
\end{scope}

\end{tikzpicture}

    \caption{Visualization of our construction of a TSP instance from a Max-Cut instance. (a) A Max-Cut instance. (b) A 2-path 2-cover of the Max-Cut instance. (c) A schematic representation of the corresponding TSP instance. The green (respectively, orange edges) represent the attachments between the 1-side (respectively, 2-side) of the \vg gadgets (black squares) and the \eg gadgets (red circles). The blue edges show the circular connection of the \eg gadgets.}
    \label{fig:schematic-construction}
\end{figure}
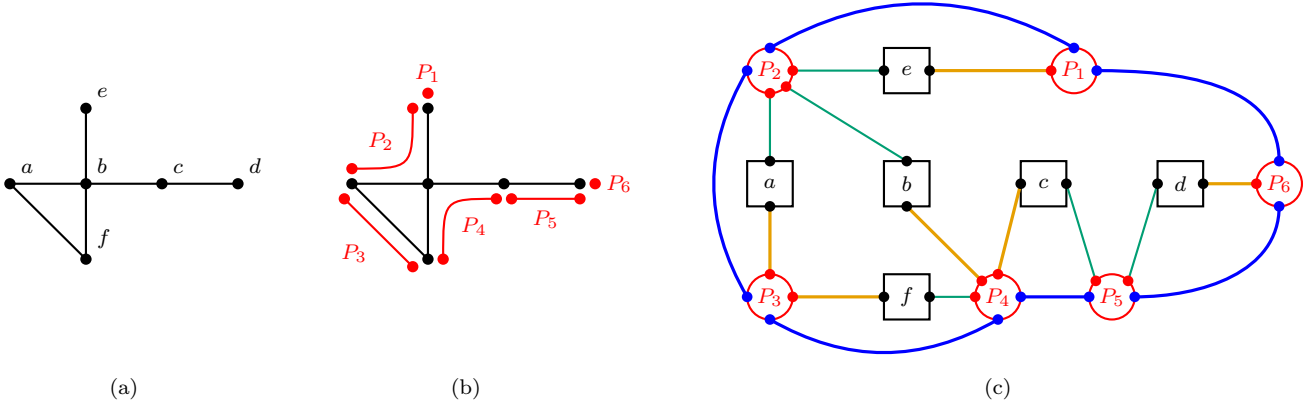

\begin{figure}[ht!]
    \centering
    \tikzset{
      vertex/.style={fill, circle, radius=0.8mm},
      pvportal/.style={rectangle, draw, fill=white, inner sep=1.5pt},
      ppportal/.style={rectangle, draw, fill=black, inner sep=2pt},
    }
    \begin{tikzpicture}[scale=0.69]
        \def\defvertices{
            \coordinate[label=above left: $\br{X}_1$] (Xb1) at (0,0);
            \coordinate (a) at (1,0);
            \coordinate[label=above right: $\br{X}_2$] (Xb2) at (2,0);
            \coordinate (b) at (1,2);
            \coordinate (c) at (1,3);
            \coordinate[label=above: $X_1 \equiv X_2$] (X) at (1,4);
            \coordinate[label=above: $Z$] (Z) at (1,5);
            \coordinate[label=below: $\br{Z}$] (Z') at (1,-1);
            \coordinate[] (R) at (-1,4);
            \coordinate[] (R') at (-1,0);
            \coordinate[] (S) at (3,4);
            \coordinate[] (S') at (3,0);
        }
        \def\drawvertices{
            \fill[] (R) circle (0.8mm);
            \fill[] (R') circle (0.8mm);
            \fill[] (S) circle (0.8mm);
            \fill[] (S') circle (0.8mm);
            \fill[] (a) circle (0.8mm);
            \fill[] (b) circle (0.8mm);
            \fill[] (c) circle (0.8mm);
            \node[pvportal] at (X) {};
            \node[pvportal] at (Xb1) {};
            \node[pvportal] at (Xb2) {};
            \node[ppportal] at (Z) {};
            \node[ppportal] at (Z') {};
        }

        \defvertices
        \draw[thick] (Xb1) -- (a) -- (Xb2) -- (b) -- (Xb1) (b) -- (c) -- (X);
        \draw[thick, mygreen] (Z') to[bend left = 25] (R') -- (R) to[bend left = 25] (Z);
        \draw[thick, myorange] (Z) to[bend left = 25] (S) -- (S') to[bend left = 25] (Z');
        \draw[thick, mygreen] (R) -- (X) (R') -- (Xb1);
        \draw[thick, myorange] (X) -- (S) (S') -- (Xb2);
        \drawvertices
        \draw (1,-2) node {(a)};
  
        \begin{scope}[shift={(6,0)}]
            \defvertices
            \draw[dashed] (Xb1) -- (b);
            \draw[very thick, red] (Xb1) -- (a) -- (Xb2) -- (b) -- (c) -- (X);
            \draw[very thick, red] (S') to[bend left = 25] (Z') to[bend left = 25] (R') (R) to[bend left = 25] (Z) to[bend left = 25] (S) (S) -- (S');
            \draw[dashed] (R) -- (R');
            \draw[very thick, red] (R) -- (X) (R') -- (Xb1);
            \draw[dashed] (X) -- (S) (S') -- (Xb2);
            \drawvertices
            \draw (1,-2) node {(b)};
        \end{scope}

        \begin{scope}[shift={(12,0)}]
            \defvertices
            \draw[dashed] (Xb2) -- (b);
            \draw[very thick, red] (Xb2) -- (a) -- (Xb1) -- (b) -- (c) -- (X);
            \draw[very thick, red] (S') to[bend left = 25] (Z') to[bend left = 25] (R') (R) to[bend left = 25] (Z) to[bend left = 25] (S) (R) -- (R');
            \draw[dashed] (S) -- (S');
            \draw[dashed] (R) -- (X) (R') -- (Xb1);
            \draw[very thick, red] (X) -- (S) (S') -- (Xb2);
            \drawvertices
            \draw (1,-2) node {(c)};
        \end{scope}
    \end{tikzpicture}    
    \caption{
        A very simple example of the constructed TSP instance for the Max-Cut instance that consists of only a single vertex. 
        (a) Black edges belong to the buoy gadget, while the green and orange edges belong to the \eg-1 gadgets attached to the 1- and 2-side of the buoy gadget, respectively. 
        (b) A tour where the 1-side of the buoy gadget is engaged.
        (c) A tour where the 2-side is engaged.
        Note that these two tours differ by a 4-swap.
    }
    \label{fig:simple_example}
\end{figure}
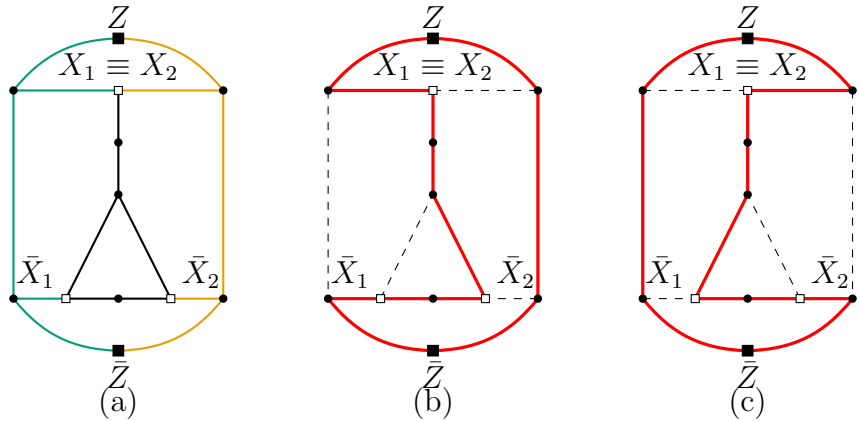

We can then establish a one-to-one correspondence between a cut of $H$ and a \defi{standard tour} of the constructed TSP instance, which can intuitively be described as follows.
For each $H$-vertex, if it is in the 1-set (resp., 2-set), then in the tour, the corresponding \vg gadget is engaged at the 1-side (resp., 2-side).
We then consider the tour as a concatenation of many paths, each of which is a path between two PP-portals of a \eg gadget.
This path traverses through all internal vertices of the \eg gadget (i.e., vertices that are not portals) as well as all vertices in each related \vg gadget that is engaged at the side of this \eg gadget.
Since each \vg gadget is engaged at exactly one side, and since each PV-portal of a \eg gadget is identified with a vertex of a \vg gadget, every vertex in the TSP instance is traversed exactly once by the tour.
See \cref{fig:simple_example} for an illustration of a very simple example.

Finally, we argue there is also a one-to-one correspondence between an improving flip sequence of the Max-Cut instance and an improving $k$-swap sequence that involves only standard tours in the constructed TSP instance.
We then complete the proof by showing that for the initial TSP solution corresponding to the initial solution in a modified Michel-Scott construction presented in~\cite{HHH2024}, we only encounter standard tours during any execution of the \kopt algorithm.

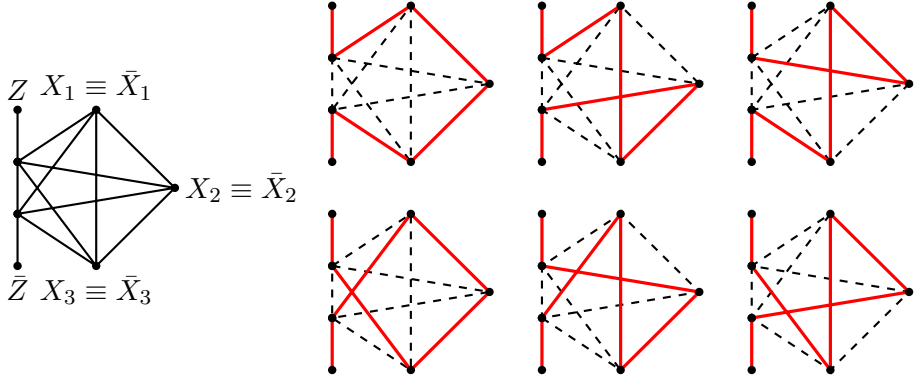
\begin{figure}[t!]
    \centering
       \begin{tikzpicture}[scale=0.69]
        \small
        \def\defvertices{
        \coordinate[] (X) at (1.5,3);
        \coordinate[] (Y) at (3,1.5);
        \coordinate[] (T) at (1.5,0);
        \coordinate[] (Z) at (0,3);
        \coordinate[] (Z') at (0,0);
        \coordinate[] (R) at (0,2);
        \coordinate[] (R') at (0,1);
        }
        \def\drawvertices{
        \fill[] (R) circle (0.8mm);
        \fill[] (R') circle (0.8mm);
        \fill[] (R) circle (0.8mm);
        \fill[] (R') circle (0.8mm);
        \fill[] (X) circle (0.8mm);
        \fill[] (Y) circle (0.8mm);
        \fill[] (T) circle (0.8mm);
        \fill[] (Z) circle (0.8mm);
        \fill[] (Z') circle (0.8mm);
        }

        \begin{scope}[shift={(0,-2)}]
        \defvertices
        \draw[thick] (Z) -- (R) -- (R') -- (Z') (R) -- (X) (R') -- (X) -- (Y) -- (R) (R') -- (Y) -- (T) -- (R) (R') -- (T) (X) -- (T);
        \coordinate[label=above: $X_1 \equiv \br{X}_1$] () at (1.5,3);
        \coordinate[label=right: $X_2 \equiv \br{X}_2$] () at (3,1.5);
        \coordinate[label=below: $X_3 \equiv \br{X}_3$] () at (1.5,0);
        \coordinate[label=above: $Z$] () at (0,3);
        \coordinate[label=below: $\br{Z}$] () at (0,0);
        \drawvertices
        \end{scope}
        
        \begin{scope}[shift={(6,0)}]
        \defvertices
        \draw[thick, dashed] (R) -- (Y) -- (R') -- (X) -- (T) -- (R) -- (R');
        \draw[very thick, red] (Z) -- (R) -- (X) -- (Y) -- (T) -- (R') -- (Z');
        \drawvertices
        \end{scope}
        
        \begin{scope}[shift={(10,0)}]
        \defvertices
        \draw[thick, dashed] (R) -- (Y) -- (X) -- (R') -- (T) -- (R) -- (R');
        \draw[very thick, red] (Z) -- (R) -- (X) -- (T) -- (Y) -- (R') -- (Z');
        \drawvertices
        \end{scope}

        \begin{scope}[shift={(14,0)}]
        \defvertices
        \draw[thick, dashed] (R) -- (X) -- (R') -- (Y) -- (T) -- (R) -- (R');
        \draw[very thick, red] (Z) -- (R) -- (Y) -- (X) -- (T) -- (R') -- (Z');
        \drawvertices
        \end{scope}

        \begin{scope}[shift={(6,-4)}]
        \defvertices
        \draw[thick, dashed] (R) -- (X) -- (T) -- (R') -- (Y) -- (R) -- (R');
        \draw[very thick, red] (Z) -- (R) -- (T) -- (Y) -- (X) -- (R') -- (Z');
        \drawvertices
        \end{scope}

        \begin{scope}[shift={(10,-4)}]
        \defvertices
        \draw[thick, dashed] (R') -- (T) -- (R) -- (R') -- (Y) -- (X) -- (R);
        \draw[very thick, red] (Z) -- (R) -- (Y) -- (T) -- (X) -- (R') -- (Z');
        \drawvertices
        \end{scope}

        \begin{scope}[shift={(14,-4)}]
        \defvertices
        \draw[thick, dashed] (R) -- (R') -- (T) -- (Y) -- (R) -- (X) -- (R') ;
        \draw[very thick, red] (Z) -- (R) -- (T) -- (X) -- (Y) -- (R') -- (Z');
        \drawvertices
        \end{scope}
        \end{tikzpicture}
    \caption{If we simply let two vertices of each side of a \eg-3 gadget coincide, then there are potentially six different paths in the gadget when all related vertices are engaged.}
    \label{fig:path_gadget_go_wrong}
\end{figure}

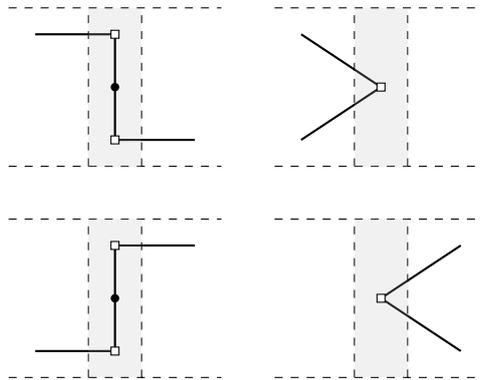
\begin{figure}[t]
    \centering
    \begin{tikzpicture}[scale = 0.7, 
        pvportal/.style={rectangle, draw, fill=white, inner sep=1.5pt}
    ]
        \def\defvertices{
            \coordinate[] (top) at (0,2);
            \coordinate[] (mid) at (0,1);
            \coordinate[] (bot) at (0,0);
        }
        \def\drawbox{
            \draw[dashed] (-0.5, -0.5) -- (-0.5, 2.5) (0.5, -0.5) -- (0.5, 2.5) (-2, -0.5) -- (2, -0.5) (-2, 2.5) -- (2, 2.5);
            \fill[mygray, opacity=0.2] (-0.5,-0.5) -- (-0.5,2.5) -- (0.5,2.5) -- (0.5,-0.5) -- (-0.5,-0.5) ;
        }
        \def\drawvertices{
            \draw[thick] (top) -- (bot);
            \fill[] (mid) circle (0.8mm);
            \node[pvportal] at (top) {};
            \node[pvportal] at (bot) {};
        }
        \defvertices
        \draw[thick] (top) -- (-1.5, 2) (1.5, 0) -- (bot); 
        \drawbox
        \drawvertices

        \begin{scope}[shift={(0,-4)}]
            \defvertices
            \draw[thick] (top) -- (1.5, 2) (-1.5, 0) -- (bot); 
            \drawbox
            \drawvertices
        \end{scope}

        \begin{scope}[shift={(5,0)}]
            \defvertices
            \draw[thick] (mid) -- (-1.5, 2) (-1.5, 0) -- (mid); 
            \drawbox
            \node[pvportal] at (mid) {};
        \end{scope}

        \begin{scope}[shift={(5,-4)}]
            \defvertices
            \draw[thick] (mid) -- (1.5, 2) (1.5, 0) -- (mid); 
            \drawbox
            \node[pvportal] at (mid) {};
        \end{scope}
    \end{tikzpicture}
    \caption{Irregular behavior (left) and regular behavior (right). 
    The gray boxes represent \ng gadgets (variants of \vg gadgets used in the construction for $k=2.5$). 
    The open boxes represent the other gadgets attached to the \ng gadgets. 
    Top and bottom figures represent the two ways that can be seen in a standard tour.}
    \label{fig:regular}
\end{figure}

\subsection{Construction for \texorpdfstring{\boldmath$k=2.5$}{k=2.5}}
\label{subsec:const_summary_2_5}
At first glance, it seems we can easily adapt the above construction to establish the all‑exp property for $k = 2.5$.
In particular, we can use a single vertex as the \vg gadgets (i.e., all four PV-portals of the \vg gadget coincide), and for the \eg gadgets, we make the two vertices of each side coincide.
For example, in \cref{fig:simple_example}, if we contract the buoy gadget to a single vertex, then the two tours in the figure differ exactly by a 2.5-swap.
However, there is a serious issue with this approach: There is a certain uniqueness property in the subtours of the \eg gadgets (\cref{lem:edge_unique_subtours} below) upon which the construction heavily depends.
However, this property no longer holds when the vertices of each side coincide.
See \cref{fig:path_gadget_go_wrong} for an illustration.

In order to circumvent this, we make a few changes in our construction.
Firstly, we have a new way to model the vertices.
Before, we required that in a standard tour, the two edges coming out of a vertex gadget have to stay within the same path gadget; we call this behavior ``regular''.
Now, we sometimes require the two edges to stay in two different gadgets, which we call ``irregular''.
See \cref{fig:regular}.
Secondly, because of the irregularity above, the gadgets to model the $H$-edges also have to change.
In order to distinguish them with the \eg gadget we saw earlier, we call these gadgets \defi{\sg gadgets}, since they model induced stars in $H$.
Note that besides the stars on one, two, and three $H$-vertices similar to the gadgets seen earlier, we now also have a brand new gadget to model a star on four $H$-vertices.
This new gadget does not work in general but is designed for specific vertices in a specific Max-Cut instance.
In particular, we use the fact that we know precisely how each of these vertices and its neighbors change in any improving flip sequence of this Max-Cut instance.
The Max-Cut instance is a further modification of the modified Michel-Scott construction mentioned in the previous subsection.

\section{Construction for \texorpdfstring{\boldmath$k \in \{3, 4\}$}{k in \{3,4\}}}
\label{sec:construction_3}
In this section, we explain the reduction for $k \in \{3, 4\}$ in full detail, including the description and some basic properties of the gadgets and the overall construction.

\subsection{Base gadgets}
While the \vg gadgets and the \eg gadgets used in the construction are different in purposes and design, they share many common terminologies.
Accordingly, we define a base gadget as follows as an umbrella term for both types of gadgets.

\begin{definition}[Base gadget]
    A \defi{base gadget} is a simple edge weighted graph.
    Each vertex of a base gadget is either a \defi{portal} or an \defi{internal vertex}.
    Some portals are called \defi{PV-portals}, while some are called \defi{PP-portals}.
    PP-portals always have degree one, while there is no degree constraint on PV-portals.
    A \defi{side} of a base gadget $B$ is a set of PV-portals, and each PV-portal belongs to at least one side.
    Unless otherwise stated, a side consists of two vertices $X^B_i$ and $\bar{X}^B_i$, which we also call the \defi{$i$-side} of $B$.
    
    A \defi{subtour} of a base gadget is a spanning subgraph 
    that comprises of vertex-disjoint paths such that only a portal can be an endpoint of a path, and PP-portals have degree one in the subgraph.
    (Note that a path may have length zero; that is, it consists of only one portal.)
    A base gadget has a set of \defi{standard subtours} that will be defined more concretely in Definitions~\ref{def:vertex-gadget} and~\ref{def:edge_gadget}.    
\end{definition}

    The PP-portals (respectively, PV-portals) are called this way, because they will be used to connect a \eg gadget with another \eg gadget (respectively, a \eg gadget with a \vg gadget).

\subsection{\Vg gadgets}

\begin{definition}
\label{def:vertex-gadget}
    A \defi{\vg gadget} $\cv$ is a base gadget with two sides, $\{X^{\cv}_1, \br{X}^{\cv}_1\}$ and $\{X^{\cv}_2, \br{X}^{\cv}_2\}$ with $X^{\cv}_1 \not\equiv \br{X}^{\cv}_1$ and $X^{\cv}_2 \not\equiv \br{X}^{\cv}_2$, and no PP-portals that satisfies the two following conditions.
    It has two standard subtours: a unique \defi{odd subtour} that is an $X^{\cv}_1\br{X}^{\cv}_1$-path and a unique \defi{even subtour} that is an $X^{\cv}_2\br{X}^{\cv}_2$-path.
    It is an \defi{$(\ell)$-\vg gadget}, if changing from the odd subtour to the even subtour involves $\ell$ edges.
\end{definition}

Note that our construction from Max-Cut/Flip to TSP/\kopt for different values of $k$ only differs in the choice of the \vg gadgets.
In particular, we will need (0)-\vg gadgets for $k=3$ and (2)-\vg gadgets for $k=4$.
We now discuss the existence of these gadgets.

Firstly, we have a \vg gadget that is simply a 2-path.

\begin{definition}
    A \defi{stick gadget} is a \vg gadget $\beta$ that consists of only a 2-path $(a, b, c)$, and $X^{\beta}_1 = a = X^{\beta}_2$ and $\br{X}^{\beta}_1 = c = \br{X}^{\beta}_2$.
\end{definition}

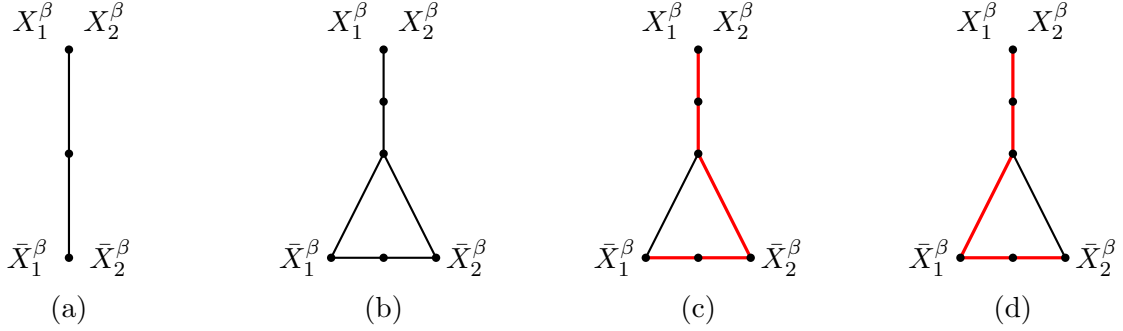
\begin{figure}[t!]
    \centering
    
\begin{tikzpicture}[scale=0.69,
    vertex/.style={fill, circle, radius=0.8mm}
]
  \small
  \def\defvertices{%
    \coordinate[label=left: $\br{X}^{\beta}_1$] (Xb1) at (0,0);
    \coordinate (a) at (1,0);
    \coordinate[label=right: $\br{X}^{\beta}_2$] (Xb2) at (2,0);
    \coordinate (b) at (1,2);
    \coordinate (c) at (1,3);
    \coordinate[label=above: $X^{\beta}_1 \quad X^{\beta}_2$] (X) at (1,4);
  }
  \newcommand{\drawverts}{%
    \foreach \v in {Xb1,a,Xb2,b,c,X} { \fill[vertex] (\v) circle; }
  }

  \begin{scope}
    \coordinate (top) at (1,4);
    \coordinate (mid) at (1,2);
    \coordinate (bot) at (1,0);
    \draw[thick] (top) -- (mid) -- (bot);
    \foreach \v in {top,mid,bot} { \fill[vertex] (\v) circle; }
    \coordinate[label=above: $X^{\beta}_1\quad X^{\beta}_2$] (dummy) at (top);
    \node[label=left: $\br{X}^{\beta}_1$] at (bot) {};
    \node[label=right: $\br{X}^{\beta}_2$] at (bot) {};
    \draw (1,-1) node {(a)};
  \end{scope}

  \begin{scope}[shift={(6,0)}]
    \defvertices
    \draw[thick] (Xb1) -- (a) -- (Xb2) -- (b) -- (Xb1) (b) -- (c) -- (X);
    \drawverts
    \draw (1,-1) node {(b)};
  \end{scope}

  \begin{scope}[shift={(12,0)}]
    \defvertices
    \draw[thick] (Xb1) -- (b);
    \draw[very thick, red] (Xb1) -- (a) -- (Xb2) -- (b) -- (c) -- (X);
    \drawverts
    \draw (1,-1) node {(c)};
  \end{scope}

  \begin{scope}[shift={(18,0)}]
    \defvertices
    \draw[thick] (Xb2) -- (b);
    \draw[very thick, red] (Xb2) -- (a) -- (Xb1) -- (b) -- (c) -- (X);
    \drawverts
    \draw (1,-1) node {(d)};
  \end{scope}
\end{tikzpicture}

    \caption{(a) Stick gadget. (b) Buoy gadget. (c) Subtour in a buoy gadget that is an $X^{\beta}_1\br{X}^{\beta}_1$-path and an $X^{\beta}_2\br{X}^{\beta}_1$-path. (d) Subtour in a buoy gadget that is an $X^{\beta}_2\br{X}^{\beta}_2$-path and an $X^{\beta}_1\br{X}^{\beta}_2$-path.}
    \label{fig:buoy_gadget}
\end{figure}

\cref{fig:buoy_gadget}(a) shows a stick gadget. 
Note that any subtour of the stick gadget must have the form $(a,b,c)$.
However, to facilitate the arguments later, we consider that the gadget has four distinct subtours: $(X^{\beta}_1, b, \br{X}^{\beta}_1)$, $(X^{\beta}_2, b, \br{X}^{\beta}_2)$, $(X^{\beta}_1, b, \br{X}^{\beta}_2)$, and $(X^{\beta}_2, b, \br{X}^{\beta}_1)$.
The first two are standard, while the last two are not.

\begin{observation}
    A stick gadget is a (0)-\vg gadget.
\end{observation}

Secondly, we have the following gadget.

\begin{definition}
    A \defi{buoy gadget} is a \vg gadget $\beta$ as depicted in \cref{fig:buoy_gadget}(b) where $X^{\beta}_1$ and $X^{\beta}_2$ coincide.
\end{definition}

Similar to the stick gadget, we also consider that this gadget has four distinct subtours: an $X^{\beta}_1\br{X}^{\beta}_1$-path, an $X^{\beta}_2\br{X}^{\beta}_1$-path, an $X^{\beta}_2\br{X}^{\beta}_2$-path, and an $X^{\beta}_1\br{X}^{\beta}_2$-path.
The first two subtours share the same underlying path as depicted in~\cref{fig:buoy_gadget}(c), while the last two share another underlying path as shown in~\cref{fig:buoy_gadget}(d).
It is easy to check that there is no other subtour and that the following observation holds.

\begin{observation}
    A buoy gadget is a (2)-\vg gadget.
\end{observation}

Note that since we want to extend the construction in this section to $k \geq 5$ in Section~\ref{sec:k_5}, in the remainder of this section and Section~\ref{sec:correctness_3}, we argue based on general $(\ell)$-\vg gadgets and not the stick and buoy gadgets specifically.

\subsection{\Eg gadgets}

\begin{figure}[t!]
    \centering
    \begin{subfigure}[t]{\textwidth}
    \centering
        \begin{tikzpicture}[scale=0.69]
            \def\defvertices{
                \coordinate[label=right: $X_1$] (X) at (1,1);
                \coordinate[label=right: $\br{X}_1$] (X') at (1,0);
                \coordinate[label=left: $Z$] (Z) at (-1,1);
                \coordinate[label=left: $\br{Z}$] (Z') at (-1,0);
                \coordinate[] (S) at (0,1);
                \coordinate[] (S') at (0,0);
            }
            \def\drawvertices{
                \fill[] (S) circle (0.8mm);
                \fill[] (S') circle (0.8mm);
                \fill[] (X) circle (0.8mm);
                \fill[] (X') circle (0.8mm);
                \fill[] (Z) circle (0.8mm);
                \fill[] (Z') circle (0.8mm);            }

            \defvertices
            \draw[thick] (Z) -- (S) -- (X) (X') -- (S') -- (Z') (S) -- (S');
            \drawvertices

            \begin{scope}[shift={(5,0)}]
                \defvertices
                \draw[very thick, red] (Z) -- (S) -- (X) (X') -- (S') -- (Z');
                \draw[thick, dashed] (S) -- (S');
                \drawvertices
                \draw (0, -1) node {(1)};
            \end{scope}

            \begin{scope}[shift={(10,0)}]
                \defvertices
                \draw[very thick, red] (Z) -- (S) -- (S') -- (Z');
                \draw[thick, dashed] (X) -- (S) (S') -- (X');
                \drawvertices
                \draw (0, -1) node {(0)};
            \end{scope}            
        \end{tikzpicture}
    \caption{\Eg-1 gadget and the two standard subtours}
    \end{subfigure}

    \vspace{1em}
    
    \begin{subfigure}[t]{\textwidth}
    \centering
       \begin{tikzpicture}[scale=0.69]
        \small
        \def\defvertices{
        \coordinate[label=right: $X_1$] (X) at (1,3);
        \coordinate[label=right: $\br{X}_2$] (Y') at (1,0);
        \coordinate[label=left: $Z$] (Z) at (0,3);
        \coordinate[label=right: $\br{X}_1$] (X') at (1,2);
        \coordinate[label=right: $X_2$] (Y) at (1,1);
        \coordinate[label=left: $\br{Z}$] (Z') at (0,0);
        \coordinate[label=left: $Y$] (R) at (0,2);
        \coordinate[label=left: $\br{Y}$] (R') at (0,1);
        }
        \def\drawvertices{
        \fill[] (R) circle (0.8mm);
        \fill[] (R') circle (0.8mm);
        \fill[] (X) circle (0.8mm);
        \fill[] (X') circle (0.8mm);
        \fill[] (Y) circle (0.8mm);
        \fill[] (Y') circle (0.8mm);
        \fill[] (Z) circle (0.8mm);
        \fill[] (Z') circle (0.8mm);
        }
        
        \defvertices
        \draw[thick] (Z) -- (R) -- (R') -- (Z') (R) -- (X) (R') -- (X') -- (Y) -- (R) (R') -- (Y');
        \drawvertices

        \begin{scope}[shift={(3.5,0)}]
        \defvertices
        \draw[thick, dashed] (X') -- (R') -- (R) -- (Y);
        \draw[very thick, red] (Z) -- (R) (R') -- (Z') (R) -- (X) (R') -- (Y') (X') -- (Y);
        \drawvertices
        \draw (0.5, -1) node {(1,1)};
        \end{scope}
        
        \begin{scope}[shift={(7,0)}]
        \defvertices
        \draw[thick, dashed] (Y) -- (X') -- (R') -- (R) -- (X);
        \draw[very thick, red] (Z) -- (R) (R') -- (Z') (R) -- (Y) (R') -- (Y');
        \drawvertices
        \draw (0.5, -1) node {(0,1)};
        \end{scope}
        
        \begin{scope}[shift={(10.5,0)}]
        \defvertices
        \draw[thick, dashed] (Y') -- (R') -- (R) -- (Y) -- (X');
        \draw[very thick, red] (Z) -- (R) (R') -- (Z') (R) -- (X) (R') -- (X');
        \drawvertices
        \draw (0.5, -1) node {(1,0)};
        \end{scope}
        
        \begin{scope}[shift={(14,0)}]
        \defvertices
        \draw[thick, dashed] (X') -- (R') (R) -- (Y)  (R) -- (X) (R') -- (Y') (X') -- (Y);
        \draw[very thick, red] (Z) -- (R) -- (R') -- (Z');
        \drawvertices
        \draw (0.5, -1) node {(0,0)};
        \end{scope}

        \begin{scope}[shift={(-3.5,0)}]
            \node [shape=rectangle,align=center](table1) at (0.5,1.5) {
            \footnotesize
            \setlength{\tabcolsep}{2pt}
            \begin{tabular}{lc}
                Edge & Weight  \\ 
                \midrule
                $Y\br{Y}$ & $\sigma_{12}$ \\
                $\br{X}_1X_2$ & $\sigma_{12}$ \\
                $YX_2$ & $\delta_{12}$ \\
                $\br{Y}\br{X}_1$ & $\delta_{12}$
            \end{tabular}
            };
        \end{scope}
    \end{tikzpicture}
     \caption{\Eg-2 gadget and the four standard subtours}
    \end{subfigure}

    \vspace{1em}

    \begin{subfigure}[t]{\textwidth}
        \centering
    \centering
       \begin{tikzpicture}[scale=0.69]
        \small
        \def\defvertices{
        \coordinate[label=above: $X_1$] (X) at (1,3);
        \coordinate[label=above: $\br{X}_1$] (X') at (2,3);
        \coordinate[label=above: $X_2$] (Y) at (3,2);
        \coordinate[label=below: $\br{X}_2$] (Y') at (3,1);
        \coordinate[label=below: $X_3$] (T) at (2,0);
        \coordinate[label=below: $\br{X}_3$] (T') at (1,0);
        \coordinate[label=above: $Z$] (Z) at (0,3);
        \coordinate[label=below: $\br{Z}$] (Z') at (0,0);
        \coordinate[] (R) at (0,2);
        \coordinate[] (R') at (0,1);
        }
        \def\drawvertices{
        \fill[] (R) circle (0.8mm);
        \fill[] (R') circle (0.8mm);
        \fill[] (R) circle (0.8mm);
        \fill[] (R') circle (0.8mm);
        \fill[] (X) circle (0.8mm);
        \fill[] (X') circle (0.8mm);
        \fill[] (Y) circle (0.8mm);
        \fill[] (Y') circle (0.8mm);
        \fill[] (T) circle (0.8mm);
        \fill[] (T') circle (0.8mm);
        \fill[] (Z) circle (0.8mm);
        \fill[] (Z') circle (0.8mm);
        }
        
        \defvertices
        \draw[thick] (Z) -- (R) -- (R') -- (Z') (R) -- (X) (R') -- (X') -- (Y) -- (R) (R') -- (Y') -- (T) -- (R) (R') -- (T') (X') -- (T);
        \coordinate[label=left: $Y$] () at (0,2);
        \coordinate[label=left: $\br{Y}$] () at (0,1);
        \drawvertices
        
        \begin{scope}[shift={(4.5,0)}]
        \defvertices
        \draw[thick, dashed] (X') -- (R') -- (R) -- (Y) (Y') -- (R')  (R) -- (T) -- (X');
        \draw[very thick, red] (Z) -- (R) -- (X) (T') -- (R') -- (Z') (X') -- (Y) (Y') -- (T);
        \drawvertices
        \draw (1.5, -1.5) node {(1,1,1)};
        \end{scope}
        
        \begin{scope}[shift={(9,0)}]
        \defvertices
        \draw[thick, dashed] (X') -- (R') -- (R) -- (X) (Y') -- (R')  (R) -- (T) -- (X') -- (Y);
        \draw[very thick, red] (Z) -- (R) -- (Y) (T') -- (R') -- (Z')  (Y') -- (T);
        \drawvertices
        \draw (1.5, -1.5) node {(0,1,1)};
        \end{scope}

        \begin{scope}[shift={(13.5,0)}]
        \defvertices
        \draw[thick, dashed] (X') -- (R') -- (R) -- (Y)  -- (X') (Y') -- (R')  (R) -- (T)  (Y') -- (T);
        \draw[very thick, red] (Z) -- (R) -- (X) (T') -- (R') -- (Z') (X') -- (T) ;
        \drawvertices
        \draw (1.5, -1.5) node {(1,0,1)};

        \end{scope}

        \begin{scope}[shift={(18,0)}]
        \defvertices
        \draw[thick, dashed] (X') -- (R') -- (R) -- (Y) (T') -- (R')  (R) -- (T) -- (X') (Y') -- (T);
        \draw[very thick, red] (Z) -- (R) -- (X) (Y') -- (R') -- (Z') (X') -- (Y) ;
        \drawvertices
        \draw (1.5, -1.5) node {(1,1,0)};
        \end{scope}

        \begin{scope}[shift={(4.5,-6)}]
        \defvertices
        \draw[thick, dashed] (Y') -- (R') -- (R) -- (Y) (T') -- (R') -- (X') (X) -- (R) -- (T) -- (X') -- (Y) (Y') -- (T);
        \draw[very thick, red] (Z) -- (R) -- (R') -- (Z')  ;
        \drawvertices
        \draw (1.5, -1.5) node {(0,0,0)};
        \end{scope}

        \begin{scope}[shift={(9,-6)}]
        \defvertices
        \draw[thick, dashed] (Y') -- (R') -- (R) -- (Y) (T') -- (R')  (R) -- (T) -- (X') -- (Y) (Y') -- (T);
        \draw[very thick, red] (Z) -- (R) -- (X) (X') -- (R') -- (Z')  ;
        \drawvertices
        \draw (1.5, -1.5) node {(1,0,0)};
        \end{scope}

        \begin{scope}[shift={(13.5,-6)}]
        \defvertices
        \draw[thick, dashed] (X') -- (R') -- (R) -- (X)  (T') -- (R')  (R) -- (T) -- (X') -- (Y) (Y') -- (T);
        \draw[very thick, red] (Z) -- (R) -- (Y) (Y') -- (R') -- (Z')  ;
        \drawvertices
        \draw (1.5, -1.5) node {(0,1,0)};
        \end{scope}
        
        \begin{scope}[shift={(18,-6)}]
        \defvertices
        \draw[thick, dashed] (X') -- (R') -- (R) -- (X) (Y') -- (R')  (R) -- (Y) (X') -- (Y) (Y') -- (T) -- (X');
        \draw[very thick, red] (Z) -- (R) -- (T) (T') -- (R') -- (Z')  ;
        \drawvertices
        \draw (1.5, -1.5) node {(0,0,1)};
        \end{scope}

        \begin{scope}[shift={(0,-6)}]
            \node [shape=rectangle,align=center](table1) at (1.2,1.5) {
            \footnotesize
            \setlength{\tabcolsep}{2pt}
            \begin{tabular}{lc}
                Edge & Weight  \\ 
                \midrule
                $\br{X}_1X_2$ & $\sigma_{12}$ \\
                $\br{X}_1X_3$ & $\delta_{12}+\delta_{23}$ \\
                $\br{X}_1\br{Y}$ & $\delta_{12}+\sigma_{23} $\\
                $X_2Y$ & $\delta_{12}  $ \\
                $\br{X}_2X_3$ & $\sigma_{23} $ \\
                $\br{X}_2\br{Y}$ & $ \delta_{23}$ \\ 
                $X_3Y$ & $\sigma_{12}+\delta_{23} $ \\
                $Y\br{Y}$ & $\sigma_{12}+\sigma_{23}  $ 
            \end{tabular}
            };
        \end{scope}
        \end{tikzpicture}
        \caption{\Eg-3 gadget and the eight standard subtours}
    \end{subfigure}
    
    \caption{\Eg gadgets. Subtour edges are in red. Unless specified, edges have weight zero. We omit the superscripts on the vertices and weight parameters for better readability.}
    \label{fig:edge_gadgets}
\end{figure}
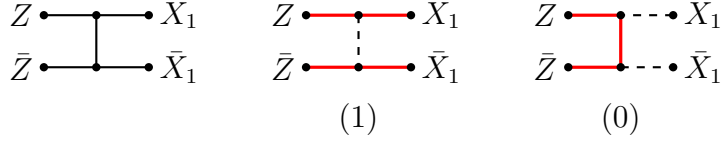
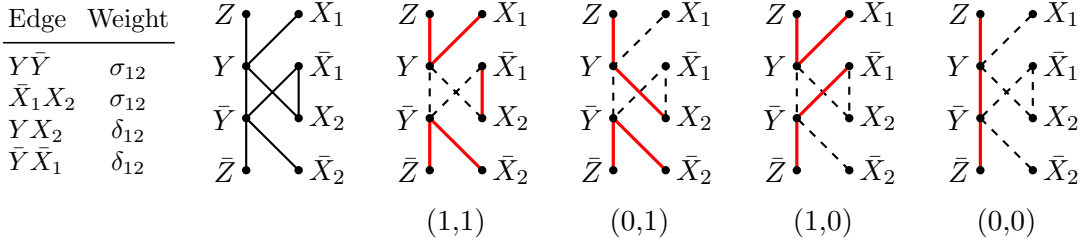
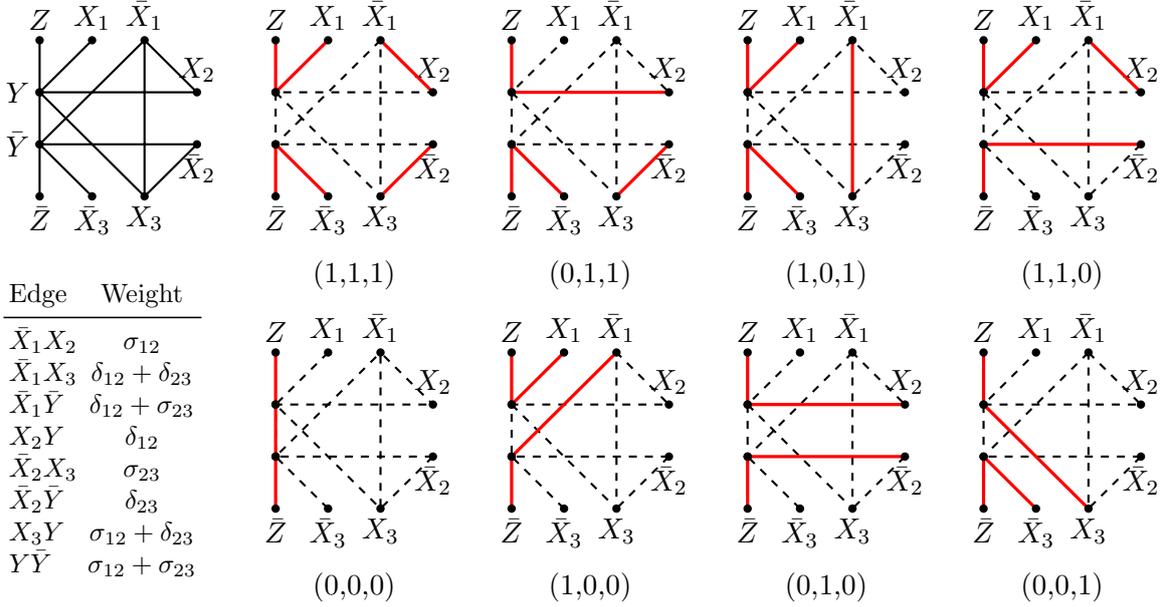

\begin{definition}
\label{def:edge_gadget}
    For $t \in \{1,2,3\}$, a \defi{\eg-$t$ gadget} $\ce$ is a base gadget as depicted in \cref{fig:edge_gadgets} with two PP-portals $Z^{\ce}$, $\br{Z}^{\ce}$ and $t$ sides
     $\{X^{\alpha}_1, \bar{X}^{\alpha}_1\}, \dots, \{X^{\alpha}_t, \bar{X}^{\alpha}_t\}$.
    The weights of the edges depend on $2(t-1)$ parameters. 
    For all $i\in [t-1]$, we have two parameters, \defi{agreeing weight} $\sigma^{\ce}_{i(i+1)}$ and \defi{disagreeing weight} $\delta^{\ce}_{i(i+1)}$ as shown in the figure. 
    For a vector $\vect{s} \in \bin^t$, an \defi{$\vect{s}$-subtour} is a subtour in which both $X^{\ce}_i$ and $\br{X}^{\ce}_i$ have degree $\vect{s}[i]$ for every $i \in [t]$. Then the standard subtours of the \eg-$t$ gadget are exactly the $\vect{s}$-subtours for all $\vect{s} \in \bin^t$.    
\end{definition}

\begin{lemma}
\label{lem:edge_unique_subtours}
    For some $t \in \{1,2,3\}$, let $\ce$ be a \eg-$t$ gadget.
    Then for each $\vect{s} \in \bin^t$, $\ce$ has exactly one $\vect{s}$-subtour as depicted in \cref{fig:edge_gadgets}.
    Further, the total edge weight of the $\vect{s}$-subtour is equal to
    \[
    \sum_{\substack{i \in [t-1]:\\ \vect{s}[i] =  \vect{s}[i+1]}} \sigma^{\ce}_{i(i+1)} + \sum_{\substack{i \in [t-1]:\\ \vect{s}[i] \neq  \vect{s}[i+1]}} \delta^{\ce}_{i(i+1)}.
    \]
\end{lemma}
\begin{proof}
    Since it is clear in this proof, we omit all superscripts $\alpha$ in all vertex and weight labels.

    Firstly, if $t = 1$, then since the degrees of $Z$ and $\br{Z}$ in a subtour have to be one, it is easy to see that the two subtours depicted in \cref{fig:edge_gadgets}(a) are the only subtours of this gadget; one of them is a (0)-subtour and the other a (1)-subtour.
    Their total edge weights are zero.
    Hence, the lemma holds in this case.

    Secondly, if $t = 2$, since $Z$ and $\br{Z}$ must have degree one, it follows that $ZY$ and $\br{Z}\br{Y}$ must be present in any subtour.
    Since neither of $Y$ and $\br{Y}$ can be an endpoint of a path, each of them must be incident to another edge in the path.
    There are five cases for the incident edges of these two vertices: (i) $Y\br{Y}$, (ii) $YX_1$ and $\br{Y}\br{X}_1$, (iii) $YX_2$ and $\br{Y}\br{X}_2$, (iv) $YX_1$ and $\br{Y}\br{X}_2$, and (v) $YX_2$ and $\br{Y}\br{X}_1$.
    For each of the first four cases, there is only one way to complete the subtour such that $X_i$ and $\br{X}_i$ have the same degree for $i \in \{1, 2\}$.
    They are depicted in \cref{fig:edge_gadgets}(b).
    For case (v), $\br{Y}\br{X}_1$ implies that $X_1$ must have degree~1 as 
    $\br{X}_1$ and $X_1$ must have the same degree, but then $Y$ would have degree three.
    Hence, case (v) cannot happen.
    It is then easy to check that the weight of the $(1,1)$- and the $(0,0)$-subtour is $\sigma_{12}$ 
    while it is $\delta_{12}$ for the $(0,1)$- and $(1,0)$-subtour.  

    Lastly, we consider $t = 3$.
    Let $\tau$ be an $\vect{s}$-subtour for some $\vect{s} \in \{0,1\}^3$.
    If $\vect{s}[3] = 0$, then $X_3$ and $\br{X}_3$ have degree zero in $\tau$.
    Note that when we remove $X_3$ and $\br{X}_3$ from $\alpha$, we obtain a path-2 gadget $\alpha'$; further, $\tau$ restricted to $\alpha'$ is an $\vect{s}'$-subtour, where $\vect{s}'$ agrees with $\vect{s}$ on the dimensions other than 3.
    Therefore, using the result for $t = 2$, we obtain that there is a unique $\vect{s}$-subtour when $\vect{s}[3] = 0$.
    Similarly, we also have that the $\vect{s}$-subtour is unique when $\vect{s}[1] = 0$ or $\vect{s}[2] = 0$.
    The remaining case is when $\vect{s} = (1,1,1)$.
    Since $X_1$ and $\br{X}_3$ must have degree one in this subtour, two paths of the subtour have to be $(Z, Y, X_1)$ and $(\br{Z}, \br{Y}, \br{X}_3)$.
    Then, there is only one way to complete the subtour, such that the other PV-portals also have degree one; that is, we have the paths $(\br{X}_1, X_2)$ and $(\br{X}_2, X_3)$.
    Hence, we obtain the uniqueness of the standard subtours, and the standard subtours have to be as depicted in \cref{fig:edge_gadgets}(c).
    It is then easy to check with these eight subtours that the weights are as claimed in the lemma.
\end{proof}

\begin{lemma}\label{lem:eg_2_non_standard_sides_in_ns_tours}
    For some $t \in \{1,2,3\}$, let $\ce$ be a \eg-$t$ gadget and let $\zeta$ be a non-standard subtour in $\ce$. If no portal has degree two in $\zeta$ then there are exactly two $i\in [t]$ with $\left|d_\zeta({X^{\ce}_i})-d_\zeta({\br{X}^{\ce}_i})\right|=1$.    
\end{lemma}
\begin{proof}
    If $\zeta$ is a subtour in which no portal has degree two, then every PV-portal has either degree one or degree zero. Since there has to be an even number of vertices with odd degree in any subtour, $d_\zeta(Z)=d_\zeta(\br{Z})=1$ and all internal vertices of $\ce$ have degree two, the number of PV-portals with degree one in $\zeta$ has to be even. Hence, the number of sides $i\in[t]$ with $\left|d_\zeta({X^{\ce}_i})-d_\zeta({\br{X}^{\ce}_i})\right|=1$ has to be even. Thus, since $t\leq3$ it is either two or zero. 
    As $\zeta$ is a non-standard subtour, there has to be at least one $i\in[t]$ with $\left|d_\zeta({X^{\ce}_i})-d_\zeta({\br{X}^{\ce}_i})\right|=1$. Hence it has to be two. 
\end{proof}

\begin{lemma}
    \label{lem:eg_involved_edges}
	Let $t \in \{1,2,3\}$, and let $\vect{s}_1$ and $\vect{s}_2$ be two distinct vectors in $\bin^t$.
	If the Hamming distance between $\vect{s}_1$ and $\vect{s}_2$ is one, then the swap between the $\vect{s}_1$- and $\vect{s}_2$-subtours in a \eg-$t$ gadget involves exactly three edges.
	Otherwise, it involves more than three edges.
\end{lemma}	
    
\begin{proof}
    By \cref{lem:edge_unique_subtours}, the $\vect{s}_1$- and $\vect{s}_2$-subtours are unique and as depicted in \cref{fig:edge_gadgets}.
    Hence, we have the number of involved edges in a swap between $\vect{s}_1$- and $\vect{s}_2$-subtours as listed in \cref{tab:swap_edge_subtours}.
    The lemma then follows.
\end{proof}

	\begin{table}[ht]
	\centering
	\caption{Number of involved edges in a swap between the $\vect{s}_1$-subtour and $\vect{s}_2$-subtour of a \eg-$t$ gadget}
	\begin{tabular}{c|p{0.7\textwidth}|c}
	$t$ & $\{\vect{s}_1, \vect{s}_2\}$ & \# involved edges \\
	\hline
	1 & \{0, 1\} & 3 \\\hline
	2 & \{00, 01\}, \{00, 10\}, \{01, 11\}, \{10, 11\}& 3 \\\hline
	2 & \{00, 11\}, \{01, 10\} & 4 \\\hline
	3 & \{000, 001\}, \{000, 010\}, \{000, 100\}, \{001, 011\}, \{001, 101\},
	      \{010, 011\}, \{010, 110\}, \{100, 101\}, \{100, 110\}, \{011, 111\}, 
	    \{101, 111\}, \{110, 111\} &	3 \\\hline
	3 & \{000, 011\}, \{000, 101\}, \{000, 110\}, \{001, 010\}, \{001, 100\},
        \{001, 111\}, \{010, 100\},  \{011, 101\}, \{100, 111\}, \{101, 110\} & 4 \\\hline
    3 & \{000, 111\}, \{001, 110\}, \{010, 101\}, \{100, 011\} & 5 \\\hline
    3 & \{010, 111\}, \{011,110\} & 6 \\
	\end{tabular}
    \label{tab:swap_edge_subtours}
	\end{table}

\subsection{Construction of the TSP instance}
\label{sec:construction}
To prove the all-exp property for TSP/\kopt we start with a Max-Cut instance $(H,w)$ such that $H$ is 2-path 2-coverable. We fix a 2-path 2-cover $\C$ of $H$.
Then we construct the corresponding TSP instance $(G,c)$  with the following steps.

\paragraph{Assigning \vg gadgets.}
We assign a $(2k-6)$-\vg gadget to each $H$-vertex.
In particular, for $k=3$, we use the stick gadget, and for $k=4$, we use the buoy gadget.
For convenience, we denote by $\cv(v)$ the \vg gadget assigned to an $H$-vertex $v$.

\paragraph{Assigning \eg gadgets.}
We assign one \eg gadget to each path in $\C$.
In particular, for each path $P$ in $\C$, we fix an endpoint as the start and the other as the end.
Suppose the order of the vertices in $P$ from start to end is $(v_1, \dots, v_t)$, for some $t \in \{1,2,3\}$.
We assign a \eg-$t$ gadget $\ce(P)$ to the path $P$.
For $i \in \{1,2,3\}$, we attach the $i$-side of $\ce(P)$ to a side of $\cv(v_i)$ such that for every \vg gadget, its two sides are attached to exactly two distinct \eg gadgets.
The \defi{label} $\labv_{P}$ of $\ce(P)$ is a vector in $\bins^t$ such that for $i \in [t]$, $\ce(P)$ is attached to the $\labv_P[i]$-side of $v_i$. 
We say that the $H$-vertices and the $H$-edges in the path $P$ are \emph{related} to $\ce(P)$.

We now assign the agreeing and disagreeing weights for a \eg-$t$ gadget $\ce(P)$.
If $t = 1$, there is nothing to do.
If $t \in \{2,3\}$, for each $i \in [t-1]$, we assign the weights as follows.
If $\labv_{P}[i]=\labv_P[i+1]$, then $\sigma^{\ce(P)}_{i(i+1)} = 0$ and $\delta^{\ce(P)}_{i(i+1)} = -w(v_i v_{i+1})$; otherwise, $\sigma^{\ce(P)}_{i(i+1)} = -w(v_i v_{i+1})$ and $\delta^{\ce(P)}_{i(i+1)} = 0$.
(Recall that $w(v_i v_{i+1})$ refers to the weight of the $H$-edge $v_i v_{i+1}$.)

\paragraph{Identifying PV-portals.}
Let $P$ be a path $(v_1, \dots, v_t)$ in $\C$ for some $t \in \{1,2,3\}$.
For $i \in [t]$, we identify the portals $X^{\ce(P)}_i$ and $\br{X}^{\ce(P)}_i$ of the \eg gadget $\ce(P)$ with the portals $X^{\cv(v_i)}_{\labv_P[i]}$ and $\br{X}^{\cv(v_i)}_{\labv_P[i]}$ of the \vg gadget $\cv(v_i)$, respectively.
After the identification, we also refer to these vertices as $X^{\ce(P)\cv(v_i)}$ and $\br{X}^{\ce(P)\cv(v_i)}$.
(Note that we do not need the subscripts, since the pair $\ce(P)$ and $\cv(v_i)$ is sufficient to deduce the subscripts of the original portals.)

\paragraph{Identifying PP-portals.}
We arrange the \eg gadgets in an arbitrary cyclic order.
We then identify the PP-portal $Z^{\ce(P)}$ of a \eg gadget $\ce(P)$ with the PP-portal $\br{Z}^{\ce(P')}$ of the succeeding \eg gadget $\ce(P')$ in the order.
In other words, after the identification, each PP-portal is adjacent to two vertices, one in each \eg gadget it is part of.

\begin{figure}[ht!]
    \centering
    \hspace*{-6pt}
    \begin{tikzpicture}[every node/.style={scale=0.8}, scale=0.45]
    \tikzset{
      circ/.style={circle, draw, fill=black, inner sep=1.2pt},
      square/.style={rectangle, draw, fill=white, inner sep=2.5pt},
      blacksquare/.style={rectangle, draw, fill=black, inner sep=2.5pt},
      reddashed/.style={myred, dashed},
      firstdashed/.style={myorange, dashed},
      seconddashed/.style={mygreen, dashed},
      rededge/.style={myred, very thick},
      firstedge/.style={myorange, very thick},
      secondedge/.style={mygreen, very thick},
      blueedge/.style ={myred,very thick},
      vertexedge/.style = {black, very thick}
    }
    
        \foreach \x in {0,...,5} 
          \foreach  \y in {0,...,2} {
             \coordinate (v\x\y) at (\x * 6 + \y,0);  
           }
    
        \foreach \x/\y/\z in {1/18/1, 2/9/1, 3/3/-1, 4/15/-1, 5/27/-1, 6/30/1} {
            \coordinate (x\x) at (\y , 4 * \z);
            \coordinate (y\x) at (\y + 2, 4 * \z);
            \coordinate (a\x) at (\y , 6 * \z);
            \coordinate (b\x) at (\y + 2, 6 * \z);
        }
    
        \coordinate (m12) at ($(v12)!0.5!(v20)$);
        \coordinate (m14) at ($(v12)!0.5!(v40)-(0,0.98)$);
    
        \coordinate (tlf) at (0,6);
        \coordinate (tln) at (-2,4);
        \coordinate (trf) at (32,6);
        \coordinate (trn) at (34,4);
        \coordinate (blf) at (0,-6);
        \coordinate (bln) at (-2,-4);
        \coordinate (brf) at (32,-6);
        \coordinate (brn) at (34,-4);
        
        \fill[mygray, opacity=0.2] (a1) -- (v30) to[bend left = 25] (v32) -- (b1) -- (a1);
        \fill[mygray, opacity=0.2] (a2) -- (x2) -- (v00) to[bend left = 25] (v02) to[bend left=12.5] (v20) to[bend left = 25] (v22) to[bend left=12.5] (v30) to[bend left = 25] (v32) -- (y2) -- (b2) -- (a2);
        \fill[mygray, opacity=0.2] (a3) -- (x3) -- (v00) to[bend right = 25] (v02) to[bend right=12.5] (v10) to[bend right = 25] (v12) -- (y3) -- (b3) -- (a3);
        \fill[mygray, opacity=0.2] (a4) -- (x4) -- (v10) to[bend right = 25] (v12) -- (v20) to[bend right = 25] (v22) to[bend right=12.5] (v40) to[bend right = 25] (v42) -- (y4) -- (b4) -- (a4);
        \fill[mygray, opacity=0.2] (a5) -- (x5) -- (v40) to[bend right = 25] (v42) to[bend right=12.5] (v50) to[bend right = 25] (v52) -- (y5) -- (b5) -- (a5); 
        \fill[mygray, opacity=0.2] (a6) -- (v50) to[bend left = 25] (v52) -- (b6) -- (a6); 
    
        \draw[rededge] (a1) -- (x1) -- (y1) -- (b1) (a2) -- (x2) (y2) -- (b2) (a3) -- (x3) -- (y3) -- (b3) (a4) -- (x4) (y4) -- (b4) (a5) -- (x5) (y5) -- (b5) (a6) -- (x6) (y6) -- (b6);
        \draw[reddashed] (x2) -- (y2) (x4) -- (y4) (x5) -- (y5) (x6) -- (y6);
    
        \draw[vertexedge] (v00) -- (v02) (v10) -- (v12) (v20) -- (v22) (v30) -- (v32) (v40) -- (v42) (v50) -- (v52);
        
        \draw[secondedge, dashed] (x1) -- (v30) (y1) -- (v32);
        
        \draw[firstedge] (x2) -- (v00);
        \draw[firstdashed] (y2) -- (v02);
        \draw[firstdashed] (x2) -- (v20) (y2) -- (v22);
        \draw[firstdashed] (x2) -- (v30);
        \draw[firstedge] (y2) -- (v32);
        \draw[firstedge] (v02) to[bend left=12.5] (v30);
        \draw[firstdashed] (v02) to[bend left=12.5] (v20) (v22) to[bend left=12.5] (v30);
        
        \draw[seconddashed] (x3) -- (v00) (y3) -- (v02);
        \draw[seconddashed] (x3) -- (v10) (y3) -- (v12);
        \draw[seconddashed] (v02) to[bend right=12.5] (v10);
        
        \draw[firstedge] (x4) -- (v10);
        \draw[firstdashed] (y4) -- (v12);
        \draw[seconddashed] (x4) -- (v20); 
        \draw[secondedge] (y4) -- (v22);
        \draw[seconddashed] (x4) -- (v40) (y4) -- (v42);
        \draw[firstedge] (v12) -- (m12);
        \draw[secondedge] (m12) -- (v20);
        \draw[seconddashed] (v22) to[bend right=12.5] (v40);
        \draw[firstdashed] (v12) to[bend right=5] (m14);
        \draw[seconddashed] (m14) to[bend right=5] (v40);

        \draw[firstedge] (x5) -- (v40) (y5) -- (v42);
        \draw[firstdashed] (x5) -- (v50) (y5) -- (v52);
        \draw[firstdashed] (v42) to[bend right=12.5] (v50);
        
        \draw[secondedge] (x6) -- (v50) (y6) -- (v52);
    
        \draw[blueedge] (b2) -- (a1) (b1) -- (a6) (b3) -- (a4) (b4) -- (a5) (a2) -- (tlf) (b6) -- (trf) (a3) -- (blf) (b5) -- (brf) (tln) -- (bln) (trn) -- (brn) (tlf) to[bend right=50] (tln) (bln) to[bend right=50] (blf) (brf) to[bend right=50] (brn) (trn) to[bend right=50] (trf);
    
        \foreach \x in {0,...,5} {
            \node[square] at (v\x0) {};
            \node[circ] at (v\x1) {};
            \node[square] at (v\x2) {};
        }
    
        \foreach \x in {1,...,6} {
            \node[circ] at (x\x) {};
            \node[circ] at (y\x) {};
        }

        \node[blacksquare] at ($(b3)!0.5!(a4)$) {};
        \node[blacksquare] at ($(b4)!0.5!(a5)$) {};
        \node[blacksquare] at ($(b2)!0.5!(a1)$) {};
        \node[blacksquare] at ($(b1)!0.5!(a6)$) {};
        \node[blacksquare] at ($(tln)!0.5!(bln)$) {};
        \node[blacksquare] at ($(trn)!0.5!(brn)$) {};
    
        \draw (1,-0.5) node {$a$};
        \draw (7,0.35) node {$f$};
        \draw (13,0.5) node {$b$};
        \draw (19,-0.35) node {$e$};
        \draw (25,0.5) node {$c$};
        \draw (31,-0.5) node {$d$};
    
        \draw ($(x1)!0.5!(b1)$) node {$P_1$};
        \draw ($(x2)!0.5!(b2)$) node {$P_2$};
        \draw ($(x3)!0.5!(b3)$) node {$P_3$};
        \draw ($(x4)!0.5!(b4)$) node {$P_4$};
        \draw ($(x5)!0.5!(b5)$) node {$P_5$};
        \draw ($(x6)!0.5!(b6)$) node {$P_6$};
    
    \end{tikzpicture}
    \caption{The full construction of the TSP instance as represented in \cref{fig:schematic-construction}(c) for $k = 3$. Black horizontal 2-paths are the stick gadgets. Shaded areas indicate the \eg gadgets. The green and orange edges indicate whether the \eg gadgets are attached to the 1- or the 2-side of the related \vg gadgets, respectively. The black squares are the PP-portals, and the white squares are the PV-portals. Note that each PP-portal is shared by two \eg gadgets. The solid edges form a tour; this tour corresponds to the cut where $V_1 = \{b, d\}$ and $V_2 = \{a, c, e, f\}$.}
    \label{fig:reduction_full}
\end{figure}
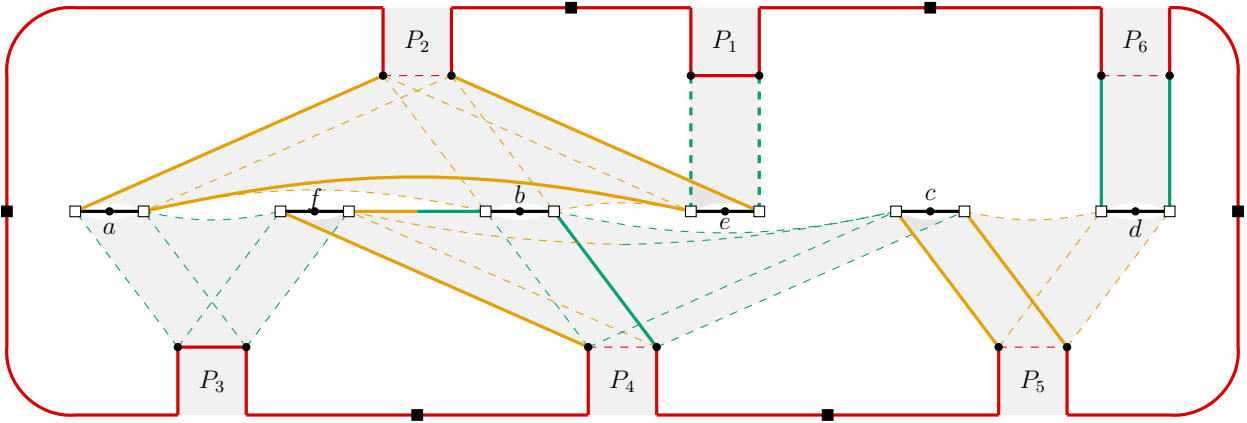

\paragraph{Completing the graph.}
Let $G$ and $c$ be the resulting graph and weight function after the four steps above; see \cref{fig:reduction_full} for an illustration of $G$.
As a TSP instance requires a complete graph, we add the remaining edges 
with very high weight so that they do not interfere with the construction so far. 
A possible choice for this high weight is to take the sum of the absolute values 
of all weights of the edges we have defined so far.
For ease of exposition we choose the weight $\infty$ and call the resulting graph~$G_{\infty}$. 
If we start with a tour with a finite total weight, the \kopt algorithm will never visit a tour that uses an edge with weight~$\infty$.
Hence, for the remaining of the reduction, we will argue based only on~$G$.

\begin{lemma}
\label{lem:G_simple}
    If $H$ has girth at least five, and if the PV-portals in the \vg gadgets are incident to only internal vertices of the gadget, then the graph $G$ is simple (i.e., with neither parallel edges nor loops). 
\end{lemma}
\begin{proof}

By definition, no base gadget contains loops or parallel edges, so any loops or parallel edges in $G$ can only arise from the two portal identification steps. 
    
    A loop arises in the identification steps exactly when two adjacent vertices are identified.
    A parallel edge can only arise, when two vertices with a common neighbor are identified with each other, or there are two edges $XY$ and $X'Y'$ such that $X,X',Y,Y'$ are four distinct vertices before the identification steps and $X$ and $X'$ as well as $Y$ and $Y'$ are identified with each other in the steps. 
   
    Let us first consider the case where all portals in every base gadget are distinct vertices.
    Before identification, every PP‑portal is adjacent only to an internal vertex of a \eg gadget. Since internal vertices are never identified with other vertices and we only identify PP-portals in distinct \eg gadgets, identifying PP‑portals cannot create loops or parallel edges.

    Now consider the identification of PV-portals.
    In a \vg gadget, a PV-portal is by assumption adjacent only to internal vertices, while in a \eg gadget, a PV-portal may also be adjacent to other PV-portals.
    Because internal vertices are not identified with any other vertices, and PV-portals of \eg gadgets are not identified with each other (but with distinct PV-portals of \vg gadgets), it follows that no parallel edge or loop arises from the identification of a PV-portal in this case. 
  
    Next, consider the case where different portals in a base gadget can be the same vertex.  
    Note first that all portals in the \eg gadgets are distinct and that in \vg gadgets only PV-portals belonging to different sides can coincide.
    Therefore nothing changes for the PP-portals and thus their identification cannot create parallel edges or loops. 

    For PV-portals the only thing that changes is that two PV-portals $X,Y$ in two different \eg gadgets $\ce_X, \ce_Y$ can now be identified with the same vertex $T$ in a \vg gadget $\cv$. 
    Since the three vertices that are identified with each other are all in different gadgets, this cannot give rise to any loops.

    There are two ways in which a parallel edge could arise here. 
    First, there could be parallel edges that would also arise in the case where all portals are distinct vertices. We know from above that such parallel edges do not arise. 
    Second, there could be another pair of vertices $X'$ and $Y'$ such that $XX'$ and $YY'$ are edges in $\ce_X$ and $\ce_Y$, respectively,
    and $X'$ and $Y'$ are identified with each other.

    As no PP-portal is adjacent to a PV-portal, $X'$ and $Y'$ have to be PV-portals and thus they have to be identified with a common vertex in a \vg gadget $\cv'$.
    Since PV-portals in the same side in the same \eg gadget are not adjacent, $X$ and $X'$ must belong to different sides of their common \eg gadget, and so must $Y$ and $Y'$.    
   
    The two \vg gadgets $\cv$ and $\cv'$ are hence distinct.
    This means that the paths in $H$ corresponding to the two \eg gadgets $\ce_X$ and $\ce_Y$ have two $H$-vertices in common (i.e., the two $H$-vertices corresponding to $\cv$ and $\cv'$).
    Hence there has to be a cycle of length at most four in $H$. We have a contradiction to the assumption on the girth of $H$.
\end{proof}
  
\section{Correctness of the construction for \texorpdfstring{\boldmath$k \in \{3, 4\}$}{k in \{3,4\}}}
\label{sec:correctness_3}
This section presents the correctness proof of the construction from Section~\ref{sec:construction_3}.
We first show that a family of 2-path 2-coverable Max-Cut instances with high girth exists 
in which an initial solution is exponentially many steps away from any local optimum (Section~\ref{sec:michel_scott}).
Then we introduce a subset of tours of $G$, which we call the standard tours (Section~\ref{subsec:standard_tours_3}), and  establish a correspondence between improving flip sequences in $H$ and improving swap sequences in $G$ that involve only standard tours (Section~\ref{subsec:corr_3}).
Finally, we prove \cref{thm:main} by showing that if we start with the standard tour corresponding to the initial cut of the Max-Cut instance $(H,w)$, we only encounter standard tours in any improving swap sequences of $(G,c)$ (Section~\ref{subsec:thmmain_proof}).

\subsection{The modified Michel-Scott construction}
\label{sec:michel_scott}
In this section, we present a family of Max-Cut instances $(H,w)$ such that $H$ has girth at least nine and is 2-path 2-coverable and we show that there exists an initial cut that is exponentially far away from any local optimum.
In fact, such a family is constructed in \cite{HHH2024} as a modification of the construction by Michel and Scott~\cite{Michel_Max_Cut_4}.

For completeness, we reproduce the construction, starting from the original Michel-Scott construction.
Since we only want to show structural properties of the unweighted graph, we omit the description of the weights. 
The graph for an instance in \cite{HHH2024} is constructed as follows.
\begin{itemize}
    \item The graph $F_0$ consists of a path $v_{0,1} v_{0,8}$.
    \item For $i \in [n]$, the graph $F_i$ contains a path of eight new vertices $v_{i,1}, v_{i,2}, v_{i,3}, v_{i,4}, v_{i,5}, \allowbreak v_{i,6}, v_{i,7}, \allowbreak v_{i,8}$ that appear in the path in that order.
    Next, we connect $F_i$ to $F_{i-1}$ as follows: We add edges connecting $v_{i-1,1}$ to $v_{i,2}, v_{i,4}$, and $v_{i,6}$, and add edges connecting $v_{i-1,8}$ to $v_{i,3}, v_{i,5}$, and $v_{i,7}$.
\end{itemize}

The final graph~$H_n$ consists of all the graphs $F_0, \dots, F_n$ and the connecting edges, as well as two new vertices $v'_1$ and $v'_2$ and two new edges $v_{n,1} v'_1$ and $v'_1 v'_2$. 

\begin{figure}[htb!]
	\centering
      \includegraphics[width=\linewidth, page=1]{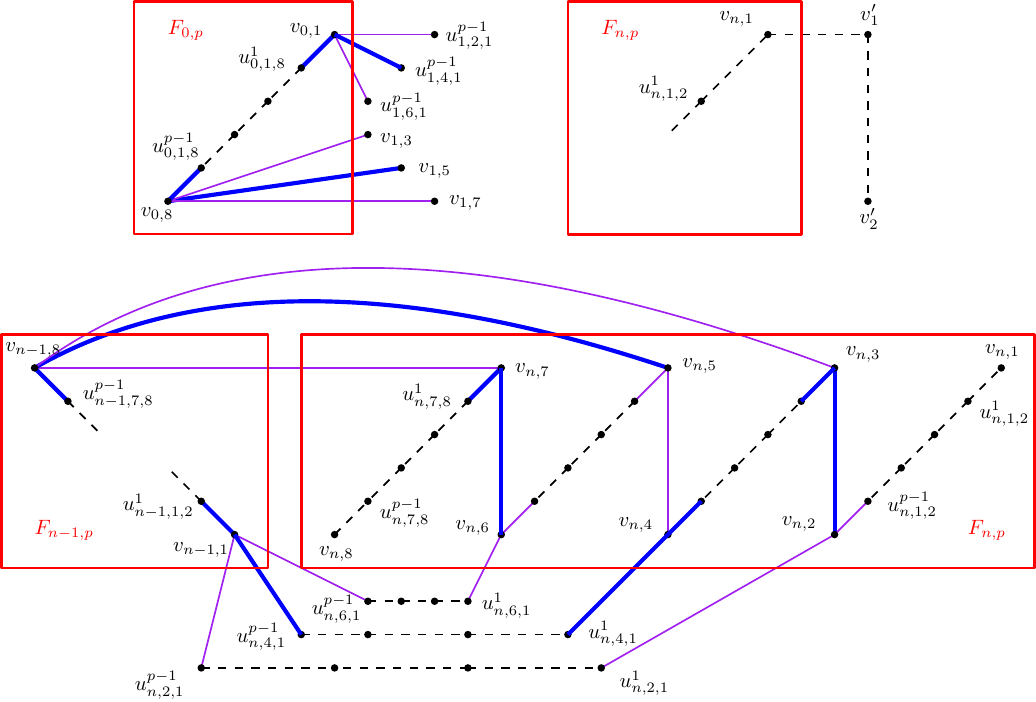}
      \caption{$H_{n,p}$ with a 2-path 2-cover. The paths of length two are either solid thick blue paths or solid thin purple paths. The dashed edges indicate paths of length one. 
     The two 0-paths  $(v'_2)$ and  $(v_{n,8})$  are not shown in this figure. 
      Note that the relative positions of $v_{n-1,1}$ and $v_{n-1,8}$ in $F_{n-1,p}$ are flipped 
      compared to those of $v_{n,1}$ and $v_{n,8}$ in $F_{n,p}$.
      }
      \label{fig:edge_gadget_assignment}
\end{figure}

The graph $H_n$ has girth~$4$. To obtain an instance with larger girth the following modification 
was described in \cite{HHH2024} (see~\cref{fig:edge_gadget_assignment}).
Let $p$ be an odd positive integer.
Consider the following pairs of vertices: $(v_{0,1}, v_{0,8})$, and for $i \in [n]$, 
$(v_{i,1}, v_{i,2})$, $(v_{i,3}, v_{i,4})$, $(v_{i,5}, v_{i,6})$, $(v_{i,7}, v_{i,8})$, 
$(v_{i,2}, v_{i-1,1})$, $(v_{i,4}, v_{i-1,1})$, $(v_{i,6}, v_{i-1,1})$.
For every pair of vertices $(v_{i,j}, v_{i',k'})$ above, we delete the edge $v_{i,j}, v_{i',k'}$ and replace it by a path of length $p$ $(v_{i,q}, u^{1}_{i,q,q'}, \dots, u^{p-1}_{i,q,q'}, v_{i',q'})$.
Let $H_{n,p}$ be the resulting graph, and for $i \in \{0, \dots, n\}$, let $F_{i,p}$ be the graph obtained from $F_i$ by the same replacement operation above.

\begin{lemma}[{\cite[Lemma 12]{HHH2024}}]
\label{lem:michel_scott_modified}
    For fixed odd $p\in\mathbb{N}$, there exists a weight function $w: H_{n,p} \to \mathbb{Q}$ and an initial cut $\gamma_{n,p}$ of $H_{n,p}$ such that there is a unique maximal improving flip sequence from $\gamma_{n,p}$ with length exponential in the number of vertices of $H_{n,p}$. 
\end{lemma}

We now show that the above graph $H_{n,p}$ has the desired property for the construction in our current paper.

\begin{lemma}
\label{lem:michel_scott_condi}
    For $p \geq 3$, $H_{n,p}$ has girth $p + 3$ and is 2-path 2-coverable.
\end{lemma}

\begin{proof}
    Consider the graph $\tilde{H}$ obtained from $H_{n,p}$ by removing all $p$-paths that are added in the process of creating $H_{n,p}$ (i.e., $\tilde{H}$ contains exactly all edges in $E(H_n) \cap E(H_{n,p})$.
    Note that $\tilde{H}$ is a forest with the following subtrees that are not isolated vertices: the path $(v_{n,1}, v'_1, v'_2)$ and the unions of three paths $(v_{i-1,8}, v_{i,3}, v_{i,2})$, $(v_{i-1,8}, v_{i,5}, v_{i,4})$, and $(v_{i-1,8}, v_{i,7}, v_{i,6})$ for $i \in [n]$.
    Hence, to obtain a cycle in $H_{n,p}$, we need to add at least one $p$-path.
    If we need to add two $p$-paths, then the cycle has length at least $2p$.
    If we add exactly one $p$-path, then it is easy to see that this $p$-path must be the one connecting $v_{i,j}$ and $v_{i,j+1}$ for some $i \in [n]$ and $j \in \{3,5\}$.
    We then obtain a cycle of length $p + 3$.
    Since $p \geq 3$, it follows that the smallest cycle in $H_{n,p}$ has length $p + 3$.

    To conclude the proof, we show a 2-path 2-cover of $H_{n,p}$ as follows; see \cref{fig:edge_gadget_assignment} for an illustration.
    We have the following 2-paths in the cover:
    \begin{itemize}
        \item For $i \in [n]$, $(v_{i,2}, v_{i,3}, u^{1}_{i,3,4}), (v_{i,4}, v_{i,5}, u^{1}_{i,5,6})$, $(v_{i,6}, v_{i,7}, u^{1}_{i,7,8})$;
        \item For $i \in [n]$,  $(u^{p-1}_{i,1,2}, v_{i,2}, u^{1}_{i,2,1}), (u^{p-1}_{i,3,4}, v_{i,4}, u^{1}_{i,4,1}), (u^{p-1}_{i,5,6}, v_{i,6}, u^{1}_{i,6,1})$;
        \item For $i \in \{2, \dots, n\}$, $(u^{p-1}_{i,2,1}, v_{i-1,1}, u^{p-1}_{i,6,1})$, $(u^{p-1}_{i,4,1}, v_{i-1,1}, u^{1}_{i-1,1,2})$, $(v_{i,3}, v_{i-1,8}, v_{i,7})$, and $(v_{i,5}, v_{i-1,8}, u^{p-1}_{i-1,7,8})$; 
        \item $(u^{p-1}_{1,2,1}, v_{0,1}, u^{p-1}_{1,6,1}), (u^{p-1}_{1,4,1}, v_{0,1}, u^{1}_{0,1,8})$, $(v_{1,3}, v_{0,8}, v_{1,7})$, and $(v_{1,5}, v_{0,8}, u^{p-1}_{0,1,8})$.
    \end{itemize} 
    The other edges in $H_{n,p}$ are covered by 1-paths.
    Finally, there are two 0-paths: $(v_2')$ and $(v_{n,8})$.
    It can be verified that the set of paths described above form a 2-path 2-cover.
\end{proof}

\subsection{Standard tours}
\label{subsec:standard_tours_3}
Let $\tau$ be a subgraph of $G$.
For a gadget $\zeta$, the \defi{subtour} of $\zeta$ in $\tau$ is the subgraph of $\zeta$ that contains all common edges of $\zeta$ and $\tau$, i.e., it is the graph $(V(\zeta), E(\zeta) \cap E(\tau))$.
We say $\tau$ is \defi{standard}, if the subtour of every gadget in $\tau$ is standard.

Note that the even and odd subtours of a \vg gadget may have the same underlying path, such as when it is a stick gadget.
For clarity, we stipulate the following for each $H$-vertex $v$ and a standard tour $\tau$ in such case.
Let $\ce(P)$ be one of the \eg gadgets related to $v$ and let $\vect{s} \in \bin^t$ be such that $\ce(P)$ has the $\vect{s}$-subtour in $\tau$. 
Further suppose $v$ is attached to the $i$-side of $\ce(P)$. 
Then $\tau$ uses the odd subtour to traverse $\cv(v)$ if and only if we have $\labv_P[i] \equiv \vect{s}[i] \pmod{2}$. 
Otherwise it uses the even subtour. 
Note that this definition is independent on which one of the two related path gadgets is chosen, since the parity of both sides of the equivalence relation is exchanged for the two path gadgets. 

The following lemma shows that this characterization of the even/odd subtour in a tour can also be used when the underlying graph of the even and odd subtours are distinct.

\begin{lemma}
\label{lem:vx_fix_edges}
    Let $\tau$ be a standard tour in $G$, and let $\ce$ be a \eg-$t$ gadget for some $t \in \{1,2,3\}$.
    Let $P = (v_1, \dots, v_t)$ be a path in $\C$, and let $\vect{s} \in \bin^t$ be such that $\ce(P)$ has the $\vect{s}$-subtour in $\tau$.
    Then for $i \in [t]$, $\cv(v_i)$ has the odd subtour in $\tau$, if and only if $\labv_P[i] \equiv \vect{s}[i] \pmod{2}$.
\end{lemma}
\begin{proof}
	If the even and odd subtours of $\cv(v)$ share the same underlying path, (e.g., when $\cv(v)$ is a stick gadget), then the lemma follows by the stipulation of odd and even tours above.
    Otherwise, if $\vect{s}[i] = 1$, this implies that for $X^{\cv(v)}_{\labv_P[i]}$ and $\br{X}^{\cv(v)}_{\labv_P[i]}$, exactly one of their incident edges in $\tau$ is in $\ce(P)$.
	Hence, they have exactly one incident edge in $\cv(v)$.
	Since $\tau$ is standard, this means that the subtour of $\cv(v)$ in $\tau$ is a $X^{\cv(v)}_{\labv_P[i]}\br{X}^{\cv(v)}_{\labv_P[i]}$-path.
	Similarly, if $\vect{s}[i] = 0$, then the subtour of $\cv(v)$ in $\tau$ is \emph{not} an  $X^{\cv(v)}_{\labv_P[i]}\br{X}^{\cv(v)}_{\labv_P[i]}$-path.
	Since the subtour of $\tau$ in $\cv(v)$ is standard, the lemma then follows.
\end{proof}

We say a standard tour~$\tau$ in $G$ \defi{corresponds} to a cut~$\gamma$ of $H$, if for every $H$-vertex $v$, $v$ is in the even set of $\gamma$ if and only if the corresponding \vg gadget has the even subtour in $\tau$.

\begin{lemma}
\label{lem:standard_tour_for_every_cut}
    For every cut of $H$, there exists a corresponding standard tour in $G$.
\end{lemma}
\begin{proof}
    Let $\gamma$ be a cut of $H$.
    We construct the following subgraph $\tau$ of $G$.
    For every $H$-vertex $v$, if $v$ is in the 1-set of $\gamma$, we add the odd subtour of $\cv(v)$ into $\tau$.
    For every path $P = (v_1, \dots, v_t)$ in $\C$ for some $t \in \{1, 2, 3\}$, let $\vect{s}_P \in \bin^t$ be the vector that satisfies $\vect{s}_P[i] \equiv \labv_P[i] \pmod{2}$ if and only if $v_i$ is in the 1-set of $\gamma$.
    Then we add to $\tau$ the $\vect{s}_P$-subtour of $\ce(P)$, which always exists by \cref{lem:edge_unique_subtours}.

    Next, we show that $\tau$ is a tour of $G$.
    Consider a path $P = (v_1, \dots, v_t)$ in $\C$.
    By \cref{lem:edge_unique_subtours} and the definition of the standard subtours of \vg gadgets, it is easy to verify that in $\tau$, there is a path connecting the two PP-portals of $\ce(P)$ that visits exactly all internal vertices of $\ce(P)$ and all the vertices of $\cv(v_i)$ for $i \in [t]$ such that $\vect{s}_P[i] = 1$.
    Now consider an $H$-vertex $v$. 
    Suppose it appears in the $i_1$-th and $i_2$-th positions in paths $P_1$ and $P_2$ in $\C$.
    Then by construction, $\labv_{P_1}[i_1] \neq \labv_{P_2}[i_2]$, and hence, $\vect{s}_{P_1}[i_1] \neq \vect{s}_{P_2}[i_2]$.
    All of the above, combined with the cyclic connection of the \eg gadgets via the PP-portals, imply that $\tau$ is a tour.

    Then by construction, $\tau$ is standard and corresponds to $\gamma$.
    This concludes the proof of the lemma.
\end{proof}

Lemmas~\ref{lem:edge_unique_subtours}, \ref{lem:vx_fix_edges}, and~\ref{lem:standard_tour_for_every_cut} imply that the correspondence between cuts of $H$ and standard tours of $G$ is one-to-one.

\subsection{Correspondence between flip sequences and swap sequences that involve only standard tours}
\label{subsec:corr_3}
In this section, we prove the correspondence between flip sequences for $(H,w)$ and swap sequences for $(G, c)$ that contain only standard tours.

\begin{lemma}
\label{lem:corr_obj}
    Let $\tau$ be the standard tour in $G$ that corresponds to a cut $\gamma$ of $H$.
    Then the weight of $\tau$ and the value of $\gamma$ sum up to 0. 
\end{lemma}
\begin{proof}
	Let $P = (v_1, \dots, v_t)$ be a path in the 2-path 2-cover $\C$ of $H$. 
    Let $\vect{s} \in \bin^t$ be the vector such that $\ce(P)$ has the $\vect{s}$-subtours in $\tau$.
    For $i \in [t-1]$, let $\psi_i$ be the agreeing weight $\sigma^{\ce(P)}_{i(i+1)}$ if $\vect{s}[i] = \vect{s}[i+1]$ and the disagreeing weight $\delta^{\ce(P)}_{i(i+1)}$ otherwise. 
    By the choice of the agreeing and disagreeing weights in the construction of $(G,c)$, it can be easily verified that if $\vect{\ell}[i] - \vect{\ell}[i+1] \equiv \vect{s}[i] - \vect{s}[i+1] \pmod{2}$, then $\psi_i = 0$; otherwise $\psi_i = -w(v_i v_{i+1})$.
    Further, by \cref{lem:vx_fix_edges}, $v_iv_{i+1}$ is not in the cut-set (i.e., $v_i$ and $v_{i+1}$ are in the same set of the cut), if and only if $\vect{\ell}[i] - \vect{\ell}[i+1] \equiv \vect{s}[i] - \vect{s}[i+1] \pmod{2}$.
    The preceding two sentences imply that the contribution of the edge $v_iv_{i+1}$ to the value of the cut $\gamma$ is equal to $-\psi_i$.
    The lemma then follows from a combination of this statement, \cref{lem:edge_unique_subtours}, and the assumption that each $H$-edge is in exactly one path of $\C$.
\end{proof}

\begin{lemma}
\label{lem:corr_nbr}
    Two cuts $\gamma_1$ and $\gamma_2$ differ by a flip, if and only if the two corresponding standard tours $\tau_1$ and $\tau_2$ differ by a $k$-swap.
\end{lemma}
\begin{proof}
    First, suppose $\gamma_1$ and $\gamma_2$ differ in an $H$-vertex $x$.
    The tours $\tau_1$ and $\tau_2$ differ exactly in the following gadgets:
    \begin{itemize}
        \item The \vg gadget $\cv(x)$ has the odd subtour in one tour and the even subtour in the other.
        \item For $i \in \{1,2\}$, the \eg gadget $\ce_i$ that is attached to the $i$-side of $\cv(x)$ has some $\vect{s}_{i1}$-subtour and $\vect{s}_{i2}$-subtour in $\tau_1$ and $\tau_2$, respectively, such that the Hamming distance between $\vect{s}_{i1}$ and $\vect{s}_{i2}$ is one.
    \end{itemize}
    Since the \vg gadget is a $(2k-6)$-\vg gadget, together with \cref{lem:eg_involved_edges}, we conclude that a swap to transform $\tau_1$ to $\tau_2$ involves exactly $2k$ edges.
    Hence, the two tours differ by a $k$-swap.

    Last, we show that if $\gamma_1$ and $\gamma_2$ differ by more than one flip, then $\tau_1$ and $\tau_2$ differ by a $k'$-swap with $k' > k$.
    By assumption, $\gamma_1$ and $\gamma_2$ differ in at least two $H$-vertices $x$ and $y$.
    Since every edge is in exactly one path in $\C$, $x$ and $y$ share at most one \eg gadget.
    Hence, there are at least three \eg gadgets whose subtours in $\tau_1$ and $\tau_2$ differ.
    Since changing between two (standard) subtours of a $(2k-6)$-vertex gadget involves $2k-6$ edges, and since changing between two standard subtours of a \eg gadget involves at least three edges by \cref{lem:eg_involved_edges}, the number of involved edges to change between $\tau_1$ and $\tau_2$ is at least $2k+3$ edges.
    Hence, they cannot differ by a $k$-swap.

    The lemma then follows.
\end{proof}

Then the previous two lemmas imply the following.

\begin{lemma}
\label{lem:correspondence}
    There is a one-to-one correspondence between improving flip sequences for $(H,w)$ and improving $k$-swap sequences for $(G,c)$ that contain only standard tours.
\end{lemma}
\begin{proof}
    Consider the transition graph $T$ for $(H,w)$ where the cuts of $H$ define the vertices, and the improving flips define the directed edges.
    Similarly, for $(G,c)$, let $T'$ be the graph whose vertices are the standard tours of $G$ and whose directed edges are defined by an improving $k$-swap between two standard tours.
    Lemmas~\ref{lem:corr_obj} and~\ref{lem:corr_nbr} and the one-to-one correspondence between cuts of $H$ and standard tours of $G$ imply that $T$ and $T'$ are isomorphic.
    Observe that improving flip sequences for $(H,w)$ and improving $k$-swap sequences that contain only standard tours for $(G,c)$ correspond to directed paths in $T$ and $T'$, respectively.
    The lemma then follows.
\end{proof}

\subsection{The all-exp property for \texorpdfstring{\boldmath$k \geq 3$}{k >= 3}}
\label{subsec:thmmain_proof}
The only missing ingredient for the main result (\cref{thm:main}) is the lemma below that will be useful to argue that if $H$ has a suitable girth $g$, then any $k$-swap sequence from a standard tour of $G$ can only contain standard tours.

\begin{lemma}
\label{lem:s_to_ns_tour}
    Suppose $H$ has girth $g$, and the \vg gadgets in $G$ are either the stick gadgets or the buoy gadgets.
    Let $\tau$ and $\tau'$ be tours of $G$ such that $\tau$ is standard while $\tau'$ is not.
    Then the swap transforming $\tau$ to $\tau'$ involves at least $g/2$ edges.
\end{lemma}
\begin{proof}
    For a \eg gadget $\ce$ and a tour $\bar{\tau}$, we say a PV-portal of $\ce$ is \defi{active} if its degree in the subtour of $\ce$ in $\bar{\tau}$ is one.
    We say the $i$-side of $\ce$ is \defi{$\bar{\tau}$-consistent} if ${X^{\ce}_i}$ and $\br{X}^{\ce}_i$ are either both active or both inactive.
    We also extend this notion to \vg gadgets as follows: Suppose the $i$-side of a \vg gadget $\cv$ is attached to the $j$-side of a \eg gadget $\ce$; then this $i$-side of $\cv$ is \emph{$\bar{\tau}$-consistent} if the $j$-side of $\ce$ is $\bar{\tau}$-consistent.
    It is easy to check the following:
    \begin{itemize}
        \item[(*)] For any tour $\bar{\tau}$, the two sides of a stick gadget or a buoy gadget are either both $\bar{\tau}$-consistent or both $\bar{\tau}$-inconsistent.
    \end{itemize} 

    Further, observe that in a stick gadget or a buoy gadget, a PV-portal is adjacent to an internal vertex of degree two.
    Therefore, in any tour of $G$, a PV-portal always has at least one incident edge in a \vg gadget, and hence, its degree in the subtour of a \eg gadget can be either zero or one.
    Combined with \cref{lem:eg_2_non_standard_sides_in_ns_tours}, this implies that
    \begin{itemize}
        \item[(**)] A \eg gadget that has a non-standard subtour in $\tau'$ has exactly two $\tau'$-inconsistent sides.
    \end{itemize} 
    
    We then create the following auxiliary graph $F$ whose vertices are the gadgets used in the construction.
    We add an edge between a \eg gadget $\ce$ and \vg gadget $\cv$ if exactly one of the vertices $X^{\ce\cv}$ and $\br{X}^{\ce\cv}$ is active in $\tau'$ (i.e., the corresponding sides in the two gadgets are $\tau'$-inconsistent).
    By (*) above, each \vg gadget has either degree zero or degree two in $F$.
    By (**) above, for a \eg gadget, if it has a non-standard subtour in $\tau'$ then it has degree two in $F$; otherwise, it has degree zero.
    Since the non-standard tours of the stick and buoy gadgets share the same underlying path with their standard tours, at least one \eg gadget must have a non-standard subtour in order for $\tau'$ to be a non-standard tour; that is, there is at least one edge in $F$. 
    Then by the degree constraint above, $F$ is a collection of cycles.

    Next, consider a cycle $C = \big(\ce(P_1), \cv(v_1), \dots, \ce(P_{\ell}), \cv(v_{\ell})\big)$ of $F$.
    By construction, for any $i \in [\ell]$, there is a $v_{i-1}v_{i}$-path that is a subpath of $P_{i}$ (we define $v_0 = v_{\ell}$).
    Further, each $H$-edge appears exactly in one path of $\C$, the paths $P_i$ are edge-disjoint.
    Since each path $P_i$ has length at most two, the preceding sentences imply that $(v_1, \dots, v_{\ell})$ is a subsequence of a cycle $K$ in $H$ and $\ell \geq |K|/2$.
    Since $H$ has girth at least $g$, we have $\ell \geq g/2$.
    This means that there are at least $g/2$ \eg gadgets with non-standard subtours in $\tau'$.
    Since $\tau$ is standard, the swap to transform $\tau$ to $\tau'$ must involve at least one edge in each of the \eg gadgets above.
    The lemma then follows.
\end{proof}

\thmmain*
\begin{proof}
    We define $H = H_{n,15}$ (i.e., $p = 15$) and let $w$ and $\gamma_{n,p}$ be the weight function and the initial cut as guaranteed in \cref{lem:michel_scott_modified}.
    By \cref{lem:michel_scott_condi}, $H$ is 2-path 2-coverable and has girth 18.
    Hence, we can apply the construction above to obtain a graph $G$ with weight $c$.
    Let $\tau_{n,p}$ be the standard tour of $G$ corresponding to $\gamma_{n,p}$.
    By \cref{lem:s_to_ns_tour}, any swap to transform a standard tour to a non-standard tour involves more than eight edges and hence cannot be a 4-swap.
    Since $k \leq 4$, this implies that any improving $k$-swap starting from the standard tour $\tau_{n,p}$ can only visit standard tours.
    Then combining this with Lemmas \ref{lem:correspondence} and \ref{lem:michel_scott_modified} yields the all-exp property for $k \in \{3, 4\}$.
\end{proof}

\section{Alternative proof for \texorpdfstring{\boldmath$k \geq 5$}{k >= 5}}
\label{sec:k_5}
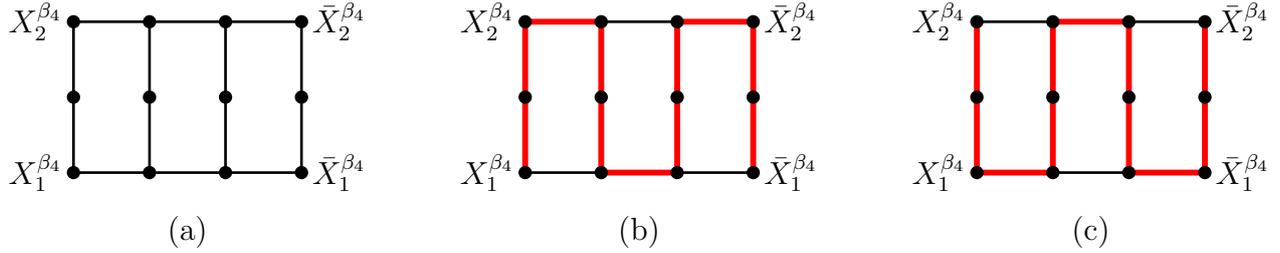
\begin{figure}
    \tikzset{redline/.style ={draw=red, line width = 2.2}}
    \tikzset{blackline/.style ={draw=black, line width = 1}}
    \pgfdeclarelayer{back}
    \pgfsetlayers{back,main}

    \def\defvertices{    
    \foreach \x in {1,...,4} 
      \foreach  \y in {1,2,3} {
         \fill (\x, \y) circle(2.5pt);
         \coordinate (a\x\y) at (\x,\y);}  
    \coordinate[label=left: $X^{\beta_4}_1$] (dummy) at (a11);
    \coordinate[label=left: $X^{\beta_4}_2$] (dummy) at (a13);
    \coordinate[label=right: $\br{X}^{\beta_4}_1$] (dummy) at (a41);
    \coordinate[label=right: $\br{X}^{\beta_4}_2$] (dummy) at (a43);}

    \hspace*{-19pt}
    \begin{tabular}{c@{~~~~~~~}c@{~~~~~~~}c}
    \begin{tikzpicture}
    \defvertices
    \foreach \x in {1,...,4} 
      \draw[blackline] (a\x1) -- (a\x3);
 
    \draw[blackline] (a11) -- (a41) (a13) -- (a43);
    \end{tikzpicture}
&
    \begin{tikzpicture}
    \defvertices
    \begin{pgfonlayer}{back}   
    \foreach \x in {1,...,4} 
      \draw[redline] (a\x1) -- (a\x3);
    \draw[redline] (a13) -- (a23) (a33)--(a43) (a21) -- (a31);
    \draw[blackline] (a11) -- (a21) (a31)--(a41) (a23) -- (a33);
    \end{pgfonlayer}
    \end{tikzpicture}
&
    \begin{tikzpicture}
    \defvertices
    \begin{pgfonlayer}{back}   
    \foreach \x in {1,...,4} 
      \draw[redline] (a\x1) -- (a\x3);
    \draw[blackline] (a13) -- (a23) (a33)--(a43) (a21) -- (a31);
    \draw[redline] (a11) -- (a21) (a31)--(a41) (a23) -- (a33);
    \end{pgfonlayer}
    \end{tikzpicture}\\ (a) & (b) & (c)
    \end{tabular}

    \caption{The XOR gadget of order four (a) and its two standard subtours ((b) and (c)).} 
    \label{fig:xor_gadget}
\end{figure}

Recall that the all-exp property for $k \geq 5$ is already shown in~\cite{HHH2024}.
By combining different \vg gadgets with our construction in Section~\ref{sec:construction_3}, we can provide an alternative proof for these values of $k$ to that in~\cite{HHH2024}.
In particular, we use the following XOR gadgets, which are introduced in the same paper~\cite{HHH2024} and generalized from~\cite{pap1978}; see \cref{fig:xor_gadget} for an illustration.

\begin{definition}[XOR gadget]
\label{def:xor_gadget}
    Let $q\geq 1$ be an integer.
    The \defi{XOR gadget} $\beta_q$ of order~$q$ is a graph containing two paths $(a_1, \dots, a_{q})$ and $(b_1, \dots, b_{q})$, and for $i \in \{1, \dots, q\}$, there is a path of length two with $a_i$ and $b_i$ as endpoints. If $q$ is even, we define $X^{\beta_q}_1$, $X^{\beta_q}_2$, $\br{X}^{\beta_q}_1$, and $\br{X}^{\beta_q}_2$ by $a_1,b_1,a_q$ and $b_q$, respectively. Otherwise, we define them as $a_1,b_1,b_q$ and $a_q$, respectively.
\end{definition}

Note that the stick gadget is the XOR-gadget of order one.

It can be easily verified that for all $q\geq 3$ the XOR-gadget of order $q$ has only standard subtours. 
Changing between the even and the odd subtour involves exactly the edges on the two paths of length $q-1$. Thus we get the following result.  

\begin{observation}\label{obs:strict_vertex_gadgets}
    For $q\geq3$, the XOR gadget of order $q$ is a $(2q-2)$-vertex gadget and has only standard subtours.
\end{observation}

\begin{theorem}
\label{thm:all_exp_5}
    TSP/\kopt has the all-exp property for $k \geq 5$.    
\end{theorem}
\begin{proof}
We use the same construction as in the proof of \cref{thm:main}, except that we use the XOR-gadget of order $k-2$ as the \vg gadget for all $H$-vertices.

By \cref{obs:strict_vertex_gadgets}, these \vg gadgets only have standard subtours.
This implies that for any tour in $G$ (i.e., the graph of the TSP instance obtained from the construction), the subtours of all \eg gadgets in this tour have to be standard.
In other words, all tours of $G$ are standard.

Finally, observe that the proof of the correspondences in Lemmas~\ref{lem:corr_obj} and~\ref{lem:corr_nbr} only rely on the fact that the \vg gadgets in the construction are $(2k-6)$-vertex gadgets (as described in Section~\ref{sec:construction}.
Since the XOR-gadgets of order $k-2$ are indeed $(2k-6)$-vertex gadget, by \cref{obs:strict_vertex_gadgets}, the correspondence in \cref{lem:correspondence} still holds.
The theorem then follows.
\end{proof}

\begin{remark}
    The proof of \cref{thm:all_exp_5} above asserts that every tour of the TSP instance has to be standard.
    This implies that the reduction for $k \geq 5$ preserves the transition graph.
    In other words, the transition graph of the Max-Cut/Flip instance and that of the corresponding TSP/\kopt instance are the same.
    Note that this preservation does not hold for our earlier reduction for $k \in \{3,4\}$ or that for $k = 2.5$ later.
\end{remark}

\section{Construction for \texorpdfstring{\boldmath$k = 2.5$}{k = 2.5}}
\label{sec:construction_2_5}
In this section, we first present the new modified Michel-Scott construction for Max-Cut/Flip in Section~\ref{sec:michel_scott_further_modified}.
After that, we discuss the new gadgets and construction used for $k = 2.5$.
Due to the substantial differences compared to the construction in the previous section, for clarity and simplicity, we provide a fresh description of the gadgets and the construction for this section, instead of redefining \vg and \eg gadgets. 
Similar to the treatments of the \sg gadgets as discussed in Section~\ref{subsec:const_summary_2_5}, we use a new term ``\ng gadget'' to call the gadgets modeling the $H$-vertices instead of \vg gadgets.
However, we keep the terms PV- and PP-portals.

\subsection{Further modified Michel-Scott construction}
\label{sec:michel_scott_further_modified}
\begin{figure}[ht!]
    \centering
    \includegraphics[width=\linewidth, page=3]{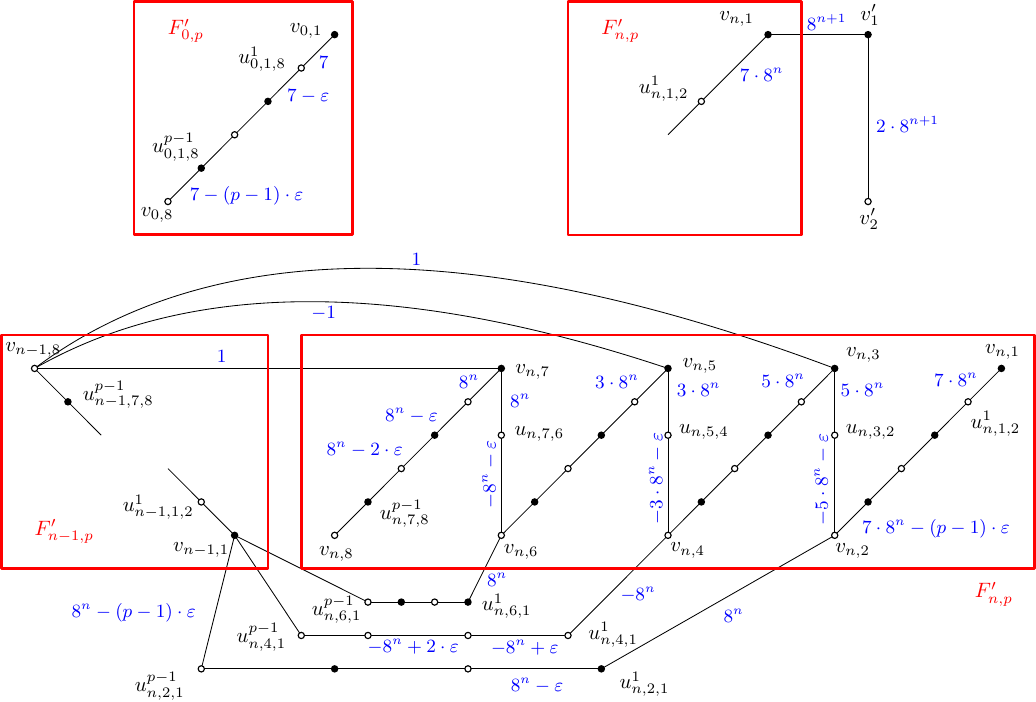}
    \caption{Further modified Michel-Scott construction. The colors of the vertices indicate which sets of the initial cut $\gamma'_{n,p}$ they are in.
    Note that the relative positions of $v_{n-1,1}$ and $v_{n-1,8}$ in $F'_{n-1,p}$ are flipped compared to that of $v_{n,1}$ and $v_{n,8}$ in $F'_{n,p}$.}
    \label{fig:michel_scott_2_5}
\end{figure}

Our further modification to the construction mentioned in Section~\ref{sec:michel_scott} is the additions of vertices between $v_{i,j}$ and $v_{i,j-1}$ for $i \in [n]$ and $j \in \{3,5,7\}$.
More specifically, recall the graph $H_{n,p}$ and the subgraphs $F_{i,p}$ for $i \in \{0,\dots,n\}$ as defined in Section~\ref{sec:michel_scott}.
The graph $H'_{n,p}$ is obtained from $H_{n,p}$ by removing each edge $v_{i,j}v_{i,j-1}$ for $i \in [n]$ and $j \in \{3,5,7\}$ and adding a new vertex $u_{i,j,j-1}$ and the path $(v_{i,j}, u_{i,j,j-1}, v_{i,j-1})$.
See \cref{fig:michel_scott_2_5} for an illustration.

\paragraph{Weights.}
Note that the weights of the original edges in $H_{n,p}$ remain the same, and the weights of the added edges are very close to that of the removed edges.
More specifically, let $\varepsilon$ be a small constant (say, $2^{-n}$).
The edges on the path $(v_{0,1}, u^1_{0,1,8}, \dots, u^{p-1}_{0,1,8}, v_{0,8})$ have weights $(7, 7 - \varepsilon, \dots, 7-(p-1)\cdot\varepsilon)$ in that order.
For $i \in [n]$,
\begin{itemize}
    \item For $j \in \{1,3,5,7\}$, the edges on the path $(v_{i,j}, u^1_{i,j,j+1}, \dots, u^{p-1}_{i,j,j+1}, v_{i,j+1})$ have weights $\big((8-j)\cdot 8^n, (8-j)\cdot 8^n - \varepsilon, \dots, (8-j)\cdot 8^n - (p-1)\cdot\varepsilon\big)$;
    \item For $j \in \{3,5,7\}$, the edges $v_{i,j}u_{i,j,j-1}$ and $u_{i,j,j-1}v_{i,j-1})$ have weights $(8-j)\cdot 8^n$ and $(8-j)\cdot 8^n - \varepsilon$, respectively;
    \item For $j \in \{2,6\}$, the edges on the path $(v_{i,j}, u^1_{i,j,1}, \dots, u^{p-1}_{i,j,1}, v_{i-1,1})$ have weights $\big(8^n, 8^n - \varepsilon, \dots, 8^n - (p-1)\cdot\varepsilon\big)$;
    \item The edges on the path $(v_{i,4}, u^1_{i,4,1}, \dots, u^{p-1}_{i,4,1}, v_{i-1,1})$ have weights $\big(-8^n, -8^n + \varepsilon, \dots, -8^n + (p-1)\cdot\varepsilon\big)$;
    \item The edges $v_{i-1,8}v_{i,3}$, $v_{i-1,8}v_{i,5}$, and $v_{i-1,8}v_{i,7}$ have weights 1, -1, and 1, respectively.
\end{itemize}
Lastly, the edges $v_{n,1}v'_1$ and $v'_1v'_2$ have weights $8^{n+1}$ and $2\cdot 8^{n+1}$, respectively.
We denote by $w'_{n,p}$ as the weight function described above.

\paragraph{Initial cut.}
We describe the initial cut $\gamma'_{n,p}$ by specifying which vertices belong to the even and odd sets of the cut.
For $i\in [n]$ and $j \in \{1, \dots, 8\}$, the vertex $v_{i,j}$ is in the even set if and only if $j$ is even.
$v'_1$ is in the odd set, and $v'_2$ is in the even set.
For $i \in [n]$ and $j \in \{3,5,7\}$, the vertex $u_{i,j,j-1}$ is in the even set.
For $i\in[n]$ and for each pair $(j,j')$ among $(1,2),(3,4),(5,6),(7,8),(2,1)$, 
and $(6,1)$, the vertices along path $(v_{i,j}, u^1_{i,j,j'}, \dots, u^{p-1}_{i,j,j'},v_{i,j'}$ alternate between the two sets; that is, $u^t_{i,j,j'}$ is in the same set as $v_{i,j}$ if and only if $t$ is even.
The same holds for $i = 0, j = 1, j'= 8$.
Finally, for $i \in [n]$, the vertices $u^1_{i,4,1}, \dots, u^{p-1}_{i,4,1}$ are in the even set.

For two appropriate vertices $v_{i,j}$ and $v_{i',j'}$, we denote by $v_{i,j} \rightsquigarrow v_{i',j'}$ the sequence $v_{i,j} u^1_{i,j,j'} \dots u^{p-1}_{i,j,j'} v_{i',j'}$.

\begin{lemma}
\label{lem:michel_scott_2_5}
Any maximal improving flip sequence of $(H'_{n,p}, w'_{n,p})$ from $\gamma'_{n,p}$ has length exponential in $n$.
In particular, its subsequence induced on $V(H'_{n,p}) \setminus \{u_{i,3,2}, u_{i,5,4}, u_{i,7,6} \mid i \in [n]\}$ is the sequence $L'_n$ defined recursively as follows:
$L'_0 = v_{0,1} \rightsquigarrow v_{0,8}$, and for $i \in [n]$,
\[
    L'_i = v_{i,1} \rightsquigarrow v_{i,2} \rightsquigarrow L'_{i-1} \rightsquigarrow v_{i,3} \rightsquigarrow v_{i,4} \rightsquigarrow L'_{i-1} \rightsquigarrow v_{i,5} \rightsquigarrow v_{i,6} \rightsquigarrow L'_{i-1} \rightsquigarrow v_{i,7} \rightsquigarrow v_{i,8}.
\]
\end{lemma}

The proof is similar to the proof of the exponential running time in~\cite{Michel_Max_Cut_4, HHH2024}.

\begin{proof}
    Given a cut, we say a vertex is \defi{happy} if flipping that vertex is not improving.
    We first characterize when each vertex in $H'_{n,p}$ is happy, based on which sets of the cut it and its neighbors are in, as follows.
    \begin{itemize}
        \item For $i \in [n]$, $v_{i-1,1}$ is happy, if and only if either (i) it is in the different set than $u^{p-1}_{i,2,1}$, $u^{p-1}_{i,4,1}$, and $u^{p-1}_{i,6,1}$ or (ii) it is in the same set as $u^{p-1}_{i,4,1}$ and in the different set than at least one of $u^{p-1}_{i,2,1}$ and $u^{p-1}_{i,6,1}$ (i.e., the remaining neighbor has no influence).
        \item For $i \in [n]$ and $j \in \{2,4,6\}$, $v_{i,j}$ is happy, if and only if it is in the different set than $u^{p-1}_{i,j-1,j}$ (i.e. other neighbors have no influence).
        \item For $i \in [n]$ and $j \in \{3,7\}$, $v_{i,j}$ is happy, if and only if either (iii) it is in the different set of the cut than $u_{i,j,j-1}$ and $u^1_{i,j,j+1}$ or (iv) it is in the same set as only one of $u_{i,j,j-1}$ and $u^1_{i,j,j+1}$ (i.e. $v_{i-1,8}$ is in the other set).
        For $j = 5$, (iv) is replaced by that it is in the different set than exactly one of $u_{i,j,j-1}$ and $u^1_{i,j,j+1}$ (i.e., $v_{i-1,8}$ is in the same set as $v_{i,j}$).
        \item For $i \in \{0, \dots, n\}$, $v_{n,8}$ is happy, if and only if it is in the different set than $u^{p-1}_{i,7,8}$ or $u^{p-1}_{0,1,8}$, whichever is applicable (i.e., other neighbors do not matter).
        \item For a degree-two vertex $v$, note that only one neighbor influences the happiness of $v$.
        In particular, let $e$ be the adjacent edge that has the higher absolute value of the weight.
        If the weight of $e$ is positive, then $v$ is happy as long as $e$ is in the cut.
        Otherwise, $v$ is happy as long as $e$ is not in the cut.
        \item $v'_1$ and $v'_2$ are happy as long as the edge $v'_1v'_2$ is in the cut.
    \end{itemize}

    Based on the analysis above, observe that $v'_1$ and $v'_2$ are never flipped.

    We prove by induction.
    For the base case, it is easy to see that $L_0$ is the only improving sequence for $(H'_{0,p}, w'_{0,p})$ from $\gamma'_{0,p}$.
    For the inductive step, suppose this is true for $(H'_{n-1,p}, w'_{n-1,p})$ and $\gamma'_{n-1,p}$.
    Observe that for $\gamma'_{n,p}$, only $v_{n,1}$ is not happy.
    Hence, we have to flip $v_{n,1}$ and subsequently all vertices on the path from $v_{n,1}$ to $v_{n,2}$.
    At this point, there are two unhappy vertices: $u_{n,3,2}$ and $u^1_{n,2,1}$.
    On the one hand, flipping $u_{n,3,2}$ does not make $v_{n,3}$ or $v_{n,2}$ unhappy; hence as long as the other vertices in the neighborhood of $v_{n,3}$ and the edge $v_{n,2}u^{p-1}_{n,1,2}$ do not change, we can flip $u_{n,3,2}$ any time, but flipping it does not result in any subsequent flips.
    After flipping $u^1_{n,2,1}$, we can sequentially flip other vertices from the path from $v_{n,2}$ to $v_{n-1,1}$.
    At this point, observe that the vertices in $F'_{n,p}$ only connect to the other part of the graphs (other than $v'_1$) via $v_{n-1,1}$ and $v_{n-1,8}$.
    As we flip $v_{n-1},1$, by the inductive hypothesis, we now go through at least the sequence $L'_{n-1}$ (and potentially some other vertices of the form $u_{i,3,2}, u_{i,5,4}, u_{i,7,6}$).
    Note that this sequence $L'_{n-1}$ only visits $v_{n-1,1}$ once at the beginning and $v_{n-1,8}$ once at the end.
    Hence, as we go through $L'_{n-1}$, there is no change in the neighborhood of the vertices in $F'_{n,p}$, and hence, we cannot flip any vertex there except for $u_{n,3,2}$.
    When we finish $L'_{n-1}$, $v_{n-1,8}$ is now in the odd set.
    Observe that at this point, if we have not flipped $u_{n,3,2}$, then it is the only unhappy vertex.
    Hence, we have to flip it.
    After $u_{n,3,2}$ is flipped, $v_{n,3}$ is now unhappy, since it is in the same set as both $v_{n-1,8}$ and $u_{n,3,2}$.
    We then have to flip $v_{n,3}$ and subsequently all vertices in the sequence $v_{n,3} \rightsquigarrow v_{n,4}$.
    Continuing with a similar argument as before, we then obtain the lemma statement.
\end{proof}

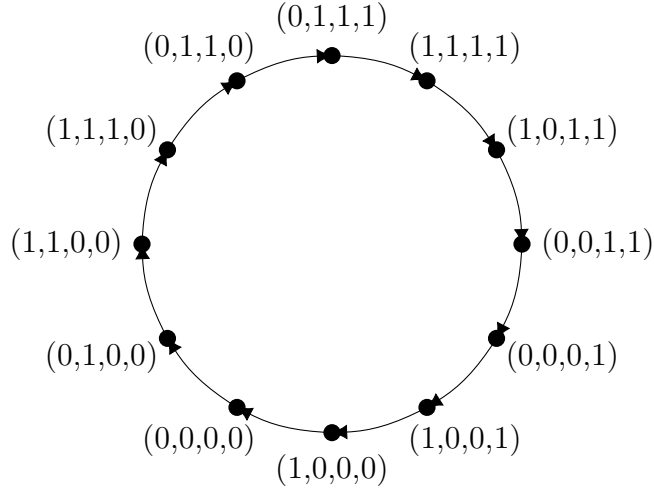
\begin{figure}[t]
    \centering
        \begin{tikzpicture}        
          \def\radius{2.5} 
        
          \foreach \i in {1,...,12} {
            \coordinate (P\i) at ({\radius*cos(30*\i)}, {\radius*sin(30*\i)});
            \node[circle, fill=black, inner sep=0.8mm] at (P\i) {};
          }

          \foreach \i/\j in {1/{(1,0,1,1)} , 2/{(1,1,1,1)} , 3/{(0,1,1,1)} , 4/{(0,1,1,0)} , 5/{(1,1,1,0)} , 6/{(1,1,0,0)} , 7/{(0,1,0,0)} , 8/{(0,0,0,0)} , 9/{(1,0,0,0)} , 10/{(1,0,0,1)} , 11/{(0,0,0,1)} , 12/{(0,0,1,1)}}{
            \node at ({1.4*\radius*cos(30*\i)}, {1.2*\radius*sin(30*\i)}) {\j};
          }
          \foreach \i/\j in {1/2, 2/3, 3/4, 4/5, 5/6, 6/7, 7/8, 8/9, 9/10, 10/11, 11/12, 12/1} {
            \draw[-{Latex[length=2mm,width=2mm]}] (P\j) to[bend left=12.5] (P\i);
          }
        \end{tikzpicture}
    \caption{The cyclic sequence $\Pi$}
    \label{fig:sequence_pi}
\end{figure}

In our construction, we make use of the subsequence described in the lemma above, especially the state of $(v_{i-1,8}, v_{i,3}, v_{i,5}, v_{i,7})$ we encounter along the sequence for $i \in [n]$.
More specifically, for some vertices $u_1, u_2, u_3, u_4$ of $H'_{n,p}$, the \defi{state} of $(u_1, u_2, u_3, u_4)$ in a cut $\gamma$ of $H'_{n,p}$ is a length-four binary vector $(a_1, a_2, a_3, a_4)$, such that for $j \in [4]$, $a_i = 1$ if and only if $u_i$ is in the 1-set of the cut $\gamma$.
Further, let $\Pi$ denote the cyclic sequence of length-four binary vectors as depicted in \cref{fig:sequence_pi}.
Then \cref{lem:michel_scott_2_5} implies the following.

\begin{observation}
\label{obs:states}
    For $i \in [n]$, the cyclic sequence $\Pi$ describes the states of $(v_{i-1,8}, v_{i,3}, v_{i,5}, \allowbreak v_{i,7})$ that we encounter in an improving flip sequence of $(H'_{n,p}, w'_{n,p})$ from $\gamma'_{n,p}$.
    That is, if in a cut in the improving flip sequence, we encounter a state $(a_1, a_2, a_3, a_4)$ of $(v_{i-1,8}, v_{i,3}, v_{i,5}, v_{i,7})$, then as we continue along the flip sequence, the next (distinct) state of $(v_{i-1,8}, v_{i,3}, v_{i,5}, v_{i,7})$ we encounter is the successor of $(a_1, a_2, a_3, a_4)$ in the sequence $\Pi$.
\end{observation}

Finally, we also rely on the following fact that $H'_{n,p}$ has sufficiently large girth (specifically in \cref{lem:corr_nbr_2_5}).

\begin{lemma}
\label{lem:girth_2_5}
    For $p \geq 3$, $H'_{n,p}$ has girth at least $p + 3$.
\end{lemma}
\begin{proof}
    Note that $H'_{n,p}$ is obtained from $H_{n,p}$ by subdividing some edges.
    Hence, the girth of $H'_{n,p}$ is at least that of $H_{n,p}$, which is $p+3$ by \cref{lem:michel_scott_condi}.
    The lemma then follows.
\end{proof}

For the remaining of this section, we define $H:=H'_{n,p}$ and $w := w_{n,p}$ for some $p \geq 3$ that we will specify later.

\subsection{\Ng gadgets}
To avoid confusion with the behavior of the \vg gadgets in the previous construction, we use the term \ng gadgets in this construction, and they come in two variants, as follows.

	A base gadget $\cn$ is a \defi{regular \ng gadget}, if it consists of only a single vertex $X^{\cn}$, which is a PV-portal.
	It is an \defi{irregular \ng gadget}, if it is a 2-path whose endpoints are two PV-portals $X^{\cn}$ and $\br{X}^{\cn}$; the middle vertex is an internal vertex.

There is only one subtour for each gadget, and we consider it standard.

\subsection{\Sg gadgets}
\begin{figure}[t!]
    \centering
    \begin{subfigure}[t]{\textwidth}
    \centering
        \begin{tikzpicture}[scale=0.69]
            \def\defvertices{
                \coordinate[label=right: $X_1$] (X) at (1,1);
                \coordinate[label=right: $\br{X}_1$] (Y) at (1,0);
                \coordinate[label=left: $Z$] (Z) at (-1,0.5);
                \coordinate[] (S) at (0,0.5);
            }
            \def\drawvertices{
                \fill[] (S) circle (0.8mm);
                \fill[] (X) circle (0.8mm);
                \fill[] (Z) circle (0.8mm);
                \fill[] (Y) circle (0.8mm);            }

            \defvertices
            \draw[thick] (Z) -- (S) -- (X) (S) -- (Y);
            \drawvertices

            \begin{scope}[shift={(5,0)}]
                \defvertices
                \draw[very thick, red] (Z) -- (S) -- (X);
                \draw[thick, dashed] (S) -- (Y);
                \drawvertices
                \draw (0, -1) node {(1)};
            \end{scope}

            \begin{scope}[shift={(10,0)}]
                \defvertices
                \draw[very thick, red] (Z) -- (S) -- (Y);
                \draw[thick, dashed] (S) -- (X);
                \drawvertices
                \draw (0, -1) node {(0)};
            \end{scope}            
        \end{tikzpicture}
    \caption{\Sg-1 gadget and the two standard subtours}
    \end{subfigure}

    \vspace{1em}

    \begin{subfigure}[t]{\textwidth}
    \centering
       \begin{tikzpicture}[scale=0.69]
        \small
        \def\defvertices{
        \coordinate[label=right: $Z$] (Z) at (1,0);
        \coordinate[label=left: $X_1$] (X) at (0,2);
        \coordinate[label=left: $\br{X}_1$] (Y) at (0,1);
        \coordinate[label=right: $Y$] (R) at (1,1);
        \coordinate[label=right: $X_2$] (R') at (1,2);
        }
        \def\drawvertices{
        \fill[] (R) circle (0.8mm);
        \fill[] (R') circle (0.8mm);
        \fill[] (X) circle (0.8mm);
        \fill[] (Y) circle (0.8mm);
        \fill[] (Z) circle (0.8mm);
        }
        
        \defvertices
        \draw[thick] (Z) -- (R) -- (R') (R) -- (X) (R') -- (X) (Y) -- (R) (R') -- (Y);
        \drawvertices

        \begin{scope}[shift={(3.5,0)}]
        \defvertices
        \draw[thick, dashed] (X) -- (R) -- (Y) -- (R');
        \draw[very thick, red] (Z) -- (R) -- (R') -- (X);
        \drawvertices
        \draw (0.5, -1) node {(1,1)};
        \end{scope}
        
        \begin{scope}[shift={(7,0)}]
        \defvertices
        \draw[thick, dashed] (R') -- (X) (Y) -- (R) -- (X);
        \draw[very thick, red] (Z) -- (R) -- (R') -- (Y);
        \drawvertices
        \draw (0.5, -1) node {(0,1)};
        \end{scope}
        
        \begin{scope}[shift={(10.5,0)}]
        \defvertices
        \draw[thick, dashed] (X) -- (R') -- (R) -- (Y) -- (R');
        \draw[very thick, red] (Z) -- (R) -- (X);
        \drawvertices
        \draw (0.5, -1) node {(1,0)};
        \end{scope}
        
        \begin{scope}[shift={(14,0)}]
        \defvertices
        \draw[thick, dashed] (R') -- (X) -- (R) -- (R') -- (Y);
        \draw[very thick, red] (Z) -- (R) -- (Y);
        \drawvertices
        \draw (0.5, -1) node {(0,0)};
        \end{scope}

        \begin{scope}[shift={(-3.5,0)}]
            \node [shape=rectangle,align=center](table1) at (0.5,1) {
            \footnotesize
            \setlength{\tabcolsep}{2pt}
            \begin{tabular}{lc}
                Edge & Weight  \\ 
                \midrule
                $X_1X_2$ & $\sigma_{12}$ \\
                $X_1Y$ & $\delta_{12}$ \\
                $\br{X}_1X_2$ & $\delta_{12}$ \\
                $\br{X}_1Y$ & $\sigma_{12}$
            \end{tabular}
            };
        \end{scope}
    \end{tikzpicture}
     \caption{\Sg-2-IR gadget and the four standard subtours}
    \end{subfigure}
    
    \vspace{1em}
    
    \begin{subfigure}[t]{\textwidth}
    \centering
       \begin{tikzpicture}[scale=0.69]
        \small
        \def\defvertices{
        \coordinate[label=right: $X_2$] (X) at (1,1);
        \coordinate[label=right: $\br{X}_2$] (Y) at (1,0);
        \coordinate[label=left: $X_1$] (R) at (0,1);
        \coordinate[label=left: $\br{X}_1$] (R') at (0,0);
        }
        \def\drawvertices{
        \fill[] (R) circle (0.8mm);
        \fill[] (R') circle (0.8mm);
        \fill[] (X) circle (0.8mm);
        \fill[] (Y) circle (0.8mm);
        }
        
        \defvertices
        \draw[thick] (R) -- (X) (R') -- (X) (Y) -- (R) (R') -- (Y);
        \drawvertices

        \begin{scope}[shift={(3.5,0)}]
        \defvertices
        \draw[thick, dashed] (X) -- (R') -- (Y) -- (R);
        \draw[very thick, red] (R) -- (X);
        \drawvertices
        \draw (0.5, -1) node {(1,1)};
        \end{scope}
        
        \begin{scope}[shift={(7,0)}]
        \defvertices
        \draw[thick, dashed] (X) -- (R) (R') -- (Y) -- (R);
        \draw[very thick, red] (R') -- (X);
        \drawvertices
        \draw (0.5, -1) node {(0,1)};
        \end{scope}
        
        \begin{scope}[shift={(10.5,0)}]
        \defvertices
        \draw[thick, dashed] (R') -- (X) (Y) -- (R') (R) -- (X);
        \draw[very thick, red] (R) -- (Y);
        \drawvertices
        \draw (0.5, -1) node {(1,0)};
        \end{scope}
        
        \begin{scope}[shift={(14,0)}]
        \defvertices
        \draw[thick, dashed] (R') -- (X) -- (R) -- (Y);
        \draw[very thick, red] (R') -- (Y);
        \drawvertices
        \draw (0.5, -1) node {(0,0)};
        \end{scope}

        \begin{scope}[shift={(-3.5,0)}]
            \node [shape=rectangle,align=center](table1) at (0.5,1) {
            \footnotesize
            \setlength{\tabcolsep}{2pt}
            \begin{tabular}{lc}
                Edge & Weight  \\ 
                \midrule
                $X_1X_2$ & $\sigma_{12}$ \\
                $X_1\br{X}_2$ & $\delta_{12}$ \\
                $\br{X}_1X_2$ & $\delta_{12}$ \\
                $\br{X}_1\br{X}_2$ & $\sigma_{12}$
            \end{tabular}
            };
        \end{scope}
    \end{tikzpicture}
     \caption{\Sg-2-II gadget and the four standard subtours}
    \end{subfigure}

    \vspace{1em}

    \begin{subfigure}[t]{\textwidth}
        \centering
    \centering
       \begin{tikzpicture}[scale=0.69]
        \small
        \def\defvertices{
        \coordinate[label=left: $X_2$] (X) at (0,0.5);
        \coordinate[label=above: $X_1$] (Y) at (1,1.5);
        \coordinate[label=right: $X_3$] (Z) at (2,0.5);
        \coordinate[label=left: $\br{X}_2$] (Xb) at (0,-0.5);
        \coordinate[label=right: $\br{X}_3$] (Zb) at (2,-0.5);
        }
        \def\drawvertices{
        \fill[] (X) circle (0.8mm);
        \fill[] (Y) circle (0.8mm);
        \fill[] (Xb) circle (0.8mm);
        \fill[] (Z) circle (0.8mm);
        \fill[] (Zb) circle (0.8mm);
        }
        
        \defvertices
        \draw[thick] (Xb) -- (Z) -- (X) -- (Y) -- (Zb) -- (Xb) -- (Y) -- (Z) (X) -- (Zb);
        \drawvertices
        
        \begin{scope}[shift={(4.5,0)}]
        \defvertices
        \draw[thick, dashed] (Xb) -- (Z) -- (X)  (Y) -- (Zb) -- (Xb) -- (Y) (X) -- (Zb);
        \draw[very thick, red] (X) -- (Y) -- (Z);
        \drawvertices
        \draw (1, -1.5) node {(1,1,1)};
        \end{scope}
        
        \begin{scope}[shift={(9,0)}]
        \defvertices
        \draw[thick, dashed]  (Xb) -- (Z) -- (X) (Zb) -- (Xb) -- (Y) -- (Z) (X) -- (Zb);
        \draw[very thick, red] (X) -- (Y) -- (Zb);
        \drawvertices
        \draw (1, -1.5) node {(1,1,0)};
        \end{scope}

        \begin{scope}[shift={(13.5,0)}]
        \defvertices
        \draw[thick, dashed] (Xb) -- (Z) -- (X) -- (Y) -- (Zb) -- (Xb) (X) -- (Zb);
        \draw[very thick, red] (Xb) -- (Y) -- (Z);
        \drawvertices
        \draw (1, -1.5) node {(1,0,1)};
        \end{scope}

        \begin{scope}[shift={(18,0)}]
        \defvertices
        \draw[thick, dashed]  (Xb) -- (Z) -- (X) -- (Y) (Zb) -- (Xb) (Y) -- (Z) (X) -- (Zb);
        \draw[very thick, red] (Xb) -- (Y) -- (Zb)  ;
        \drawvertices
        \draw (1, -1.5) node {(1,0,0)};
        \end{scope}

        \begin{scope}[shift={(4.5,-5)}]
        \defvertices
        \draw[thick, dashed]  (Xb) -- (Z) (X) -- (Y) -- (Zb) -- (Xb) -- (Y) -- (Z) (X) -- (Zb);
        \draw[very thick, red] (X) -- (Z);
        \drawvertices
        \draw (1, -1.5) node {(0,1,1)};
        \end{scope}

        \begin{scope}[shift={(9,-5)}]
        \defvertices
        \draw[thick, dashed]  (Xb) -- (Z) -- (X) -- (Y) -- (Zb) -- (Xb) -- (Y) -- (Z);
        \draw[very thick, red] (X) -- (Zb);
        \drawvertices
        \draw (1, -1.5) node {(0,1,0)};
        \end{scope}
        
        \begin{scope}[shift={(13.5,-5)}]
        \defvertices
        \draw[thick, dashed]  (Z) -- (X) -- (Y) -- (Zb) -- (Xb) -- (Y) -- (Z) (X) -- (Zb);
        \draw[very thick, red] (Xb) -- (Z);
        \drawvertices
        \draw (1, -1.5) node {(0,0,1)};
        \end{scope}

        \begin{scope}[shift={(18,-5)}]
        \defvertices
        \draw[thick, dashed]  (Xb) -- (Z) -- (X) -- (Y) -- (Zb) (Xb) -- (Y) -- (Z) (X) -- (Zb);
        \draw[very thick, red] (Xb) -- (Zb);
        \drawvertices
        \draw (1.5, -1.5) node {(0,0,0)};
        \end{scope}

        \begin{scope}[shift={(0,-6)}]
            \node [shape=rectangle,align=center](table1) at (1.2,1.5) {
            \footnotesize
            \setlength{\tabcolsep}{2pt}
            \begin{tabular}{lc}
                Edge & Weight  \\ 
                \midrule
                $X_1X_2$ & $\sigma_{12}$ \\
                $X_1\br{X}_2$ & $\delta_{12} $\\
                $X_1X_3$ & $\sigma_{13}$ \\
                $X_1\br{X}_3$ & $\delta_{13}$ \\
                $X_2X_3$ & $\delta_{12}+\delta_{13}$ \\
                $X_2\br{X}_3$ & $\delta_{12}+\sigma_{13}$ \\
                $\br{X}_2X_3$ & $\sigma_{12}+\delta_{13}$ \\
                $\br{X}_2\br{X}_3$ & $\sigma_{12}+\sigma_{13}$
            \end{tabular}
            };
        \end{scope}
        \end{tikzpicture}
        \caption{\Sg-3 gadget and the eight standard subtours}
    \end{subfigure}
    
    \caption{\Sg gadgets. Edges not in a subtour are dashed. Unless specified, edges have weight zero. We omit the superscripts on the vertices and weight parameters for better readability.}
    \label{fig:star_gadgets}
\end{figure}
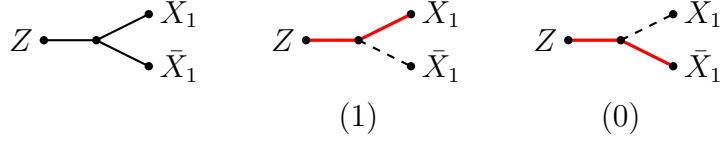
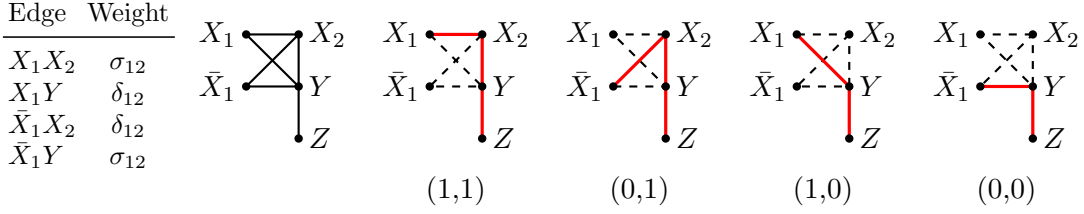
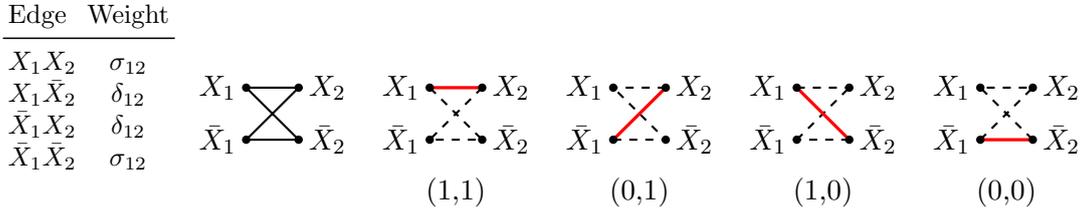
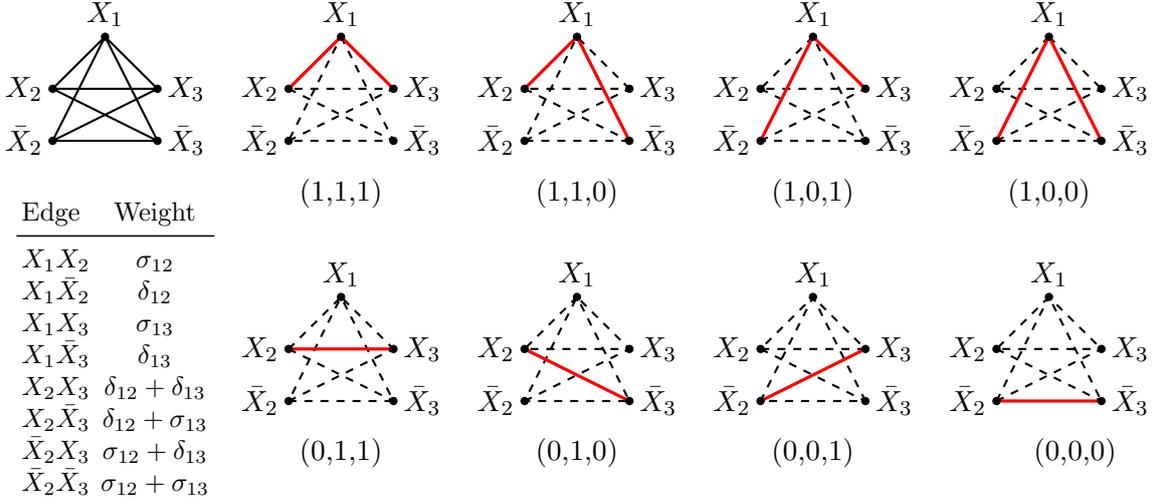

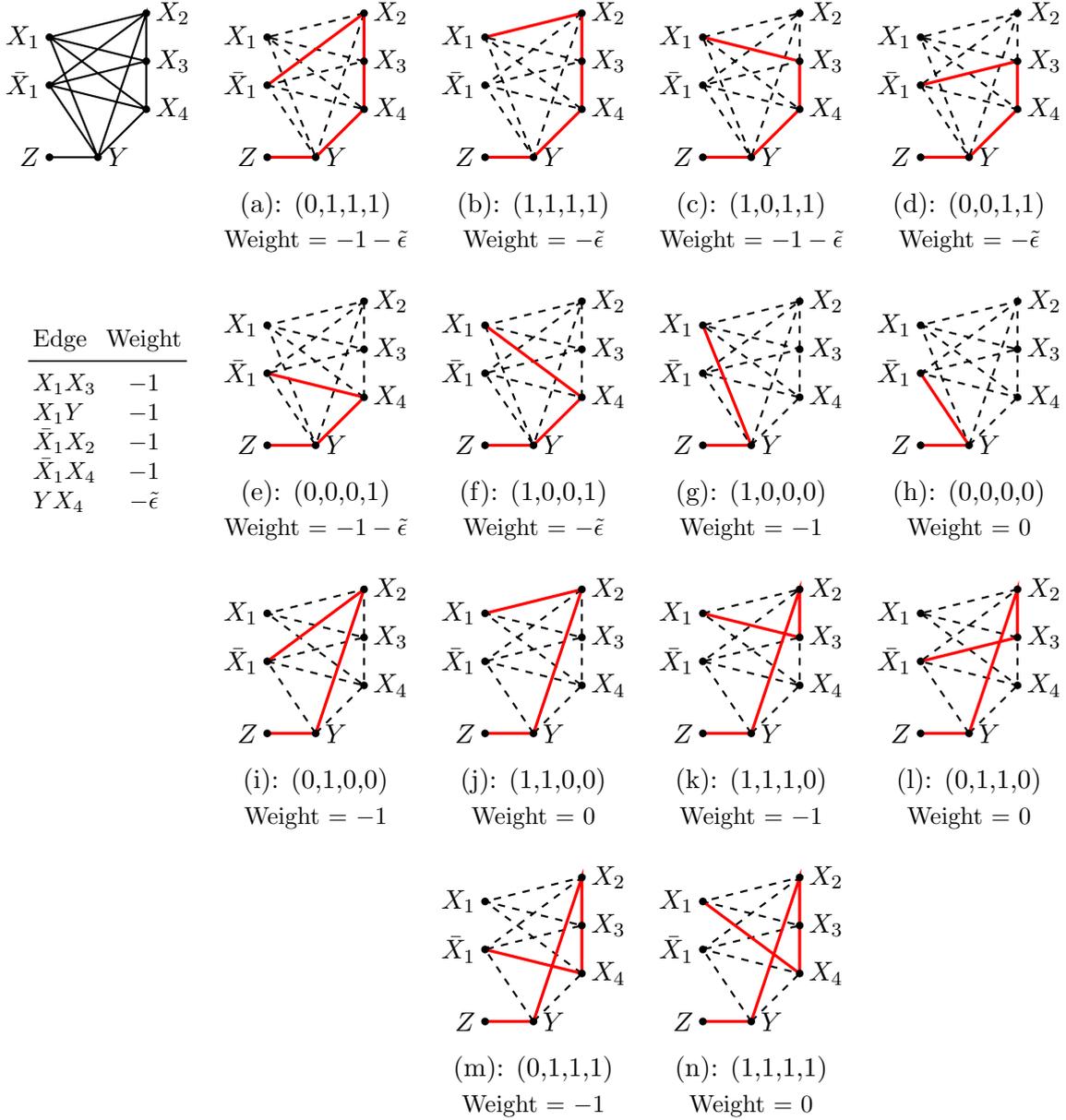
\begin{figure}[t!]
        \centering
    \centering
       \begin{tikzpicture}[scale=0.69]
        \small
        \def\defvertices{
        \coordinate[label=right: $X_4$] (X) at (2,2);
        \coordinate[label=right: $X_3$] (R) at (2,3);
        \coordinate[label=right: $X_2$] (S) at (2,4);
        \coordinate[label=left: $X_1$] (T) at (0,3.5);
        \coordinate[label=left: $\br{X}_1$] (Tb) at (0,2.5);
        \coordinate[label=left: $Z$] (Z) at (0,1);
        \coordinate[label=right: $Y$] (Y) at (1,1);
        }
        \def\drawvertices{
        \fill[] (X) circle (0.8mm);
        \fill[] (Y) circle (0.8mm);
        \fill[] (R) circle (0.8mm);
        \fill[] (Z) circle (0.8mm);
        \fill[] (S) circle (0.8mm);
        \fill[] (T) circle (0.8mm);
        \fill[] (Tb) circle (0.8mm);
        }
        
        \defvertices
        \draw[thick] (Z) -- (Y) -- (X) -- (R) -- (S) -- (Y) -- (Tb) -- (R) -- (T) -- (S) -- (Tb) -- (X) -- (T) -- (Y);
        \drawvertices
        
        \begin{scope}[shift={(4.5,0)}]
        \defvertices
        \draw[thick, dashed]  (S) -- (Y) -- (Tb) -- (R) -- (T) -- (S) (Tb) -- (X) -- (T) -- (Y);
        \draw[very thick, red] (Z) -- (Y) -- (X) -- (R) -- (S) -- (Tb);
        \drawvertices
        \draw (1, 0) node {(a): (0,1,1,1)};
        \draw (1,-0.75) node {\footnotesize Weight = $-1 - \tilde{\epsilon}$};
        \end{scope}
        
        \begin{scope}[shift={(9,0)}]
        \defvertices
        \draw[thick, dashed]  (S) -- (Y) -- (Tb) -- (R) -- (T) (S) -- (Tb) -- (X) -- (T) -- (Y);
        \draw[very thick, red] (Z) -- (Y) -- (X) -- (R) -- (S) -- (T);
        \drawvertices
        \draw (1, 0) node {(b): (1,1,1,1)};
        \draw (1,-0.75) node {\footnotesize Weight = $- \tilde{\epsilon}$};
        \end{scope}

        \begin{scope}[shift={(13.5,0)}]
        \defvertices
        \draw[thick, dashed] (R) -- (S) -- (Y) -- (Tb) -- (R) (T) -- (S) -- (Tb) -- (X) -- (T) -- (Y);
        \draw[very thick, red] (Z) -- (Y) -- (X) -- (R) -- (T);
        \drawvertices
        \draw (1, 0) node {(c): (1,0,1,1)};
        \draw (1,-0.75) node {\footnotesize Weight = $-1 - \tilde{\epsilon}$};
        \end{scope}

        \begin{scope}[shift={(18,0)}]
        \defvertices
        \draw[thick, dashed]  (R) -- (S) -- (Y) -- (Tb) (R) -- (T) -- (S) -- (Tb) -- (X) -- (T) -- (Y);
        \draw[very thick, red] (Z) -- (Y) -- (X) --(R) -- (Tb);
        \drawvertices
        \draw (1, 0) node {(d): (0,0,1,1)};
        \draw (1,-0.75) node {\footnotesize Weight = $- \tilde{\epsilon}$};
        \end{scope}

        \begin{scope}[shift={(4.5,-6)}]
        \defvertices
        \draw[thick, dashed] (X) -- (R) -- (S) -- (Y) -- (Tb) -- (R) -- (T) -- (S) -- (Tb) (X) -- (T) -- (Y);
        \draw[very thick, red] (Z) -- (Y) -- (X) -- (Tb);
        \drawvertices
        \draw (1, 0) node {(e): (0,0,0,1)};
        \draw (1,-0.75) node {\footnotesize Weight = $-1 - \tilde{\epsilon}$};
        \end{scope}

        \begin{scope}[shift={(9,-6)}]
        \defvertices
        \draw[thick, dashed] (X) -- (R) -- (S) -- (Y) -- (Tb) -- (R) -- (T) -- (S) -- (Tb) -- (X) (T) -- (Y);
        \draw[very thick, red] (Z) -- (Y) -- (X) -- (T);
        \drawvertices
        \draw (1, 0) node {(f): (1,0,0,1)};
        \draw (1,-0.75) node {\footnotesize Weight = $- \tilde{\epsilon}$};
        \end{scope}
        
        \begin{scope}[shift={(13.5,-6)}]
        \defvertices
        \draw[thick, dashed] (Y) -- (X) -- (R) -- (S) -- (Y) -- (Tb) -- (R) -- (T) -- (S) -- (Tb) -- (X) -- (T);
        \draw[very thick, red] (Z) -- (Y) -- (T);
        \drawvertices
        \draw (1, 0) node {(g): (1,0,0,0)};
        \draw (1,-0.75) node {\footnotesize Weight = $-1$};
        \end{scope}

        \begin{scope}[shift={(18,-6)}]
        \defvertices
        \draw[thick, dashed] (Y) -- (X) -- (R) -- (S) -- (Y) (Tb) -- (R) -- (T) -- (S) -- (Tb) -- (X) -- (T) -- (Y);
        \draw[very thick, red] (Z) -- (Y) -- (Tb);
        \drawvertices
        \draw (1, 0) node {(h): (0,0,0,0)};
        \draw (1,-0.75) node {\footnotesize Weight = $0$};
        \end{scope}

        \begin{scope}[shift={(4.5,-12)}]
        \defvertices
        \draw[thick, dashed] (Y) -- (X) -- (R) -- (S) (Y) -- (Tb) -- (R) -- (T) -- (S) (Tb) -- (X) -- (T);
        \draw[very thick, red] (Z) -- (Y) -- (S) -- (Tb);
        \drawvertices
        \draw (1, 0) node {(i): (0,1,0,0)};
        \draw (1,-0.75) node {\footnotesize Weight = $-1$};
        \end{scope}

        \begin{scope}[shift={(9,-12)}]
        \defvertices
        \draw[thick, dashed] (Y) -- (X) -- (R) -- (S) (Y) -- (Tb) -- (R) -- (T) (S) -- (Tb) -- (X) -- (T);
        \draw[very thick, red] (Z) -- (Y) -- (S) -- (T);
        \drawvertices
        \draw (1, 0) node {(j): (1,1,0,0)};
        \draw (1,-0.75) node {\footnotesize Weight = $0$};
        \end{scope}
        
        \begin{scope}[shift={(13.5,-12)}]
        \defvertices
        \draw[thick, dashed] (Y) -- (X) -- (R) (Y) -- (Tb) -- (R) (T) -- (S) -- (Tb) -- (X) -- (T);
        \draw[very thick, red] (Z) -- (Y) -- (S) -- (R) -- (T);
        \drawvertices
        \draw (1, 0) node {(k): (1,1,1,0)};
        \draw (1,-0.75) node {\footnotesize Weight = $-1$};
        \end{scope}        

        \begin{scope}[shift={(18,-12)}]
        \defvertices
        \draw[thick, dashed] (Y) -- (X) -- (R) (Y) -- (Tb) (R) -- (T) -- (S) -- (Tb) -- (X) -- (T);
        \draw[very thick, red] (Z) -- (Y) -- (S) -- (R) -- (Tb);
        \draw (1,-0.75) node {\footnotesize Weight = $0$};
        \drawvertices
        \draw (1, 0) node {(l): (0,1,1,0)};
        \end{scope}   

        \begin{scope}[shift={(9,-18)}]
        \defvertices
        \draw[thick, dashed] (Y) -- (X) (Y) -- (Tb) -- (R) -- (T) -- (S) -- (Tb) (X) -- (T);
        \draw[very thick, red] (Z) -- (Y) -- (S) -- (R) -- (X) -- (Tb);
        \drawvertices
        \draw (1, 0) node {(m): (0,1,1,1)};
        \draw (1,-0.75) node {\footnotesize Weight = $-1$};
        \end{scope}   

        \begin{scope}[shift={(13.5,-18)}]
        \defvertices
        \draw[thick, dashed] (Y) -- (X) (Y) -- (Tb) -- (R) -- (T) -- (S) -- (Tb) -- (X);
        \draw[very thick, red] (Z) -- (Y) -- (S) -- (R) -- (X) -- (T);
        \drawvertices
        \draw (1, 0) node {(n): (1,1,1,1)};
        \draw (1,-0.75) node {\footnotesize Weight = $0$};
        \end{scope}  

        \begin{scope}[shift={(0,-6)}]
            \node [shape=rectangle,align=center](table1) at (1.2,1.5) {
            \footnotesize
            \setlength{\tabcolsep}{2pt}
            \begin{tabular}{lc}
                Edge & Weight  \\ 
                \midrule
                $X_1X_3$ & $-1$ \\
                $X_1Y$ & $-1$ \\
                $\br{X}_1X_2$ & $-1$\\
                $\br{X}_1X_4$ & $-1$\\
		 $YX_4$ & $-\tilde{\epsilon}$
            \end{tabular}
            };
        \end{scope}
        \end{tikzpicture}
        \caption{\Sg-4 gadget and the 14 standard subtours. Edges not in a subtour are dashed. Unless specified, edges have weight zero. We omit the superscripts on the vertices for better readability.}
        \label{fig:star_gadget_4}
    \end{figure}

Next, we define the \sg gadgets.
For $t \in \{1,2,3,4\}$, a \defi{\sg-$t$ gadget} $\cs$ is a base gadget as depicted in Figures~\ref{fig:star_gadgets} and~\ref{fig:star_gadget_4}.
(Note that there are two types of \sg-$2$ gadgets, which we call \sg-2-IR gadgets and \sg-2-II gadgets.)
It has $t$ sides, which are sets of vertices defined as follows.
For $i \in [t]$, the \defi{$i$-side} consists of $X_t^{\cs}$ and $\br{X}_t^{\cs}$ (if this vertex exists).
A side is \defi{regular}, if it has one vertex; otherwise, it is \defi{irregular}.
The vertices in the sides are the PV-portals of $\cs$.
If there is only one irregular side, then $\cs$ also has a PP-portal $Z^{\cs}$; otherwise, it has none.
The weights of the edges depend on additional parameters as shown in the figures.
For the \sg-4 gadget, the additional parameter is $\tilde{\epsilon}$, which we consider to be a very small constant such that $n\tilde{\epsilon}$ is smaller than the increase in value for any improving flip in $H$ (say $\tilde{\epsilon} = 2^{-2n}$).
For a vector $\vect{s} \in \bin^t$, an \defi{$\vect{s}$-subtour} is a subtour such that the degrees of the PV-portals in the subtour are listed in \cref{tab:star_degrees}.
Then the standard subtours of the \eg-$t$ gadget are exactly the $\vect{s}$-subtours for $\vect{s} \in \bin^t$, if the subtours exist.
(Recall that the PP-portals need to have degree one in any subtour.)

\begin{table}[h!]
\begin{subtable}{0.48\textwidth}
\centering
\begin{tabular}[t]{c c c}
 & regular & irregular \\
\hline
$X^{\cs}_i$ & 0 & 0 \\
$\br{X}^{\cs}_i$ &  & 1
\end{tabular}
\caption{If $\vect{s}[i] = 0$}
\end{subtable}
\hfill
\begin{subtable}{0.48\textwidth}
\centering
\begin{tabular}[t]{c c c}
 & regular & irregular \\
\hline
$X^{\cs}_i$ & 2 & 1 \\
$\br{X}^{\cs}_i$ &  & 0
\end{tabular}
\caption{If $\vect{s}[i] = 1$}
\end{subtable}
\caption{Degrees of the PV-portals in the $i$-side in an $\vect{s}$-subtour of a star gadget $\cs$}
\label{tab:star_degrees}
\end{table}

\begin{lemma}
    \label{lem:star_unique_subtours}
    For $t \in \{1,2,3\}$, let $\cs$ be an \sg-$t$ gadget.
    Then for each $\vect{s} \in \bin^t$, $\cs$ has exactly one $\vect{s}$-subtour as depicted in \cref{fig:star_gadgets}.
    Further, the total edge weight of the subtour is equal to
    \[
    \sum_{\substack{i \in [t-1]:\\ \vect{s}[1] =  \vect{s}[i+1]}} \sigma^\cs_{1(i+1)} + \sum_{\substack{i \in [t-1]:\\ \vect{s}[1] \neq  \vect{s}[i+1]}} \delta^\cs_{1(i+1)}.
    \]
\end{lemma}
\begin{proof}
    The uniqueness of each $\vect{s}$-subtour and the total edge weight can be easily verified from the definitions of the \sg-gadget and $\vect{s}$-subtour.
\end{proof}

\begin{lemma}
    \label{lem:star_4_standard_subtours}
    The subtours depicted in \cref{fig:star_gadget_4} are all standard subtours of a \sg-4 gadget.
\end{lemma}
\begin{proof}
    As it is clear from context, we omit the superscripts in this proof.
    
    In a standard subtour, the degree requirements of all the portals dictate that there must be a path $P$ from $Z$ to either $X_1$ or $\br{X}_1$.
    Further, since $X_2, X_3$, and $X_4$ do not form a triangle in the gadget, their degree requirements in a standard subtour imply that if each of these vertices is not on the path $P$ above, then it has to be on a 0-path (i.e., it is an isolated vertex in the subtour).
    By symmetry, we assume $P$ goes from $Z$ to $X_1$; the case when $P$ goes to $\br{X}_1$ is analogous.

    Since $Y$, as an internal vertex, must have degree two in the subtour, and since $Z$ has degree one in the subtour, it follows that the edge $YZ$ has to be in the subtour.
    We consider the cases of the other adjacent edge $e$ of $Y$ in the subtour.

    \textit{Case 1: $e = YX_1$}. The path $P$ then has to be $(Z, Y, X_1)$, and we have subtour (g) in \cref{fig:star_gadget_4}.

    \textit{Case 2: $e = YX_2$}. We consider the possibility of the path $P$ after visiting $X_2$.
    If it goes to $X_1$, we have subtour (j) in \cref{fig:star_gadget_4}.
    If it goes to $X_3$ and then $X_1$, we have subtour (k).
    Finally, if it continues with $X_3, X_4, X_1$, we have subtour (n).

    \textit{Case 3: $e = YX_4$}. Using a similar case distinction as above, we obtain subtours (f), (d), and (b).
\end{proof}

We define the following changes between the standard subtours.

\begin{definition}
    Let $\cs$ be a \sg-$t$ gadget for some $t \in \{1,2,3,4\}$.
    Let $\vect{s}_1$ and $\vect{s}_2$ be two (not necessarily distinct) vectors in $\bin^t$.
    For some $i \in [t]$, we say the swap from a $\vect{s}_1$-subtour to a $\vect{s}_2$-subtour of $\cs$ is a \defi{2-change at the $i$-side}, if the $i$-th side of $\cs$ is irregular, and the swap involves exactly two adjacent edges $X^{\cs}_ix$ and $\br{X}^{\cs}_ix$ for some vertex $x$ of the gadget.
    We say the swap is a \defi{2.5-change at the $i$-side}, if the $i$-th side of $\cs$ is regular, and the swap involves exactly three edges forming a triangle such that two of these edges contain $X^{\cs}_i$ and are contained in one of the subtours.
\end{definition}

Then we have two following easy lemmas.

\begin{lemma}
    \label{lem:star_involved_edges}
    For $t \in \{1,2,3\}$, let $\cs$ be a \sg-$t$ gadget, and let $\vect{s}_1$ and $\vect{s}_2$ be two distinct vectors in $\bin^t$.
    For $i \in [t]$, $\vect{s}_1$ and $\vect{s}_2$ differ exactly at the $i$-th coordinate, if and only if the swap between the $\vect{s}_1$- and $\vect{s}_2$-subtours in a \sg-$t$ gadget is either a 2-change or a 2.5-change at the $i$-side. 
\end{lemma}

\begin{proof}
    By \cref{lem:star_unique_subtours}, all standard subtours for a \sg-1/2/3 gadget are as depicted in \cref{fig:star_gadgets}
    The lemma can then be easily verified from the figure.
\end{proof}

Similarly, from \cref{lem:star_4_standard_subtours} and \cref{fig:star_gadget_4}, we have the following observation.

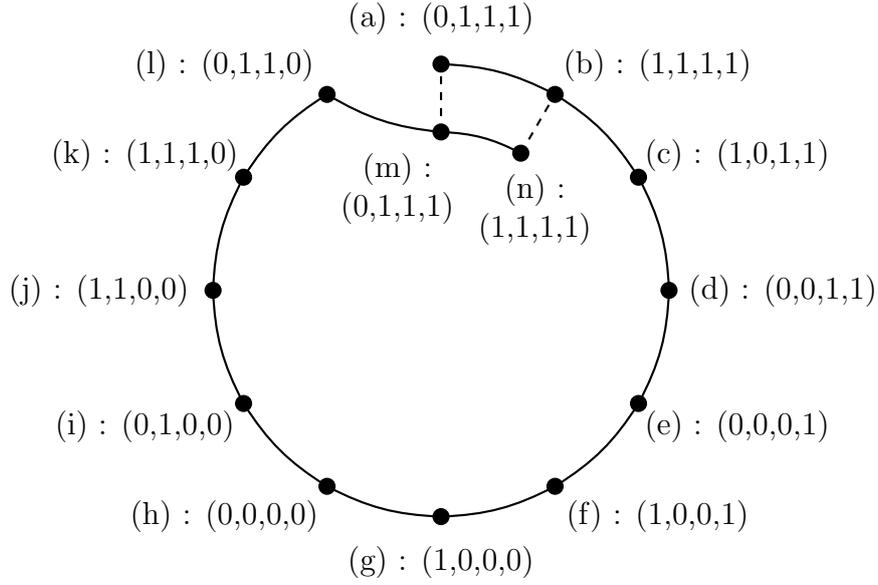
\begin{figure}[ht!]
    \centering
        \begin{tikzpicture}        
          \def\radius{3} 
        
          \foreach \i in {1,...,12} {
            \coordinate (P\i) at ({\radius*cos(30*\i)}, {\radius*sin(30*\i)});
            \node[circle, fill=black, inner sep=0.8mm] at (P\i) {};
          }

        \coordinate (P13) at ({0.7*\radius*cos(30*3)}, {0.7*\radius*sin(30*3)});
        \node[circle, fill=black, inner sep=0.8mm] at (P13) {};

        \coordinate (P14) at ({0.7*\radius*cos(30*2)}, {0.7*\radius*sin(30*2)});
        \node[circle, fill=black, inner sep=0.8mm] at (P14) {};

          \foreach \i/\j in {1/{(c) : (1,0,1,1)} , 3/{(a) : (0,1,1,1)} , 5/{(k) : (1,1,1,0)} , 6/{(j) : (1,1,0,0)} , 7/{(i) : (0,1,0,0)} , 9/{(g) : (1,0,0,0)} , 11/{(e) : (0,0,0,1)} , 12/{(d) : (0,0,1,1)}}{
            \node at ({1.5*\radius*cos(30*\i)}, {1.2*\radius*sin(30*\i)}) {\j};
          }

          \foreach \i/\j in {2/{(b) : (1,1,1,1)} , 4/{(l) : (0,1,1,0)} , 8/{(h) : (0,0,0,0)} , 10/{(f) : (1,0,0,1)}}{
            \node at ({1.9*\radius*cos(30*\i)}, {1.15*\radius*sin(30*\i)}) {\j};
          }

          \node[align=center] at ({0.5*\radius*cos(30*3.75)}, {0.45*\radius*sin(30*3)}) {(m) :\\ (0,1,1,1)};

          \node[align=center] at ({0.5*\radius*cos(30*1.15)}, {0.45*\radius*sin(30*1.75)}) {(n) :\\ (1,1,1,1)};
          
          \foreach \i/\j in {1/2, 2/3, 4/5, 5/6, 6/7, 7/8, 8/9, 9/10, 10/11, 11/12, 12/1} {
            \draw[thick] (P\j) to[bend left=12.5] (P\i);
          }

          \draw[thick] (P13) to[bend left=12.5] (P14);
          \draw[thick, dashed] (P13) -- ({\radius*cos(30*3)}, {\radius*sin(30*3)});
          \draw[thick, dashed] (P14) -- ({\radius*cos(30*2)}, {\radius*sin(30*2)});
          \draw[thick] (P13) to[bend left=12.5] ({\radius*cos(30*4)}, {\radius*sin(30*4)});
        \end{tikzpicture}
    \caption{All the standard subtours of a \sg-4 gadget, with edges describing some (but not all) pairs of subtours that differ by a 2- or 2.5-change (solid edges) or by a 2-swap (dashed edges)}
    \label{fig:star_four_improving_swaps}
\end{figure}

\begin{observation}
\label{obs:star_four_involved_edges}
    \cref{fig:star_four_improving_swaps} describes some (but not all) pairs of standard subtours of a \sg-4 gadget that differ by a 2-, 2.5-change, or a 2-swap.
    In particular, let $\vect{s}$ be a vector in $\Pi$, and let $\vect{s}'$ be the successor of $\vect{s}$ in $\Pi$.
    Then except for subtour (n), every $\vect{s}$-subtour of a \sg-4 gadget can be transformed into an $\vect{s}'$-subtour by a 2- or 2.5 change.
\end{observation}

\subsection{Construction of the TSP instance}
\label{subsec:construction_2_5}
\begin{figure}[htb!]
	\centering
      \includegraphics[width=\linewidth, page=2]{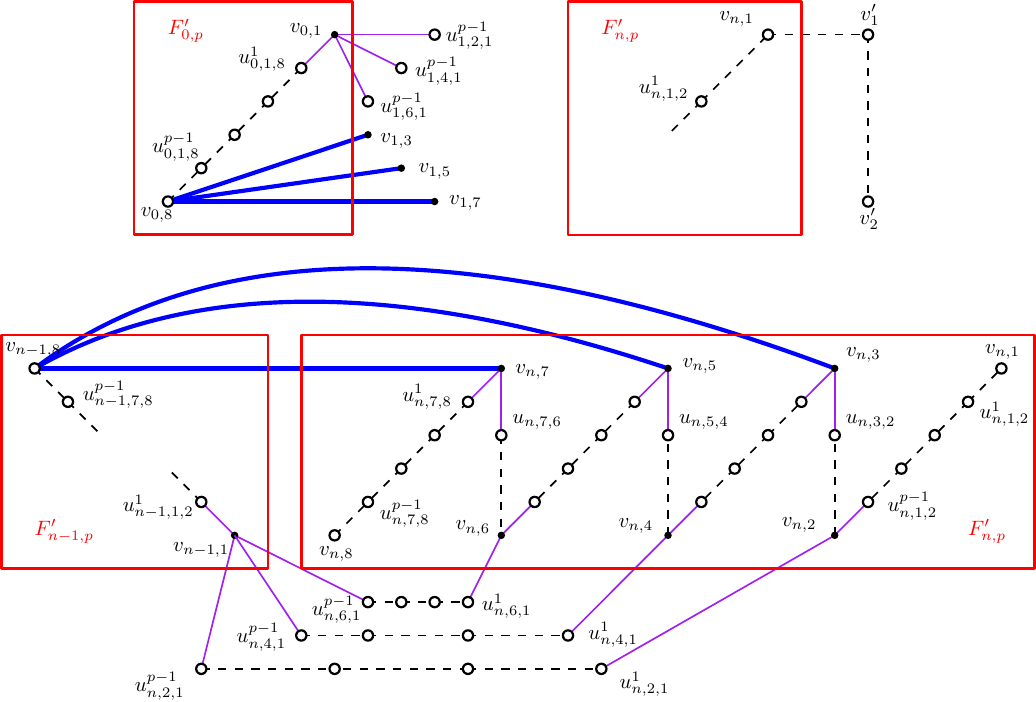}
      \caption{Assignments of \ng and \sg gadgets. Black nodes represent regular \ng gadgets, while big white nodes represent irregular \ng gadgets. Assignments of the star-2, 3, 4 gadgets are represented by black dashed edges, purple thin edges, and blue thick edges, respectively.
      The two \sg-1 gadgets assigned to  $(v'_2)$ and  $(v_{n,8})$  are not shown in this figure.
      }
      \label{fig:gadget_assignment_2_5}
\end{figure}

See \cref{fig:gadget_assignment_2_5} for an illustration on the assignment of \ng gadgets and \sg gadgets, as described below.

\paragraph{Assigning \ng gadgets.}
For $i \in [n]$, we assign a regular \ng gadget to each of $v_{i-1,1}$, $v_{i,2}$, $v_{i,3}$, $v_{i,4}$, $v_{i,5}$, $v_{i,6}$, and $v_{i,7}$.
We assign an irregular \ng gadget to each of the other $H$-vertices.
We denote by $\cn(v)$ the \ng gadget assigned to the $H$-vertex $v$.

\paragraph{Assigning \sg gadgets.}
We define the following set $\calS$ of sequences, where each sequence $(v_1, \dots, v_t)$ corresponds to a star of $t$ vertices centered at $v_1$ for some $t \in [4]$.
\begin{itemize}
    \item $(v_{i-1,8}, v_{i,3}, v_{i,5}, v_{i,7})$ for $i \in [n]$,
	\item $(v_{i-1,1}, u^{p-1}_{i,2,1}, u^{p-1}_{i,6,1})$ for $i \in [n]$, 
	\item $(v_{i-1,1}, u^{1}_{i-1,1,2}, u^{p-1}_{i,4,1})$ for $i \in \{2, \dots, n\}$,
    \item $(v_{i,j}, u^{1}_{i,j,1}, u^{p-1}_{i,j-1,j})$ for $i \in [n]$ and $j \in \{2,4,6\}$,
    \item $(v_{i,j}, u^{1}_{i,j,j+1}, u_{i,j,j-1})$ for $i \in [n]$ and $j \in \{3,5,7\}$,
	\item $(v_{0,1}, u^{1}_{0,1,8}, u^{p-1}_{1,4,1})$,
    \item $(u,v)$ for any $H$-edge $uv$ not covered by any star corresponding to a sequence described above, and
    \item two sequences of length one, $(v_{n,8})$ and $(v'_2)$.
\end{itemize}
Note that each $H$-vertex appears in exactly two sequences, and each $H$-edge appears in the star corresponding to exactly one sequence.

For a sequence $S = (v_1, \dots, v_t)$, we assign to that sequence a \sg-$t$ gadget, denoted by $\cs(S)$.
When $t = 2$, if $\cn(v_1)$ and $\cn(v_2)$ are irregular, we use a \sg-2-II gadget.
Otherwise, one of the \ng gadget is irregular, and the other regular; in that case, we use a \sg-2-IR gadget.

We also construct a label $\labv_S \in \bin^{|S|}$ for each sequence $S$.
We first define $\labv_S = (0,0,0,0)$ for all sequences $S$ of length four.
Then we assign the other labels arbitrarily, as long as the following holds: For each $H$-vertex $v$, if $v$ appears at the $i$-th position in a sequence $S$ and the $i'$-th position in another sequence $S'$, then $\{\labv_{S}[i], \labv_{S'}[i']\} = \{0,1\}$.

\begin{definition}
\label{def:two_sequences_2_5}
    For every $H$-vertex $v$, $S_0(v)$ and $S_1(v)$ are sequences of $\calS$, and $i_0(v)$ and $i_1(v)$ are indices that satisfy the following:
    $v$ appears in the $i_0(v)$-th position in $S_0(v)$ and in the $i_1(v)$-th position in $S_1(v)$, and $\labv_{S_0(v)}[i_0(v)] = 0$ and $\labv_{S_1(v)}[i_1(v)] = 1$.
\end{definition}

For each sequence $S = (v_1, \dots, v_t)$ with $t \leq 3$, we define the weights of $\cs(S)$ as follows: For $i \in [t-1]$, if $\labv_S[1] = \labv_S[i+1]$, then $\sigma_{1(i+1)} = 0$ and $\delta_{1(i+1)} = -w(v_1v_{i+1})$; otherwise, $\sigma_{1(i+1)} = -w(v_1v_{i+1})$ and $\delta_{1(i+1)} = 0$.

\paragraph{Identifying PV-portals.}
It is easy to verify that if $\cn(v)$ is regular (resp., irregular), then the $i_1(v)$-side of $\cs(S_1(v))$ and the $i_2(v)$-side of $\cs(S_2(v))$ are regular (resp., irregular).
We identify $X^{\cn(v)}$ with $X^{\cs(S_1(v))}_{i_1(v)}$ and $X^{\cs(S_2(v))}_{i_2(v)}$.
If $\cn(v)$ is irregular, we additionally identify $\br{X}^{\cn(v)}$ with $\br{X}^{\cs(S_1(v))}_{i_1(v)}$ and $\br{X}^{\cs(S_2(v))}_{i_2(v)}$.

\paragraph{Identifying PP-portals.}
Consider the following sequence $E_n$ defined recursively as follows: $E_0 = v'_2v_{0,8}$ and for $i \in [n]$, $E_i = E_{i-1}v_{i,2}v_{i,8}v_{i,6}v_{i,4}$.
Note that for each $H$-vertex $v$ in $E_n$, there is exactly one sequence $S$ in $\calS$ such that $S$ contains $v$ and $\cs(S)$ contains one PP-portal; we say this PP-portal is related to $v$.
Further, for every sequence $S$ such that $\cs(S)$ contains a PP-portal, $S$ contains exactly one $H$-vertex in $E_n$.
We identify the PP-portals accordingly to the sequence $E_n$.
In particular, for $i = 0, \dots,$ we identify the PP-portal related to the $(2i+1)$-th vertex of $E_n$ with the PP-portal related to the $(2i+2)$-th vertex.

Let $G$ and $c$ be the resulting graph and weight function.

\begin{lemma}
$G$ is a simple graph.
\end{lemma}
\begin{proof}
    Similar to \cref{lem:G_simple}, we only need to argue about the portal identification steps.
    Observe that we identify triplets of PV-portals with each other, and each PV portal is only in one such triplet.
    A triplet contains a portal in a \ng gadget and a portal in two different \sg gadgets.
    Further, PP-portals are always adjacent to internal vertices, and each PP-portal is identified with exactly one other PP-portal.
    
    From the observations above, it follows that a loop cannot arise from the identification of the portals.
    Moreover, parallel edges can only occur if two PV-portals $X$ and $X'$ are adjacent in one \sg gadget, and they are identified with two adjacent PV-portals $\tilde{X}$ and $\tilde{X}'$ in another \sg gadget.
    However, this cannot happen, since two sequences of $\calS$ have at most one $H$-vertex in common.
    The lemma then follows.
\end{proof}

\section{Correctness of the reduction for \texorpdfstring{\boldmath$k = 2.5$}{k = 2.5}}
\label{sec:correctness_2_5}
\subsection{Standard tours}
A tour $\tau$ of $G$ is \defi{standard}, if the subtour of every \sg gadget in $\tau$ is a standard subtour of the gadget.
In the following lemma, recall $S_0, S_1, i_0, i_1$ as defined in \cref{def:two_sequences_2_5}.

\begin{figure}[ht!]
    \centering
    \includegraphics[width=\linewidth, page=4]{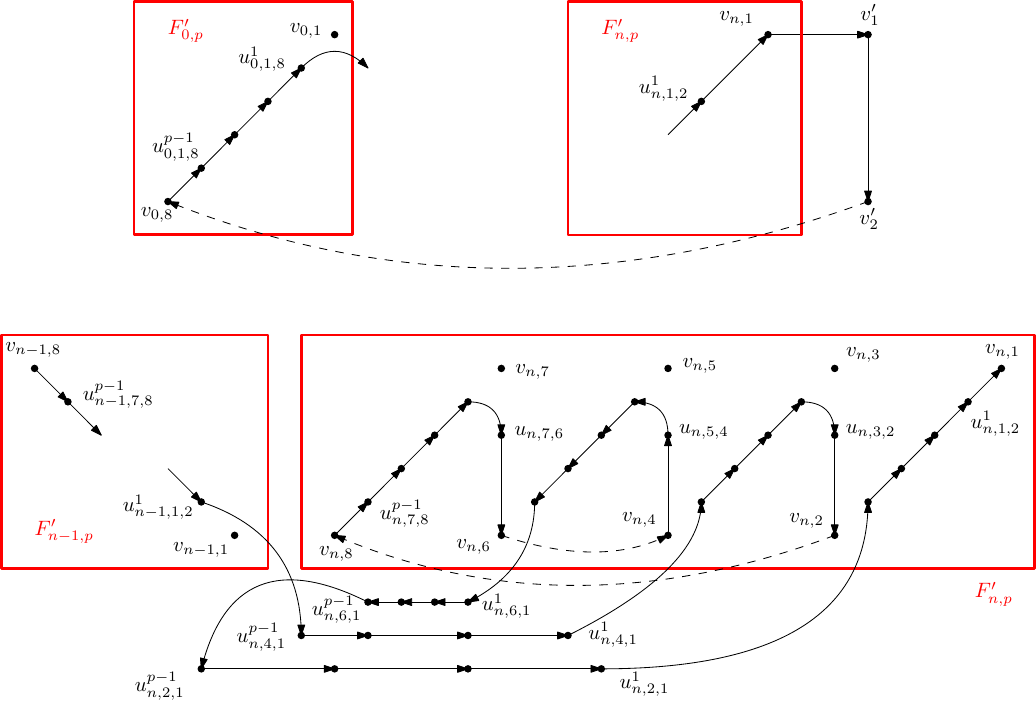}
    \caption{The cycle of $\tilde{H}$ as described in the proof of \cref{lem:standard_tour_2_5}. Solid edges and dash edges are edges of types (i) and (ii), respectively.}
    \label{fig:michel_scott_cycle}
\end{figure}

\begin{lemma}
\label{lem:standard_tour_2_5}
    $\tau$ is a standard tour of $G$, if and only if it is a subgraph of $G$ that satisfies the following properties:
    \begin{itemize}
        \item For every gadget, the induced subgraph of $\tau$ on the gadget is a standard subtour of the gadget.
        \item For every $H$-vertex $v$, suppose the \sg gadgets $\cs(S_0(v))$ and $\cs(S_1(v))$ have the $\vect{s}_0$- and $\vect{s}_1$-subtours in $\tau$, respectively.
        Then $\vect{s}_0[i_0(v)] \neq \vect{s}_1[i_1(v)]$.
    \end{itemize}
\end{lemma}
\begin{proof}
    If $\tau$ is a standard tour of $G$, then the first property is satisfied by definition.
    To see the second property, first consider the PV-portal $x$ of a regular \ng gadget.
    Since $\tau$ is a tour of $G$, $x$ has degree two in $\tau$.
    Further, since $\tau$ is a standard tour, $x$ must have degree two in one \sg gadget and degree zero in another \sg gadget.
    Next, consider the PV-portals $x$ and $\br{x}$ of an irregular \ng gadget.
    Since the internal vertex of this \ng gadget has degree two in $G$, the whole gadget has to be in $\tau$.
    Combined with the fact that $\tau$ is standard, this implies that the other incident edge of $x$ in $\tau$ is in a \sg gadget, and that of $\br{x}$ is in another \sg gadget.
    Overall, the second property in the lemma statement then follows.

    Now we show the other direction: if $\tau$ is a subgraph that satisfies the two properties, then it is a standard tour of $G$.
    Consider a \sg gadget $\cs$ in $G$ and its subtour in $\tau$.
    By assumption, this subtour is standard.
    From Lemmas~\ref{lem:star_unique_subtours}, \ref{lem:star_4_standard_subtours}, Figures~\ref{fig:star_gadgets} and~\ref{fig:star_gadget_4}, this subtour must contain a unique path of positive length. 
    This path connects either the PP-portal and a vertex of the irregular side (if $\cs$ is a \sg-1/2-IR/4 gadget) or the two irregular sides (if $\cs$ is a \sg-2-II/3 gadget).
    We call this path the \defi{primary path} of $\cs$.
    By the second property of $\tau$, observe that for every PV-portal of a regular \ng gadget, if it is isolated in a \sg gadget, then it is not isolated in the other \sg gadget that contains it.
    Further, for an irregular \ng gadget $\cn(v)$, in one of the two \sg gadgets that contain $X^{\cn(v)}$ and $\br{X}^{\cn(v)}$, $X^{\cn(v)}$ has degree one and $\br{X}^{\cn(v)}$ has degree zero, while in the other \sg gadget, $X^{\cn(v)}$ has degree zero and $\br{X}^{\cn(v)}$ has degree one.
    This implies that each PV-portal is in the primary path of exactly one \sg gadget.
    
    Motivated by the above observation, we construct the following auxiliary graph on the vertices of $H$. 
    For every \sg gadget that has two irregular sides, we add an edge between the two corresponding $H$-vertices (we call this an edge of type (i)).
    For two \sg gadgets whose PP-portals are identified, note that each of these \sg gadgets then has exactly one irregular side; we then add an edge between the $H$-vertices corresponding to these two irregular sides (we call this an edge of type (ii)).
    Let $\tilde{H}$ be the resulting graph.
    We now show that $\tilde{H}$ consists of exactly one cycle; see \cref{fig:michel_scott_cycle}.
    In particular, consider the following sequence $E'_n$ defined recursively as follows: $E'_0 = v'_2 v_{0,8} u^{p-1}_{0,1,8} \dots u^1_{0,1,8}$, and for $i \in [n]$,
    \begin{align*}
        E'_i = &E'_{i-1} u^{p-1}_{i,4,1} \dots u^1_{i,4,1} u^{p-1}_{i,3,4} \dots u^1_{i,3,4} u_{i,3,2} v_{i,2} v_{i,8} u^{p-1}_{i,7,8} \dots u^1_{i,7,8} u_{i,7,6} v_{i,6} v_{i,4} u_{i,5,4} \\ 
        &u^{1}_{i,5,6} \dots u^{p-1}_{i,5,6}u^{1}_{i,6,1} \dots u^{p-1}_{i,6,1} u^{p-1}_{i,2,1} \dots u^1_{i,2,1} u^{p-1}_{i,1,2} \dots u^1_{i,1,2}        
    \end{align*}
    Note that the sequence $E_n$ as described in Section~\ref{subsec:construction_2_5} is a subsequence of $E'_n$.
    It can then be verified by a simple induction that $E_n v_{n,1} v'_1$ is indeed a cycle in $\tilde{H}$, and there are no edges of $\tilde{H}$ outside this cycle.

    The above then implies that the primary paths of all \sg gadgets together the subtours of all irregular \ng gadgets form a cycle.
    Hence, $\tau$ is a tour.
    Since all gadgets have standard subtours in $\tau$, $\tau$ is a standard tour of $G$ as claimed.
\end{proof}

Motivated by the lemma above, we define the mapping $\phi$ that maps a standard tour $\tau$ of $G$ to a cut $\phi(\tau)$ of $H$, as follows.
For every $H$-vertex $v$, recall $S_0(v)$ and $i_1(v)$ as defined in \cref{def:two_sequences_2_5}; that is, $S_0(v)$ is the unique sequence in $\calS$ such that $v$ appears at the $i_0(v)$-th position of $S_0(v)$, and $\labv_{S_0(v)}[i_0(v)] = 0$.
Let $t := |S_0(v)|$.
Since $\tau$ is standard, there exists $\vect{s}\in\bin^t$ such that the subtour of $\cs(S_0(v))$ in $\tau$ is a $\vect{s}$-subtour.
If $\vect{s}[i_0(v)] = 1$, then $v$ is in the 1-set of the cut $\phi(\tau)$; otherwise, $v$ is in the 2-set.

Recall that $n$ is a parameter in the definition of $H$ (i.e., $H = H'_{n,p}$), and $\tilde{\epsilon}$ is a very small constant that appears in the definition of the \sg-4 gadget (see \cref{fig:star_gadget_4}).

\begin{lemma}
\label{lem:corr_obj_2_5}
    Let $\tau$ be a standard tour of $G$.
    Then the weight of $\tau$ plus the value of the cut $\phi(\tau)$ is in the range $[- n\tilde{\epsilon}, 0]$. 
\end{lemma}
\begin{proof}
    The proof of this lemma continues similar to that of the proof of \cref{lem:corr_obj}.
    In particular, each sequence in $\calS$ corresponds to a star in $H$, and every $H$-edge is contained in the star of a unique sequence in $\calS$.    
    Consider a sequence $S$ in $\calS$ of some length $t$.
    Let $A$ be the contribution of all the edges in the stars corresponding to $S$ to the value of the cut $\phi(\tau)$.
    Let $B$ be the weight from the subtour of $\cs(S)$ in $\tau$.
    If $t \leq 3$, with an analogous argument as in \cref{lem:corr_obj}, it follows that $A = -B$.
    If $t = 4$, from \cref{tab:star_gadget_4_weights}, we can see that either $A = -B$ or $A = -B - \tilde{\varepsilon}$.
    Since there are $n$ \sg-4 gadgets, the lemma then follows.
\end{proof} 

\begin{table}[ht!]
    \centering
    \begin{tabular}{c c c c c c c}
        Subtour & $v_{i-1,8}$ & $v_{i,3}$ & $v_{i,5}$ & $v_{i,7}$ & Weight of subtour & Weight of edges in cut \\
        \hline
        (a) & 0 & 1 & 1 & 1 & $-1 - \tilde{\epsilon}$ & 1 \\
        (b) & 1 & 1 & 1 & 1 & $ - \tilde{\epsilon}$ & 0 \\
        (c) & 1 & 0 & 1 & 1 & $-1 - \tilde{\epsilon}$ & 1 \\
        (d) & 0 & 0 & 1 & 1 & $ - \tilde{\epsilon}$ & 0 \\
        (e) & 0 & 0 & 0 & 1 & $-1 - \tilde{\epsilon}$ & 1 \\
        (f) & 1 & 0 & 0 & 1 & $ - \tilde{\epsilon}$ & 0 \\
        (g) & 1 & 0 & 0 & 0 & $-1$ & 1 \\
        (h) & 0 & 0 & 0 & 0 & 0 & 0 \\
        (i) & 0 & 1 & 0 & 0 & $-1$ & 1 \\
        (j) & 1 & 1 & 0 & 0 & 0 & 0 \\
        (k) & 1 & 1 & 1 & 0 & $-1$ & 1 \\
        (l) & 0 & 1 & 1 & 0 & 0 & 0 \\
        (m) & 0 & 1 & 1 & 1 & $-1$ & 1 \\
        (n) & 1 & 1 & 1 & 1 & 0 & 0
    \end{tabular}
    \caption{Comparison of the weight of each subtour of a \sg-4 gadget attached to the sequence $(v_{i-1,8}, v_{i,3}, v_{i,5}, v_{i,7})$ and the weight of the corresponding star in the cut. The labels of the subtours refer to that in \cref{fig:star_gadget_4}.}
    \label{tab:star_gadget_4_weights}
\end{table}

\begin{lemma}
\label{lem:corr_nbr_2_5}
    Let $\tau$ and $\tau'$ be tours of $G$ such that $\tau$ is standard, and $\tau'$ is obtained from $\tau$ by a 2.5-swap.
    If $H$ has girth at least nine, then $\tau'$ is also standard, and $\phi(\tau)$ and $\phi(\tau')$ either are the same or differ by a flip.
\end{lemma}
\begin{proof}
    Suppose $\tau'$ is obtained from $\tau$ by a 2-swap.
    This implies that there is a 4-cycle $K$ in $G$ such that $\tau$ contains exactly two non-adjacent edges of this cycle.
    Note that the edges in an irregular \ng gadget have to be in any tour of $G$, and hence they cannot be part of this 4-cycle $K$.
    
    Consider the case when $K$ lies entirely in a \sg gadget $\cs$.
    If $\cs$ is a \sg-2-IR/2-II/3 gadget, then $K$ must contain two vertices $x$ and $\br{x}$ of an irregular side of the gadget.
    This implies that both $x$ and $\br{x}$ have degree one in the subtour of the gadget in $\tau$, a contradiction to $\tau$ being a standard tour.
    If $\cs$ is a \sg-4 gadget, then in order to avoid the same contradiction, $K$ must not contain both $X^{\cs}_1$ and $\br{X}^{\cs}_1$.
    The only such 4-cycles are $(X^{\cs}_1, X^{\cs}_2, Y^{\cs}, X^{\cs}_4)$ and $(\br{X}^{\cs}_1, X^{\cs}_2, Y^{\cs}, X^{\cs}_4)$.
    In these cases, after performing the 2-swap, $\tau'$ satisfies the properties in \cref{lem:standard_tour_2_5} and hence is a standard tour of $G$.
    Further, note that $\phi(\tau')$ is then the same as $\phi(\tau)$.
    
    Now consider the case when $K$ lies in at least two \sg gadgets.
    Note that a common vertex of two \sg gadgets has to be a PV-portal.
    If $K$ contains no two PV-portals of the same irregular \ng gadget, then the $K$ corresponds to a cyclic sequence of \sg gadgets $(\cs_1, \dots, \cs_r)$ for some $r \in \{2,3,4\}$, such that any two adjacent \sg gadgets in the sequence share a common PV-portal.
    However, this implies a cycle of at most $2r \leq 8$ vertices in $H$, a contradiction to the assumption that $H$ has girth at least nine.
    Hence, $K$ must contain the PV-portals $x$ and $\br{x}$ of an irregular \ng gadget.
    The 4-cycle $K$ is then $(x, y, \br{x}, y')$, where $x, y$, and $\br{x}$ are in a \sg gadget, and $x, y', \br{x}$ are in another \sg gadget.
    In that case, after performing the 2-swap, $\tau'$ satisfies the properties in \cref{lem:standard_tour_2_5} and hence is a standard tour of $G$.
    Further, $\phi(\tau')$ is also obtained from $\phi(\tau)$ by flipping the $H$-vertex corresponding to the irregular \ng gadget that contains $x$ and $\br{x}$.
    
    Now suppose $\tau'$ is obtained from $\tau$ by a 2.5-swap that is not a 2-swap.
    This means there are two triangles $xyz$ and $xy'z'$ in $G$, such that $\tau$ contains two edges in one triangle and one edge in the other.
    If one of these triangles lies in more than one \sg gadget, then similar to the arguments above, this triangle must either correspond to a cycle of length at most six in $H$ or contain two PV-portals of a \ng gadget. 
    The former case contradicts the assumption on the girth of $H$, while the latter cannot occur, since such two portals cannot be connected by an edge in any gadget, and hence they cannot be part of a triangle in $G$.
    Therefore, each of $xyz$ and $xy'z'$ lies entirely in a \sg gadget.

    Now suppose both triangles lie in the same \sg gadget.
    Then this gadget has to be a \sg-3 or -4 gadget $\cs$.
    If $\cs$ is a \sg-3 gadget, then $x = X^{\cs}_1$.
    Since $\tau$ contains two edges in one triangle and one edge in the other, it follows that the two vertices of an irregular side must have degree one in $\tau$, a contradiction to $\tau$ being standard.
    If $\cs$ is a \sg-4 gadget, then we note that all the triangles in $\cs$ are $X^{\cs}_1X^{\cs}_2X^{\cs}_3$, $X^{\cs}_1X^{\cs}_2Y^{\cs}$, $X^{\cs}_1X^{\cs}_3X^{\cs}_4$, $X^{\cs}_1X^{\cs}_4Y^{\cs}$, $\br{X}^{\cs}_1X^{\cs}_2X^{\cs}_3$, $\br{X}^{\cs}_1X^{\cs}_2Y^{\cs}$, $\br{X}^{\cs}_1X^{\cs}_3X^{\cs}_4$, and $\br{X}^{\cs}_1X^{\cs}_4Y^{\cs}$.
    Therefore, we must have that either $x \in \{X^{\cs}_1, \br{X}^{\cs}_1\}$ or $X^{\cs}_1$ is in a triangle and $\br{X}^{\cs}_1$ is in the other.
    In the former case, since $x$ must have degree zero or two in $\tau$, we then have a contradiction to $\tau$ being standard.
    In the latter case, we then have that both $X^{\cs}_1$ and $\br{X}^{\cs}_1$ have degree one in $\tau$, again a contradiction.

    Hence, the two triangles lie in different \sg gadgets.
    Since two \sg gadget cannot share an edge, there are exactly two \sg gadgets, and $x$ is the only vertex shared by the two gadgets.
    So $x$ must be a PV-portal.
    Since $x$ has degree two in one of the triangles, and since $\tau$ is standard, $x$ cannot be part of an irregular side.
    In other words, $x$ is the PV-portal of a regular \ng gadget.
    It is now easy to see that after performing the 2.5-swap, $\tau'$ satisfies the properties in \cref{lem:standard_tour_2_5} and is hence a standard tour of $G$.
    Further, $\phi(\tau')$ can be obtained from $\phi(\tau)$ by flipping the $H$-vertex corresponding to the \ng gadget containing $x$.
\end{proof}

\begin{corollary}
\label{cor:corr}
    Let $\tau$ and $\tau'$ be tours of $G$ such that $\tau$ is standard, and $\tau'$ is obtained from $\tau$ by a 2.5-swap.
    Then it holds that
    \begin{enumerate}[(a)]
        \item If there is an improving flip from $\phi(\tau)$ to $\phi(\tau')$, then the swap from $\tau$ to $\tau'$ is also improving;
        \item If the swap from $\tau$ to $\tau'$ is improving, and if $\phi(\tau)$ and $\phi(\tau')$ are distinct, then the flip from $\phi(\tau)$ to $\phi(\tau')$ is improving.
    \end{enumerate}
\end{corollary}
\begin{proof}
    The statements are direct consequences of Lemmas~\ref{lem:corr_obj_2_5}, \ref{lem:corr_nbr_2_5}, and the fact that $n\tilde{\epsilon}$ is less than the increase in value for any improving flip in $H$.
\end{proof}

\subsection{The all-exp property for \texorpdfstring{\boldmath$k = 2.5$}{k = 2.5}}

We define the \defi{initial tour} $\tau_0$ as the following subgraph of $G$ constructed from $(V(G),\emptyset)$ as follows.
For every irregular \ng gadget, we add the subtour of the gadget to $\tau_0$.
For every \sg-4 gadget, we add subtour (a) as depicted in \cref{fig:star_gadget_4} to $\tau_0$.
Next, consider a sequence $S = (v_1, \dots, v_t)$ in $\calS$ for some $t \leq 3$.
Let $\vect{s}$ be a vector in $\bin^t$ such that for $i \in [t]$, $\vect{s}[i] = 0$ if and only if either $v_i$ is in the 2-set of $\gamma_{n,p}$ and $\labv_S[i] = 0$ or $v_i$ is in the 1-set of $\gamma_{n,p}$ and $\labv_S[i] = 1$.
Then for the \sg gadget $\cs(S)$, we add to $\tau_0$ the $\vect{s}$-subtour, which is unique by \cref{lem:star_unique_subtours}.
By \cref{lem:standard_tour_2_5}, the initial tour $\tau_0$ is then a standard tour of $G$.

\thmmaintwo*

\begin{proof}
    We choose $p \geq 7$ so that by \cref{lem:girth_2_5}, $H$ has girth at least ten.
    Hence, the assumption required for \cref{lem:corr_nbr_2_5} is satisfied.
    
    Let $R_r = (\tau_0, \tau_1, \dots, \tau_{r})$ be a (not necessarily maximal) improving swap sequence of $G$ from the initial tour $\tau_0$.
    By \cref{lem:corr_nbr_2_5}, $\tau_i$ is a standard subtour of $G$ for $i \in [r]$.
    Denote by $\phi(R_r)$ the sequence obtained from $\big(\phi(\tau_0), \phi(\tau_1), \dots, \phi(\tau_r)\big)$ by removing consecutive identical elements. 
    
    Firstly, we prove by induction on $r$ that $\phi(R_r)$ is an improving flip sequence of $H$.
    For the base case $r = 0$, $\big(\phi(\tau_0)\big)$ is trivially an improving flip sequence of $H$.
    For the inductive step from $r-1$ to $r$, we know by the inductive hypothesis that $R_{r-1}$ is an improving flip sequence of $H$.
    By \cref{lem:corr_nbr_2_5}, since $\tau_{r}$ is obtained from $\tau_{r-1}$ by an improving 2.5-swap, $\phi(\tau_r)$ and $\phi(\tau_{r-1})$ either are the same or differ by a flip.
    If they are the same, then $R_{r} = R_{r-1}$, and we are done.
    Otherwise, since the swap from $\tau_{r-1}$ to $\tau_{r}$ is improving, \cref{cor:corr} implies that the flip from $\phi(\tau_{r-1})$ to $\phi(\tau_r)$ is also improving.
    Hence, $R_r$ is an improving flip sequence of $H$.

    Secondly, we have the following claim.
    
    \begin{claim}
        If there is an improving flip from $\phi(\tau_r)$, then there is also an improving swap from $\tau_r$. 
    \end{claim} 
    \begin{claimproof}        
    Suppose there is an improving flip of an $H$-vertex $v$ from $\phi(\tau_r)$ to obtain a cut $\gamma'$.
    Recall $S_0(v), S_1(v), i_0(v)$, and $i_1(v)$ from \cref{def:two_sequences_2_5}.
    Since $\tau_r$ is standard, the subtours of $\cs(S_0(v))$ and $\cs(S_1(v))$ in $\tau_r$ are some $\vect{s}_0$- and $\vect{s}_1$-subtours, respectively.
    Let $\vect{s}'_0$ and $\vect{s}'_1$ be the vectors obtained from $\vect{s}_0$ and $\vect{s}_1$ by flipping the $i_0(v)$-th and $i_1(v)$-th coordinates, respectively (by flipping, we mean the operation of switching between 0 and 1). 
    If $v \notin \{ v_{i-1,8}, v_{i,3}, v_{i,5}, v_{i,7} \mid i \in [n]\}$, then let $\tau'$ be the subgraph of $G$ obtained from $\tau_r$ by replacing the subtours of $\cs(S_0(v))$ and $\cs(S_1(v))$ by the $\vect{s}'_0$- and $\vect{s}'_1$-subtours, respectively.
    By \cref{lem:star_involved_edges}, \cref{cor:corr}, and the assumption that flipping $v$ improves $\phi(\tau_r)$, it follows that $\tau'$ is well-defined and can be obtained from $\tau_r$ by an improving 2.5-swap.
    It remains to argue for the case of $v \in \{ v_{i-1,8}, v_{i,3}, v_{i,5}, v_{i,7}\}$ for some $i \in [n]$.

    Suppose that $v = v_{i-1,8}$ for some $i \in [n]$.
    Then $\cs(S_0(v))$ is a \sg-4 gadget and $\cs(S_1(v))$ is a \sg-2-II gadget.
    Since $\labv_{\cs(S_0(v))} = (0,0,0,0)$, $\vect{s}_0$ is also the state of $(v_{i-1,8}, v_{i,3}, v_{i,5}, v_{i,7})$ in $\phi(\tau_r)$.
    As shown above, the sequence $R_r$ is an improving flip sequence of $H$ from $\phi(\tau_0) = \gamma_{n,p}$.
    Hence, by \cref{obs:states}, the state $\vect{s}_0$ is in the cyclic sequence $\Pi$, and in the cut $\gamma'$ (i.e., the state obtained from $\phi(\tau_r)$ by flipping $v$), the state $\vect{s}'_0$ of $(v_{i-1,8}, v_{i,3}, v_{i,5}, v_{i,7})$ is the successor of $\vect{s}_0$ in $\Pi$. 
    Recall that $\vect{s}'_0$ and $\vect{s}_0$ differ by a flip of the first coordinate (i.e., the coordinate corresponds to $v_{i-1,8}$).
    Hence, $\vect{s}_0$ cannot be $(1,1,1,1)$, since its successor in $\Pi$, $(1,0,1,1)$, differs from it at the second coordinate.
    Therefore, the $\vect{s}_0$-subtour of $\cs(S_0(v))$ in $\tau_r$ cannot be the subtour (n) in \cref{fig:star_gadget_4}.
    Therefore, by \cref{obs:star_four_involved_edges}, there exists a $\vect{s}'_0$-subtour of $\cs(S_0(v))$ that differ by the $\vect{s}_0$-subtour in $\tau_r$ by a 2- or 2.5-change; more precisely, this has to be a 2-change at the 1-side, since $v = v_{i-1,8}$.
    Since $\cs(S_1(v))$ is a \sg-2-II gadget, by \cref{lem:star_involved_edges}, its $\vect{s}'_1$-subtour differs its $\vect{s}_1$-subtour by a 2-change.
    Let $\tau'$ be the subgraph of $G$ obtained from $\tau_r$ by replacing the subtours above for $\cs(S_0(v))$ and $\cs(S_1(v))$.
    Then $\tau'$ can be obtained from $\tau$ by a 2-swap, and we can see that $\phi(\tau') = \gamma'$.
    Hence, by \cref{cor:corr}, the 2-swap from $\tau$ to $\tau'$ is improving.

    The case when $v \in \{v_{i,3}, v_{i,5}, v_{i,7}\}$ for some $i \in [n]$ is almost analogous.
    In particular, $\cs(S_1(v))$ is a \sg-4 gadget and $\cs(S_0(v))$ is a \sg-3 gadget. 
    By a similar argument as above, if the $\vect{s}_1$-subtour of $\cs(S_1(v))$ in $\tau_r$ is not subtour (n), then we can obtain a tour $\tau'$ of $G$ from $\tau_r$ by performing a 2.5-swap, such that $\phi(\tau') = \gamma'$.
    By \cref{cor:corr}, this 2.5-swap is improving.
    If the $\vect{s}_1$-subtour is subtour (n), then we can change this subtour to subtour (b) by a 2-swap to obtain a standard tour $\tau''$ of $G$ with slightly less weight (note that in this case $\phi(\tau'') \equiv \phi(\tau_r)$); in other words, there is also an improving 2-swap from $\tau_r$. 
    This completes the proof of the claim.
    \end{claimproof}

    Lastly, the above two points imply that for every maximal improving swap sequence from $\tau_0$, there is a corresponding maximal improving flip sequence from $\phi(\tau_0) = \gamma_{n,p}$.
    Combined with \cref{lem:michel_scott_2_5}, this means that every maximal improving swap sequence from $\tau_0$ is exponential in $n$.
    The theorem then follows.
\end{proof}

\section{Conclusion}
\label{sec:conclusion}
We have shown that for $k=3$ and $k=4$ the \kopt algorithm for the traveling salesman problem has the all-exp property, thereby resolving the runtime complexity for two of its three remaining open cases. 
In doing so, we also provided a simpler construction for the previously known cases $k\geq 5$. 

Additionally, we presented a new construction establishing the all‑exp property for the 2.5‑\opt algorithm.
 A key novelty here is the introduction of irregular gadgets. 
The proofs in \cite{HHH2024} and ours employ a reduction from Max-Cut/Flip restricted to certain instances with maximum degree four. Each \kopt step in the constructed TSP-instances simulates flipping a vertex in a Max-Cut instance.  
The irregular gadgets simulate flips in a new way that reduces 
the required number of exchanged edges from three to two.

The only remaining open case is $k=2$. Although we are able to simulate flipping some vertices (including degree-four vertices) via a $2$-swap in the $2.5$-\opt construction, there seems to be no easy way to extend the construction to $k=2$. 
The main issue is that we would need to use irregular \ng gadgets for all vertices.
Consequently, some \sg gadgets would have to be related to more than two irregular \ng gadgets.
This in turn creates many obstacles.
For example, we would lose the property that there is only one ``primary" path (i.e., a path of positive length) in every subtour of a \sg gadget, since there are at least three vertices with degree one in a subtour.
If the primary paths in a \sg gadget never change the sides that contain their endpoints, then either they would become ``independent" of each other, limiting our modeling capacity, or we would need more than a 2-swap to simulate the complex interaction of three or more vertices.
If the paths may change these sides, then a $k$-\opt step may lead to multiple disconnected subtours instead of a single spanning cycle.
Hence, we would need a more ``global" argument to ensure the connectivity of the tour.
Overall, a construction for $k = 2$ would require additional new ideas.

\bibliographystyle{plain}
\bibliography{refs}

\end{document}